\documentclass[namedreferences]{solarphysics}

\usepackage[hyperref,optionalrh]{spr-sola-addons} 
\usepackage{graphicx}        
\usepackage{color}           
\usepackage{breakurl}        

\usepackage{epstopdf}

\usepackage{amssymb}
\usepackage{eurosym}
\usepackage{color}
\usepackage{bm}
\usepackage{booktabs}
\usepackage{multirow}
\usepackage{longtable}
\usepackage{arydshln}
\usepackage{algorithmic}
\usepackage{algorithm}
\usepackage[font=scriptsize]{caption}
\usepackage[utf8]{inputenc}
\usepackage[english]{babel}
\usepackage[caption=false]{subfig}
\usepackage[]{subfig}

\usepackage{hyperref}
\hypersetup{
    colorlinks=true,
    linkcolor=blue,
    filecolor=magenta,      
    urlcolor=cyan,
}

\usepackage{lscape}

\usepackage{savesym}
\savesymbol{iint}
\savesymbol{iiint}
\usepackage{amsmath}

\usepackage{enumerate}
\newcommand{\ve}[1]{\mbox{\boldmath $#1$}}
\newenvironment{ltable}
{\begin{landscape}\begin{table}}
{\end{table}\end{landscape}}

\chardef\us=`\_

\begin{document}

\begin{article}
\begin{opening}

\title{Forecasting Solar Flares Using Magnetogram-based Predictors and Machine Learning} 

\author[addressref={aff1,aff2},corref,email={cflorios@aueb.gr}]{\inits{K.}\fnm{Kostas}~\lnm{Florios}\orcid{0000-0002-8210-1125}}
\author[addressref=aff1,email={jkonto@noa.gr}]{\inits{I.}\fnm{Ioannis}~\lnm{Kontogiannis}\orcid{0000-0002-3694-4527}}
\author[addressref=aff3,email={sunpark@tcd.ie}]{\inits{S-H.}\fnm{Sung-Hong}~\lnm{Park}\orcid{0000-0001-9149-6547}}
\author[addressref=aff3,email={guerraaj@tcd.ie}]{\inits{J.A.}\fnm{Jordan A. }~\lnm{Guerra}\orcid{0000-0001-8819-9648}}
\author[addressref=aff4,email={benvenuto@dima.unige.it}]{\inits{F.}\fnm{Federico}~\lnm{Benvenuto}\orcid{0000-0002-4776-0256}}
\author[addressref={aff5},email={shaun.bloomfield@northumbria.ac.uk}]{\inits{D.S.}\fnm{D. Shaun}~\lnm{Bloomfield}\orcid{0000-0002-4183-9895}}
\author[addressref=aff1,email={manolis.georgoulis@academyofathens.gr}]{\inits{M.K.}\fnm{Manolis K.}~\lnm{Georgoulis}\orcid{0000-0001-6913-1330}}

\address[id=aff1]{Research Center for Astronomy and Applied Mathematics, Academy of Athens, Greece}
\address[id=aff2]{Department of Statistics, Athens University of Economics and Business, Greece}
\address[id=aff3]{School of Physics, Trinity College Dublin, Ireland}
\address[id=aff4]{Dipartimento di Matematica, Universit\`a di Genova, Italy}
\address[id=aff5]{Northumbria University, Newcastle upon Tyne, NE1 8ST, UK}

\runningauthor{K. Florios \textit{et al.}}
\runningtitle{Forecasting Solar Flares Using Machine Learning}

\begin{abstract}
We propose a forecasting approach for solar flares based on data from Solar Cycle 24, taken by the \textit{Helioseismic and Magnetic Imager} (HMI) on board the \textit{Solar Dynamics Observatory} (SDO) mission. In particular, we use the Space-weather HMI Active Region Patches (SHARP) product that facilitates cut-out magnetograms of solar active regions (AR) in the Sun in near-realtime (NRT), taken over a five-year interval (2012 -- 2016). Our approach utilizes a set of thirteen predictors, which are not included in the SHARP metadata, extracted from line-of-sight and vector photospheric magnetograms. We exploit several Machine Learning (ML) and Conventional Statistics techniques to predict flares of peak magnitude $>$M1 and $>$C1, within a 24\,h forecast window. The ML methods used are multi-layer perceptrons (MLP), support vector machines (SVM) and random forests (RF). We conclude that random forests could be the prediction technique of choice for our sample, with the second best method being multi-layer perceptrons, subject to an entropy objective function. A Monte Carlo simulation showed that the best performing method gives accuracy {\rm ACC}=0.93(0.00), true skill statistic {\rm TSS}=0.74(0.02) and Heidke skill score {\rm HSS}=0.49(0.01) for $>$M1 flare prediction with probability threshold 15\% and {\rm ACC}=0.84(0.00), {\rm TSS}=0.60(0.01) and {\rm HSS}=0.59(0.01) for $>$C1 flare prediction with probability threshold 35\%.
\end{abstract}
\end{opening}


\section{Introduction}
     \label{S-Introduction} 

 \noindent Solar flares are sudden brightenings that occur in the solar atmosphere and release enormous amounts of energy, over the entire electromagnetic spectrum. Flares are quite prominent in X-rays, UV, and optical lines \citep{fletcher:2011} and they are often (but not always) accompanied by eruptions that eject solar coronal plasma into the interplanetary space (coronal mass ejections, CMEs). These very intense phenomena - the largest explosions in the solar system - are associated with regions of enhanced magnetic field, called active regions (AR) and are associated, in white light, with sunspot groups. Depending on their peak X-ray intensity, as recorded by the National Oceanic and Atmospheric Administration's (NOAA) \textit{Geostationary Operational Environmental Satellite} (GOES) system, flares are categorized in classes, the strongest and most important being X, M and C (in decreasing order). Flare classification is logarithmic, with a base of 10, and is complemented by decimal sub-classes ({\it e.g.} M5.0, C3.2 {\it etc.}).

The solar flare radiation may be detrimental to infrastructures, instruments and personnel in space, therefore flare forecasting is an integral part of contemporary space-weather forecasting. Forecast mainly employs measurements of the AR magnetic field in the solar photosphere. Magnetic-field-based predictors represent AR magnetic complexity or the energy budget available to power flares. Recent developments in instrumentation have led to a regular production of such measurements offering the opportunity to produce extensive databases with properties suitable for solar flare prediction.

On the other hand, machine learning in recent years has become an increasingly popular approach for performing computer cognition tasks which were inherently possible only using human intelligence. Thus, machine learning (ML) is a subfield of artificial intelligence (AI) and it aims at using past data in order to train computers so that they can apply the accumulated knowledge to new, previously unseen, data. The acquisition of knowledge is the training phase and the application of what was learned to future scenarios is the prediction phase. Typically, ML is more interested in prediction than conventional statistics. ML can also interface with conventional statistics in a field called statistical learning \citep{Hastie:09}. Learning is either called supervised or unsupervised depending on whether it is done with a teacher or not. Supervised learning comprises regression and classification, while unsupervised learning is also called clustering. In our study, we focus on classification, where a set of input variables or predictors belongs to one of two classes (binary classification). ML is more powerful than traditional statistical techniques such as, say, generalized linear models that include probit, logit, {\it etc.} for binary classification, because it can help model more complex nonlinear relationships. An introduction to ML research can be found in several textbooks \citep{MacKay:03,Hastie:09}.

Several researchers have recently used ML techniques to effectively forecast solar flares. More often, the techniques used by researchers were: neural networks \citep{Wang:08,Yu:09,Colak:09,Ahmed:13}, support vector machines  \citep{Li:08,Yuan:10,Bobra:15,Boucheron:15}, ordinal logistic regression \citep{Song:09}, decision trees \citep{Yu:09} and relevance vector machines \citep{Al-Ghraibah:15}. Very recently, random forests have also been used \citep{Barnes:16,Liu:17}.

We use predictors calculated from near-realtime (NRT) Space-weather HMI Active Region Patches (SHARP) data combined with state-of-the-art ML and statistical algorithms in order to effectively forecast flare events for an arbitrarily chosen 24-hour forecast window. Flare magnitudes of interest are $>$M1 and $>$C1. Prediction is binary, meaning that a given flare class is considered to either happen or not within the next 24 hours after prediction. Our predictions are effective immediately, therefore with zero latency. Analysis involves a comprehensive NRT SHARP sample including all calendar days between years 2012 and 2016, at a cadence of 3 hours. Results in this work summarize the findings of the first eighteen months of the ``Flare Likelihood And Region Eruption foreCASTing'' (FLARECAST) project and, while based on ongoing work, we took every effort to present robust and unbiased results. 

The contribution of the present work is twofold:
    \begin{itemize}
        \item The utilization of novel magnetogram-based predictors in a multi-parameter solar flare prediction model. \vspace{0.1cm}
        \item The utilization of classic and novel ML techniques, such as multi-layer perceptrons (MLP), support vector machines (SVM) and especially, for one of the first times\footnote{ In June 2017, we noticed a manuscript by \cite{Liu:17} which also uses the random forest algorithm for solar flare prediction using SDO/HMI data. Nevertheless, the specific details in that paper regarding the sampling strategy and the feature extraction are very different from our choices. For example, in \cite{Liu:17} only flaring ARs (at the level $>$B1 class) were considered and the sample size was N=845, while in our paper we consider both flaring and non-flaring ARs with N=23,134.}, random forests (RF), for the forecasting of $>$M1 and $>$C1 flares.
    \end{itemize}

\noindent For the interested reader, the application code is available at \url{http://dx.doi.org/10.17632/4f6z2gf5d6.1}, along with the benchmark dataset used in this work. The run time for all methods is of the order of few minutes. 

The analysis presented here is part of the EU Horizon 2020 FLARECAST project, aiming to develop a NRT online forecasting system for solar flares. The study is organized as follows: Section 2 describes the data selected to train and test the algorithms and presents the predictors used, together with background information on the solar physics aspects of magnetogram-based calculations. Section 3 describes the ML algorithms in terms of their core principles, along with some additional remarks and comments. Section 4 is devoted to the forecast experiments and a comparison with similar published results and statistics. Section 5 presents the main conclusions and future integration of the present work in the FLARECAST operational system. Four Appendices, describing multiple complementary aspects of this work are also included.

\section{Data and Classification Predictors}

\subsection{Data}
\label{sec:data}

The \textit{Helioseismic and  Magnetic Imager} \citep[HMI;][]{hmi:2012} on board the \textit{Solar Dynamics Observatory} \citep[SDO;][]{sdo:2012}, provides regular full-disk solar observations of the three components of the photospheric magnetic field. The HMI team has created the Space Weather HMI Active Region Patches (SHARPs), which are cut-outs of solar regions-of-interest along with a set of parameters potentially useful for solar flare prediction \citep{Bobra:14}. For our analysis, we use the near-realtime (NRT), cylindrical equal area (CEA) SHARP data to calculate a set of predictors.

To associate SHARPs with flare occurrence we use the \textit{Geostationary Operational Environmental Satellite} (GOES) soft X-ray measurements. For each SHARP we search for flares within the next 24 hours by either matching the NOAA AR numbers with those of the recorded flares or by comparing the corresponding longitude and latitude ranges, considering also the differential solar rotation.
 
The algorithms of Section \ref{sec:algorithms} are tested on a sample of the 2012-2016 SHARP dataset. We consider all days in the period October 1, 2012 to January 13, 2016 and for every given day we compute the set of predictors (see Section~\ref{sec:predictors}) at a cadence of 3 hours, starting at 00:00\,UT. For our analysis, only SHARP cut-outs that correspond to NOAA ARs are considered. In this way, we get a fairly representative sample of the solar activity including several flares of interest, with a sufficiently high sampling frequency.

\subsection{Predictors}
\label{sec:predictors}

\begin{figure}
         \includegraphics[width=12cm]{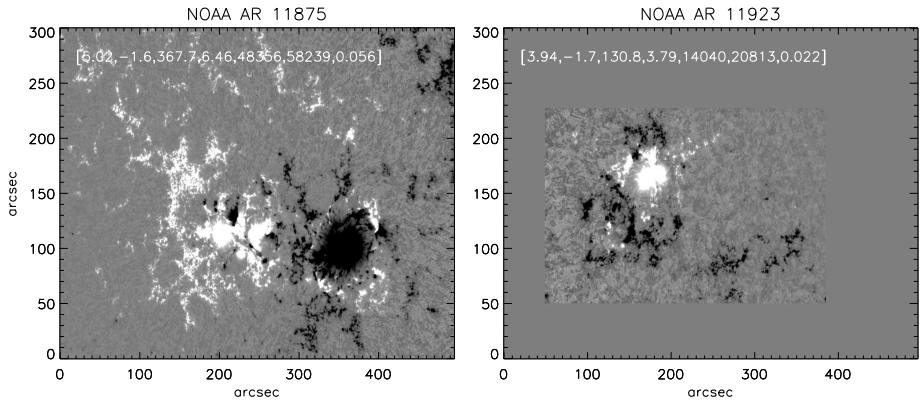}
\caption{{\footnotesize Two SHARP frames depicting AR with very different levels of flaring activity. NOAA AR\,11875 (left) produced 7 C-, 0 M- and 0 X-class flares within 24h while NOAA AR\,11923 (right) produced no flares. The two AR are scaled so as to retain their original relative size and, for comparison, vectors of the seven predictors used are included in the frames. The names of all $K=7$ predictors [{\rm logR}, {\rm FSPI}, {\rm TLMPIL}, {\rm DI}, ${\rm WL}_{{\rm SG}}$, IsinEn1, IsinEn2] are defined in Section \ref{sec:predictors}. High values of the predictors statistically indicate a powerful AR (left), with low values indicating a quiescent, flare-quiet AR (right).
 }}
\label{fig:ars2}       
\end{figure}

The set of thirteen predictors consists of both predictors already proposed in the literature and new ones, and comprises a subset of the parameter set developed for the FLARECAST project. In Figure~\ref{fig:ars2} we show two sample magnetograms to demonstrate how the predictors reflect the complexity and size of the corresponding active region. The predictors utilized for this study are the following:

\subsubsection{Magnetic Polarity Inversion Line (TLMPIL)} A magnetic polarity inversion line (MPIL) in the photosphere of an AR separates distinct patches of positive- and negative-polarity magnetic  flux. Several studies have been carried out to investigate the relationship between flare occurrence and MPIL characteristics \citep{Schrijver:2007,Falconer:2012}. We determine a specific subset of a MPIL, that has been also identified as MPIL*, with i) a strong gradient in the vertical component of the field across the MPIL and ii) a strong horizontal component of the field around the MPIL. MPIL* has been considered as the single most likely place in AR where potential magnetic instabilities, such as, say, magnetic flux cancellation and/or magnetic flux rope formation \citep{Fang:2012} can take place. Such processes seem intimately related to flares. We use the total length $L_{\rm tot}$ of MPIL* segments in active regions as an MPIL quantification parameter.

\subsubsection{Decay Index (DI)} The decay index is a quantitative measure for the torus magnetic instability in a current-carrying magnetic flux rope \citep{Kliem:2006}. It has been found that the larger the value of decay index in AR magnetic fields, the more likely it is to obtain a solar eruption involving a major solar flare \citep{Zuccarello:2015}. We developed a decay index parameter derived by the ratio $L_{{\rm hs}}$/$h_{{\rm min}}$, where $L_{{\rm hs}}$ is the length of a highly sheared portion of a MPIL and $h_{{\rm min}}$ is the minimum height at which the decay index achieves a purported critical value of 1.5. This ratio can be used to measure the degree of instability in a flux rope. Notice that if there are more than one MPIL in an AR, then we calculate the ratio $L_{{\rm hs}}$/$h_{{\rm min}}$ for every MPIL and take the peak value for a given time, that represents the highest eruptive potential of the AR.

\subsubsection{Gradient-weighted integral length of the neutral line $ (WL_{ SG})$} The gradient-weighted integral length of neutral line, ${\rm WL}_{{\rm SG}}$, is defined in \cite{2008ApJ...689.1433F} as,

\begin{equation}
{\rm WL}_{{\rm SG}} = \int (\nabla B_{z}){\rm d}l \,,
\end{equation}

\noindent
and corresponds to the line integral of the vertical-field ($B_{z}$) horizontal gradient over all neutral line (or MPIL) segments on which the potential horizontal field is greater than 150 G. This MPIL-related property has been reported to show a useful empirical association with the occurrence of solar eruptions \citep[flares, CMEs, SPEs;][]{SWE:SWE372,SWE:SWE20132} and is the main predictor used in the Magnetic Forecast  (MAG4) forecasting service, developed in the University of Alabama (\url{http://www.uah.edu/cspar/research/mag4-page}).

For these calculations of ${\rm WL}_{\rm {SG}}$, two approximations of the vertical field $B_{z}$ are used: $B_{\rm los}$ (line of sight; uncorrected) and $B_{r}$, keeping in mind that in former case, only values for regions located within $30^o$  from the central meridian are considered accurate. For each magnetogram, a MPIL mask is determined as in the calculation of MPIL characteristics, described previously. In order to select the strong-horizontal field segments of MPILs, the potential field extrapolation method developed by \cite{1981A&A...100..197A} is used. Finally, the horizontal gradient of $B_{z}$ is calculated numerically and integrated over all MPIL segments. The accuracy of the calculated values was estimated by comparing flare rates derived from our calculations of ${\rm WL}_{{\rm SG}}$ \citep[using Equation 4 along with Table 1 values in][]{SWE:SWE372} with the flare rates from the text output of MAG4.

\subsubsection{Ising Energy (IsinEn1, IsinEn2)} The Ising energy is a quantity that parameterizes the magnetic complexity of an AR \citep{ahmed2010}. For a two-dimensional distribution of positive and negative interacting magnetic elements, the Ising energy is defined as,

\begin{equation}
E_{{\rm Ising}}=-\sum_{ij}\frac{S_{i}S_{j}}{d^{2}} \,,
\label{ising}
\end{equation}

\noindent where $S_{i}$ ($S_{j}$) equals to +1 (-1) for positive (negative) pixels and $d$ is the distance between opposite polarity pairs. The interacting magnetic elements can be either the individual pixels with a minimum flux density value as in \cite{ahmed2010} or the opposite-polarity partitions, produced using a flux-partitioning scheme \citep{barnes05}. The latter variation is introduced for the first time in the FLARECAST project, with promising results and  an assessment of its merit as a predictor is underway (Kontogiannis \textit{et al.}, in preparation). The Ising energy calculation produces four predictors, two for the line-of-sight magnetic field and two for the radial magnetic field component.

\subsubsection{Fourier Spectral Power Index (FSPI)} The spectral power index, $\alpha$, corresponds to the power-law exponent in fitting the one-dimensional power spectral density $E(k)$ extracted from magnetograms by the relation,

\begin{equation}
E(k)\sim k^{-\alpha} \,.
\end{equation}

\noindent This index parameterizes the power contained in magnetic structures of spatial scales $l$ (= $k^{-1}$) belonging to the inertial range of magnetohydrodynamic (MHD) turbulence. Empirically, AR with spectral power index higher than 5/3 (Kolmogorov's exponent for turbulence) are thought to display an overall high productivity of flares \citep[\textit{e.g.} see][]{Guerra:2015}.

The spectral power index has been historically calculated from the vertical component of the photospheric magnetic field, as inferred from the line-of-sight component assuming perfectly radial magnetic fields. First, the magnetogram is processed using the fast Fourier transform (FFT). A two-dimensional power spectral density (PSD) is then obtained as,

\begin{equation}
E(k_x, k_y) = |FFT[B(x,y)]|^{2} \,.
\end{equation}

In order to express $E(k_x, k_y)$ from the Fourier $k_{x}$ and $k_{y}$ to the isotropic wavenumber $k = (k_{x}^{2} + k_{y}^{2})^{1/2}$, it is necessary to calculate $E(k)'$ -- the integrated PSD over angular direction in Fourier space. From this last step, the one-dimensional PSD is obtained as $E(k)$ = $2\pi k E(k)'$. Finally, the power-law fit is performed as a linear fit in a logarithmic representation of $E(k)$ {\it vs.} $k$ and $\alpha$ is measured for the assumed turbulent inertial range of 2-20 Mm ({\it i.e.} 0.05-0.5 Mm$^{-1}$).

\subsubsection{Schrijver's $R$ value (logR)} 

The $R$-value property quantifies the unsigned photospheric magnetic flux near strong MPILs. The presence of such MPILs indicates that twisted magnetic structures carrying electrical currents have emerged into the AR through the solar surface. Therefore, $R$ represents a proxy for the maximum free magnetic energy that is available for release in a flare. This property and its usefulness in forecasting was first investigated by \cite{Schrijver:2007}.

The algorithm for calculating $R$ is relatively simple, computationally inexpensive, and was originally developed to use line-of-sight  magnetograms from the \textit{Michelson Doppler Imager} (MDI) \citep{Scherrer:1995} on board the \textit{Solar and Heliospheric Observatory} (SoHO). First, a bitmap is constructed for each polarity in a magnetogram, indicating where the magnitude of positive and negative magnetic flux densities exceeds the threshold value of $\pm$150 Mx cm$^{-2}$. These bitmaps are then dilated by a square kernel of 3 $\times$ 3 pixels and the areas where the bitmaps overlap are defined as strong-field MPILs. This combined bitmap is then convolved with a Gaussian filter of full width at half maximum (FWHM) $\approx$15 Mm. This particular value is constrained by how far from MPILs flares are observed to occur in extreme ultraviolet images of the solar corona. Finally, the convolved bitmap is multiplied by the absolute flux value of the line-of-sight magnetogram and $R$ is calculated as the sum over all pixels. Notice that since the $R$ value was implemented by \cite{Schrijver:2007} for MDI magnetograms, the SHARP magnetograms were resampled to the spatial scale of MDI, before the kernel application and subsequent calculations.




\section{Machine Learning Algorithms and Conventional Statistics Models} 
      \label{sec:algorithms}

The ML algorithms used in this study are MLPs, SVMs and RFs. Among the hundreds of ML algorithms proposed for binary classification \citep[{\it e.g.},][]{Fernandez-Delgado:14} these three categories of algorithms are representative of three important approaches in ML: i) artificial neural networks (ANN), ii) kernel-based methods and iii) classification and regression trees. This is the reason why they were used in the present study, in order to furthermore investigate whether the usage of RFs could bring any improvements in flare prediction in comparison to SVMs and MLPs. The RFs belong to the category of {\it ensemble} methods while the MLPs utilize unconstrained optimization and SVMs use constrained optimization techniques ({\it e.g.}, quadratic programming). In general, the working principle of ML comprises the following steps: i) train the model using a training set, ii) predict using the trained model and a testing set and iii) check whether the algorithm predicted well, in what is called the validation of the overal ML procedure. For further study, the reader is referred to \citet{Vapnik:98}, \citet{MacKay:03} and \citet{Hastie:09}.

\subsection{Multi-Layer Perceptrons}

\begin{figure}
        \centering
        \includegraphics[scale=0.55]{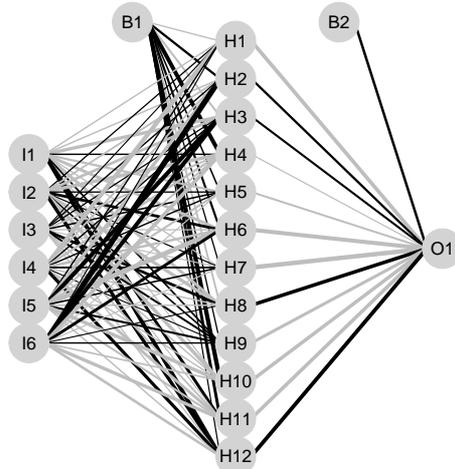}

\caption{{\footnotesize Example MLP neural network  with 6 inputs, 12 hidden nodes, 1 output and 2 biases. Bold, darker lines indicate large positive weights \ve{\omega}.   
 }}
\label{fig:NN1}       
\end{figure}

\noindent The MLP is a feed-forward network, thus it is described by the planar graph shown in Figure \ref{fig:NN1}. It contains an input layer, a hidden layer and an output layer of neurons. By the term neuron, we denote a basic processing unit where inputs are summed using specific weights and the result is squashed {\it via} an activation function. The hidden layer might actually expand in a series of hidden layers. Nevertheless, the simplest MLP networks have just one hidden layer. In principle the term hidden describes every layer which is neither the input nor the output layer, but resides in between, as presented in Figure \ref{fig:NN1}. A sufficient number of hidden nodes allows the MLP to approximate any continuous nonlinear function of several inputs with a desired degree of accuracy \citep{Hornik:89}, which is what characterizes the MLPs as universal approximators. It also holds that the greater the number of hidden nodes is, the more complex the nonlinear function that can be approximated by the neural network with a desired degree of accuracy. Usually, the number of hidden nodes does not have to be more than twice the number of input nodes (or predictors). Actually, if too many hidden nodes are utilized, then the overfitting problem arises, which means that the MLP memorizes the sample observations and generalizes badly in the prediction phase. Usually, and in this study, the optimal number of hidden neurons (called size of the MLP) is determined with a fine-tuning procedure ({\it e.g.} cross-validation approach, see Section \ref{TuningMLalgorithms}) before the training phase starts. The tuning phase is relatively time consuming, so it need not be executed every time the training starts. It can be conducted for a single realization of the training set.

An MLP network is actually a kind of a nonlinear regression (classification) technique, equivalent to a nonlinear  mapping from input \ve{I} to an output \ve{O} =  \ve{O(I; \omega, A)}. The output is a continuous function of the input and of the weights \ve{\omega}. The network is described by a given architecture \ve{A}, which typically defines the number of nodes in every layer ({\it e.g.} input, hidden and output). In general, MLP networks can be used to solve regression and classification problems. The statistical model of a MLP neural network for binary outcome, as described in the following, is based on \cite{MacKay:03}. For a recent survey on neural networks, the interested reader is referred to \cite{Prieto:16}. \\   

\subsubsection{Classification Networks} \label{Sec:ClassNets}

\noindent We consider a MLP with $l$ inputs called $I_l$ and bias $B_1$. Also the network contains a single hidden layer with $j$ hidden nodes $H_j$ and bias $B_2$. We have in general $i$ outputs $O_i$, while typically a single output is all that is needed ($i=1$). 

In the case of a classification problem, the propagation of the information from the inputs \ve{I} to the output \ve{O} is described by, 


            \begin{equation}\label{Eq:HiddenLayer}
            \begin{array}{rl}
                 { \alpha_{j}^{(1)} = \sum \limits_{l= 1}^L {\omega_{jl}^{(1)} I_l} + B_{j}^{(1)}; \;\;\;\; H_j = f(\alpha_{j}^{(1)})} \,,   \\
                 { \alpha_{i}^{(2)} = \sum \limits_{j=1}^J {\omega_{ij}^{(2)} H_j} + B_{i}^{(2)}; \;\;\;\;  O_i = g(\alpha_{i}^{(2)})} \,, \\
            \end{array} 
            \end{equation}

\noindent where, for example, $f(\alpha) = \frac {1}{(1+{\rm exp}(-\alpha))}$ and $g(\alpha) = \frac {1}{(1+{\rm exp}(-\alpha))}$. 


The index $l$ is used for the inputs $I_1, \dots, I_L$, the index $j$ is used for the hidden units and the index $i$ is used for the outputs ($i=1$). The weights \emph{$\omega_{jl}$}, \emph{$\omega_{ij}$} and biases $B_{j}$, $B_{i}$ define the parameter vector \ve{\omega} to be estimated. The nonlinear logistic function $f$ at the hidden layer (also known as activation function) helps the neural network approximate any generic continuous nonlinear function with a desirable degree of accuracy \citep{Hornik:89}. Visually, a neural network can be represented as a series of layers consisting of nodes, where every node is connected to nodes of the subsequent layer only (feed forward networks).


In the case of binary classification, the MLP is trained using a dataset of examples $D = \{ \ve{I}^{(n)}, \ve{T}^{(n)} \}$ by adjusting \ve{\omega} in order to minimize $G(\ve{\omega})$, the negative log-likelihood function,

\begin{equation}
G(\ve{\omega}) = - \Big( \sum \limits_{n=1}^N {\ve{T}^{(n)}  {\rm ln}(\ve{O}(\ve{I}^{(n)}; \ve{\omega}))  + (1-\ve{T}^{(n)}) {\rm ln}(1- \ve{O}(\ve{I}^{(n)}; \ve{\omega})) \Big)  } \,.  \;  \label{eq.Neg.LogLik}
\end{equation}

\noindent Notice that $\ve{I}^{(n)}$ is the matrix of the predictors and $\ve{T}^{(n)}$ is the vector of the targets for observation $n=1,\dots,N$. In Equation \ref{eq.Neg.LogLik}, $\ve{T}^{(n)}$ is 0 (1) for the negative (positive) class, respectively, and $\ve{O}(\ve{I}^{(n)};\ve{\omega)}$ is strictly between 0 and 1 (a probability) a fact that is ensured by Equations \ref{Eq:HiddenLayer}.

\subsection{Support Vector Machines}

The SVM variant we use is the $C$-Support Vector Classification ($C$-SVC) according to the widely used library LIBSVM \citep{LIBSVM11,Meyer:03}. 

Let us assume a vector of $K$ predictor values at observation $i$, $\ve{x}_i \in R^K$, $i=1,\dots,N$, which belongs in one of two classes, and an indicator vector $\ve{y} \in R^N$ such that $y_i \in \{1,-1\}$. 
Notice that the positive class has label $+1$ and the negative class has label $-1$. Then the $C$-SVC solves the optimization problem:

            \begin{equation}\label{exp:C-SVC}
            \begin{array}{rl}
                \mbox{minimize}    & {\frac{1}{2} \ve{\omega}^T\ve{\omega} + C \sum_{i=1}^{N} \xi_{i} } \,, \\
                                   &                        \\
                \mbox{subject to} &                        \\
                                   &  {y_{i}(\ve{\omega}^T \phi(\ve{x}_i) + b) \geq 1 - \xi_i,   \;\;\;\;\; i = 1,2,\ldots,N,      }    \\
                                   &  {\xi_i  \geq 0, \;\;\;\;\; i = 1,2,\ldots,N } \,,            \\
            \end{array} 
            \end{equation}

\noindent where $\phi(\ve{x}_i)$ is an arbitrary unknown function which maps $\ve{x}_i$ into a higher dimensional space and $C>0$ is the regularization parameter.
The optimization in $C$-SVC  model is performed by changing the decision variables: $\ve{\omega}$, $b$, $\ve{\xi}$.
Actually, LIBSVM solves the dual of $C$-SVC  which depends on a quantity:
$K(\ve{x}_i,\ve{x}_j) = \phi(\ve{x}_i)^T \phi(\ve{x}_j)$, which is called the \textit{kernel} function. While the $\phi(\ve{x}_i)$ is unknown, the kernel function is known and is equal to the inner product of $\phi(\ve{x}_i)$ with itself but for different pairs of observations $i$ and $j$. This is the so-called kernel trick of SVMs. As seen below, the kernel is a similarity measure and takes the maximum value of 1 when ${\rm dist}(\ve{x}_i,\ve{x}_j)=0$.

We have used the Radial Basis Function (RBF) (or Gaussian) kernel which is defined as $K(\ve{x},\ve{x'}) = {\rm exp}(-\gamma || \ve{x}-\ve{x'}||^2)$.
A variant of the $C$-SVC model has been used for flare prediction in \cite{Bobra:15}.

For imbalanced datasets which account for rare events (\textit{e.g.}, in our case the $>$M1 flares) some researchers e.g. \cite{Bobra:15} have used two different values for the regularization parameter $C$ in Equation \ref{exp:C-SVC}, thereby penalizing more the constraint violations for the minority class. These authors have used $C_1$ and $C_2$ with a ratio $C_2 / C_1 \in \{2,15\}$, where $C_1$ is the coefficient for the majority class (no events) and $C_2$ is the coefficient for the minority class (events). While we generally use the SVM in the original unweighted version in Equation \ref{exp:C-SVC}, in auxiliary runs we experimented also with using different values $C_1$ and $C_2$ with a ratio $C_2 / C_1 \in \{2,15,20\}$ to account for the imbalanced nature of the $>$M1 flares dataset.
 
\subsection{Random Forests}
The RF is a relatively recent ML methodology, introduced by \cite{Breiman:01}. The RF approach is an ensemble of tree predictors, where we let each tree vote for the most popular class. It has been reported \citep{Fernandez-Delgado:14} that RF offers significant performance improvement over other classification algorithms. The RF approach relies on randomness and involves the concept of split purity and the Gini index for variable selection \citep{Breiman:84}.  

According to \cite{Hastie:09}, the goal of the RF algorithm is to randomly build a set (or ensemble) of trees, by repeating the tree-formation process B times to create B trees. In particular, the algorithm: i) chooses a bootstrap sample from the training data, ii) grows a tree $T_b$ to the bootstrapped sample by applying consequently the following two substeps: Substep 1, select m variables randomly out of the M variables, and Substep 2, split the current node into two children nodes, having picked the best variable (node) from the m chosen ones. By repeating steps i) and ii) (where ii) consists of Substeps 1 -- 2), the algorithm creates a set (called ensemble) of trees $\{T_b\}_1^B$. Then, in the classification case studied in the present paper, a voting procedure for every tree $T_b$ is followed in order to obtain the class prediction of the random forest.

This is one of the first times RF is used for flare forecasting. Other related works are \cite{Liu:17} and \cite{Barnes:16}. Furthermore, three recent applications of RF in astrophysics are by \citep{Vilalta:13,Schuh:15,Granett:17}.

\subsection{Implementation of ML algorithms} 

\subsubsection{Multi-layer Perceptrons}
MLPs were implemented using the R programming language and the nnet package \citep{nnet.R}. The options used were: linout=FALSE, to ensure that sigmoid activation functions are used at the output node, entropy = TRUE, to ensure that the negative log-likelihood objective function is minimized during the training phase (and not the default Sum of Squares Error (${\rm SSE}$) criterion), and size=iNode, where ${\rm iNode}$ for both $>$M1 flares and for $>$C1 flares was chosen with a tuning procedure. 

\subsubsection{Support Vector Machines}
SVMs were implemented using the R programming language and the e1071 package \citep{e1071.R}. The options used were: probability=TRUE, in order to obtain probability estimates for every element of the training set as well as probability estimates for every element of the testing set.

\subsubsection{Random Forests}
RFs were implemented using randomForest package \citep{randomForest.R} in the R programming language. The options used were: importance = TRUE, to create importance information for every predictor, na.action=na.omit, to exclude records of predictors with missing values appearing in preliminary versions of the dataset (but lacking from the final version of the dataset). 

\subsection{Conventional Statistics Models}

\noindent Non-ML (or statistical) methods also considered are: i) linear regression (LM), ii) probit regression (PR) and iii) logit regression (LG). Although multiple linear regression is known to be redundant for binary outcomes, since it can yield probabilistic predictions outside the interval $[0,1]$, we still include it in the array of tested methods. The reason is that some practitioners still use it for binary outcomes (calling it linear probability model (LPM), see \cite{Greene:02}) and there is always interest to consider ordinary least squares (OLS) as an entry-level method for any regression analysis. An interesting article about the lack of use of probit and logit in astrophysics modeling is \cite{Desouza:15}. The statistical algorithms were implemented in the statistical programming language R using the lm and glm functions.

For a description of these well known methods the reader is referred to \citep{Greene:02,Winkelmann06}.

    
\section{Data preparation, Results and Discussion} 
      \label{Comp.Exper}

First, we implement ML predictions on $>$M1 flares. Second, we use statistical methods for the prediction of $>$M1 flares. Third, we predict $>$C1 flares with ML algorithms. Finally, we predict $>$C1 flares with the statistical algorithms. The following subsections describe these four experiments, presenting at first a single combination of training/testing set for every flare class and category of techniques.

Results are presented for the prediction step in terms of: i) skill scores profiles (SSP) of {\rm ACC}, {\rm TSS} and {\rm HSS} as functions of the probability threshold, ii) ROC curves, and iii) RD plots for all methods: (for the explanation of metrics {\rm ACC}, {\rm TSS}, {\rm HSS} as well as ROC curves and RD diagrams --  see following Section \ref{sec:metricconcepts}). Skill score profiles were created by a code we developed in R, ROC curves were created using the ROCR package \citep{ROCR.R}, while reliability diagrams were created using the verification package \citep{verification.R}.

All algorithms were implemented and run using the R programming language $3.3.2$ \cite{language.R} and the RStudio $0.99$ IDE.

    \subsection{Data Pre-processing}
    \noindent The data comprise the $K=7$ predictors [{\rm logR}, {\rm FSPI}, {\rm TLMPIL}, {\rm DI}, ${\rm WL}_{\rm {SG}}$, IsinEn1, IsinEn2] described in Section \ref{sec:predictors} and computed using either the line-of-sight magnetograms, $B_{{\rm los}}$, of SHARP data or the respective radial component, $B_{r}$ \citep{Bobra:14}. Hence, we test $K = 2 \times 6 + 1 = 13$ predictors\footnote{ This is because, for predictor ${\rm WL}_{\rm {SG}}$, we considered only the $B_{r}$ version.}. The sample comprises $N=23,134$ observations, randomly split in half into $N_1=11,567$ observations for the training, and $N_2=11,567$ observations for the testing set. The random split is performed for 200 replications and all six prediction algorithms ({\it i.e.} MLP, SVM, RF, LM, probit and logit) of Section \ref{sec:algorithms} are trained and perform on identical training and test sets.  The metrics {\rm ACC}, {\rm TSS} and {\rm HSS} of Section \ref{sec:metricconcepts} are computed always for the testing (out-of-sample) set. We have standardized all predictor variables to have mean equal to 0 and standard deviation equal to 1, because several ML algorithms involve non-linear optimization ({\it e.g.} MLPs). This helps to better train the ML algorithms and also explains the effect of every predictor variable on the studied outcome in the case of the statistical models LM, probit and logit.

    \subsection{Tuning of ML algorithms} \label{TuningMLalgorithms}
    \noindent As with any parameterized algorithm ({\it e.g.} simulated annealing, evolutionary algorithms, and other metaheuristics), the performance of ML algorithms depends on a number of crucial parameters which need to be fine tuned before the application of the ML procedure ({\it e.g.} training, testing and validation steps). The optimal tuning of ML algorithms is more or less still an open question in the ML community and always poses a big challenge for any practitioner. This choice of optimal options for the ML algorithms themselves is similar to the choice of optimal parameters for other numerical models, ({\it e.g.} MHD models), where the analyst also has to explore the optimal parameter space in several crucial parameters before conducting numerical MHD simulations. The algorithms MLP, SVM and RF have their critical hyperparameters ({\it e.g.} parameters that are critical for the forecasting performance of every algorithm) tuned {\it via} a 10-fold cross-validation study exploiting only the training set at one of its realizations. The set of plausible values for every ML algorithm is as follows: i) MLP: size (number of hidden neurons) $\in \{4,13,26\}$ and decay  (weight decay parameter) $\in \{10^{-3},10^{-2},10^{-1}\}$, ii) SVM: $\gamma$ (parameter in the RBF (or Gaussian) kernel) $\in$ $\{10^{-6}$, $10^{-5}$, $10^{-4}$, \dots, $10^{-1}\}$ and cost (regularization parameter) $\in \{10,100\}$ and iii) RF: mtry (number of variables randomly sampled as candidates at each split) $\in \{ \lfloor \sqrt{K} \rfloor = 3\}$ and ntree (number of trees to grow) $\in \{500\}$.

Actually, we have tuned only the MLP and SVM classifiers, because the default RF values mtry=3 and ntree=500 immediately provided satisfactory results. 
Tuning of the MLP and SVM was mostly needed in the $>$M1 flares case, that was found harder to predict than $>$C1 flares, but was also performed in the $>$C1 flares case.
Thus, the hyperparameters for MLP and SVM needed tuning since, for example, the default values $\gamma$ = 1 and cost = 1 for SVM provided unsatisfactory results. We have used the tune.nnet and tune.svm functions of the R package e1071 for tuning the MLP and SVM, respectively. After the tuning, both MLP and SVM improved their performance  significantly.

For the $>$M1 flares, the selected values are size = 26 and decay = 0.1 for the MLP and $\gamma$ = 0.1 and cost = 10 for the SVM. These values are used throughout the remainder of this work. For the $>$C1 flares case, the selected values are size = 4 and decay = 0.1 for the MLP and $\gamma$ = 0.001 and cost = 100 for the SVM.

    \subsection{Comparison Metrics}\label{sec:metricconcepts}

    \noindent A wide variety of metrics exist in order to characterize the quality of binary classification. Among these, no single one is fit for all purposes. There exist two types of metrics, suitable for either categorical or probabilistic classification. In the former case a strict class membership is returned from the model and in the latter case a probability of membership is returned. In this section we concentrate on categorical forecast metrics for binary classification.
    \noindent In what follows, let ${\rm ACC}$ denote accuracy, {\rm TSS} denote true skill statistic and {\rm HSS} denote Heidke skill score. The performance of algorithms is measured using a number of metrics. These are derived from the so-called contingency table or confusion matrix, a representation of which is provided in Table \ref{table:cf1}:


\begin{table}[h!]
\caption{$2 \times 2$ contingency table for binary forecasting}
\begin{tabular}{||c c c||}
 \hline
     & ACTUAL &   \\ [0.5ex]
 \hline
 PREDICT & NO & YES  \\
 NO & TN  & FN   \\
 YES & FP  & TP   \\ [1ex]
 \hline
\end{tabular}
\label{table:cf1}
\end{table}

\noindent Table \ref{table:cf1} includes true positives ({\rm TP}; events predicted and observed), true negatives ({\rm TN}; events not predicted and not observed), false positives ({\rm FP}; events predicted but not observed) and false negatives ({\rm FN}; events not predicted but observed), where $N = {\rm TP} + {\rm FP}+ {\rm FN}+ {\rm TN}$ is the sample size. From these elements:


\noindent The meaning of {\rm ACC} is the proportion correct, namely the number of correct forecasts of both event and non-event, normalized by the total sample size,

        \begin{equation}\label{def:accuracy}
            {\rm ACC} = \frac{{\rm TP} + {\rm TN}}{N} \,.  
        \end{equation}    

\noindent The {\rm} TSS \citep{Hanssen:65} compares the probability of detection ({\rm POD}) to the probability of false detection ({\rm POFD}), 


        \begin{equation} \label{def:tss}
            {\rm TSS} = {\rm POD} - {\rm POFD} =  \frac{{\rm TP}}{{\rm TP}+ {\rm FN}} - \frac{{\rm FP}}{{\rm FP}+ {\rm TN}} \,.
        \end{equation}

\noindent Moreover, the {\rm TSS} is the maximum vertical distance from the diagonal in the ROC curve, that relates the {\rm POD} and {\rm POFD} for different probability thresholds -- see Section \ref{Comp.Exper}. The {\rm TSS} covers the range from $-1$ up to $+1$, while the value of zero indicates lack of skill. Values below zero are linked to forecasts behaving in a contrarian way, namely mixing the role of the positive class with the role of the negative class. In any negative {\rm TSS} value, by exchanging the roles of YES and NO events, we can obtain the corresponding positive {\rm TSS} value which would be identical in absolute value terms with the negative {\rm TSS} value.

\noindent The {\rm HSS} \citep{Heidke:26} measures the fractional improvement of the forecast over the random forecast,

        \begin{equation}\label{def:hss}
           {\rm HSS} =   \frac{2 ({\rm TP} \times {\rm TN} - {\rm FP} \times \rm {FN})}{({\rm TP} + {\rm FN}) ({\rm FN} + {\rm TN}) + ({\rm TP}+ {\rm FP}) ({\rm FP}+ {\rm TN})} \,,   
        \end{equation}

\noindent which ranges from $- \infty$ to 1. Any negative value means that the random forecast is better, a zero value means that the method has no skill over the random forecast, and an ideal forecast method provides a {\rm HSS} value equal to 1.

\noindent The {\rm TSS} and {\rm HSS} metrics are among the most popular metrics for comparison purposes in Meteorology and Space Weather and were conceptually compared in \cite{Bloomfield:12}. In a probabilistic forecasting, such as the one for solar flares, they must be assigned a probability threshold, thus appearing as functions of this threshold.

To summarize, {\rm ACC} is the most popular classification metric, but in rare events such as flares $>$M1, the {\rm ACC} can be artificially high for the naive model which will always predict the majority class (``no event''). Thus, {\rm TSS} and {\rm HSS} are more suitable for flare prediction. Moreover, {\rm TSS} has the advantage of being invariant to the frequency of events in a sample \citep[\textit{e.g.} see][]{Bloomfield:12}. Typically, both {\rm TSS} and {\rm HSS} need to be evaluated, for a given probability threshold, in order to assess the merit of a given probabilistic forecasting model, such as the ones we develop in this study.

Regarding the probabilistic assessment of classifiers, the present study utilizes the visual approaches of Receiver Operating Characteristic (ROC) curves and Reliability Diagrams (RD) ({\it e.g.} see Section \ref{Comp.Exper}). The ROC describes the relationship between the POD and the POFD for different probability thresholds ({\it e.g.} see Figure \ref{fig:1}b). The Area Under the Curve (AUC) in the ROC has an ideal value of one. The RD describes the relationship between the returned probabilities by the model and the actual observed frequencies of the data. A binning approach is used to construct the RD, in which probabilities are assigned to intervals of arbitrary length (for example we use 20 bins of length 0.05 each). For an example of RD, see Figure \ref{fig:1}c. Also, to algebraically assess the probabilistic performance of classifiers, we use the Brier Score (BS) \citep{Brier:50} and Brier Skill Score (BSS) \citep{Wilks:11}, as well as the AUC \citep{Marzban:04}.

    \subsection{Results on $>$M1 Flare Prediction}

\subsubsection{Prediction of $>$M1 Flare Events Using Machine Learning}
\noindent Figure \ref{fig:1} shows the forecast performances of the three tested ML methods, using both binary scores (SSP [left]; ROC [middle]) and probabilistic ones (RD [right]). In particular:
\begin{enumerate}[i)]
\item
Regarding the MLPs, we notice a wide plateau with more-or-less flat profile for {\rm HSS} and less so for {\rm TSS}. This occurs because the number of hidden neurons ({\rm size}=26) is twice the number of input neurons, causing the MLP to provide probability estimates clustered around 0 and 1. The ROC curve is reasonably good, with maximum {\rm TSS}=0.726. Moreover, the RD shows a systematic over-prediction above a forecast probability of ~0.4.
\item
For the SVMs, the SSP plateau noticed in case of the MLPs is not present here, with nearly monotonically decreasing values of {\rm TSS} and {\rm HSS} appearing. The ROC curve shows a maximum {\rm TSS}=0.629, while the RD seems slightly better than for MLP, with some under-prediction below a forecast probability of 0.4 and generally large uncertainties. When we use the weighted version of the SVM, with a ratio of $C_2 / C_1 = 20$, then the ROC curve improves providing a maximum {\rm TSS}= 0.718, but the overall forecasting ability as measured by the SSP and RD remains worse than the MLP.
\item
With respect to the RFs, the SSP behaviour is such that {\rm HSS} shows a plateau around its peak value, albeit smaller than in case of MLPs, while {\rm TSS} monotonically decreases. This said, notice that the peak {\rm HSS} and {\rm TSS} values are higher in this case ({\it e.g.} {\rm TSS}=0.780 and {\rm HSS}=0.587). The ROC curve is better than that of MLPs and SVMs with a maximum {\rm TSS}=0.780. The RD, finally, appears clearly better than those of MLPs and SVMs, presenting some mild under-prediction, mainly within error bars, above a forecast probability of 0.2.
\end{enumerate}

\subsubsection{Prediction of $>$M1 Flare Events Using Statistical Models.}

\noindent Figure \ref{fig:11} shows the forecast performances of the three tested statistical methods, for $>$M1 flare prediction. In particular:

Regarding the LM, the SSP is different between {\rm TSS} and {\rm HSS}, with {\rm TSS} peaking more impulsively and for smaller probabilities and then decreasing nearly monotonically. The ROC curve shows also a significant performance with maximum {\rm TSS}=0.744 that can also be seen in the RD, which shows a very good behavior, albeit with error bars, for the entire range of forecast probabilities.

As far as the PR is concerned, a slightly improved behavior in comparison with LM can be seen here, for the SSP, the ROC curves and the RD. The RD, also, seems more reliable in this case compared to LM, although differences are mostly within error bars.

For the LG, we notice a similar behavior as in the LM and especially PR method, and the RD in this case appears as good as the PR RD.


\subsubsection{Monte Carlo Simulation for $>$M1 Flares}

        \noindent In Table \ref{tab:MonteCarloM1} we provide the average values of the skill scores {\rm ACC}, {\rm TSS} and {\rm HSS} for all prediction methods after the 200 replications of the Monte Carlo experiment regarding $>$M1 flares prediction. We notice from Table \ref{tab:MonteCarloM1} that the maximum {\rm HSS}=0.57 is obtained with the RF method for a probability threshold of 25\%. The corresponding RF score values are {\rm ACC}=0.96$\pm$0.00, {\rm TSS}=0.63$\pm$0.02 and {\rm HSS}=0.57$\pm$0.02. The second best method in Table \ref{tab:MonteCarloM1} for the same probability threshold is MLP, with {\rm ACC}=0.95$\pm$0.00, {\rm TSS}=0.56$\pm$0.02 and {\rm HSS}=0.50$\pm$0.02. Considering the threshold where the maximum {\rm TSS} is observed, we get the optimal results for method RF and threshold 10\%, with values {\rm ACC}=0.90$\pm$0.00, {\rm TSS}=0.77$\pm$0.01 and {\rm HSS}=0.42$\pm$0.01. The second best method may be considered the LM at 10\% threshold with {\rm ACC}=0.88$\pm$0.00, {\rm TSS}=0.73$\pm$0.01 and {\rm HSS}=0.35$\pm$0.01. The difference between RF and LM is statistically significant at 0.01\% level as shown in Table \ref{tab:t-tests} at row 1. For the range of thresholds 10\% to 25\% the method RF yields increasing values of {\rm HSS} and decreasing values of {\rm TSS}. For example, an appealing forecasting model could be RF with threshold 15\% and metrics {\rm ACC}=0.93$\pm$0.00, {\rm TSS}=0.74$\pm$0.02 and {\rm HSS}=0.49$\pm$0.01 in Table \ref{tab:MonteCarloM1}, but this would depend on the needs and requirements of a given decision maker.

\clearpage

\begin{figure}[h]
\centering
\subfloat[Subfigure 1 list of figures text][MLP, SSP]{
\includegraphics[width=0.295\textwidth]{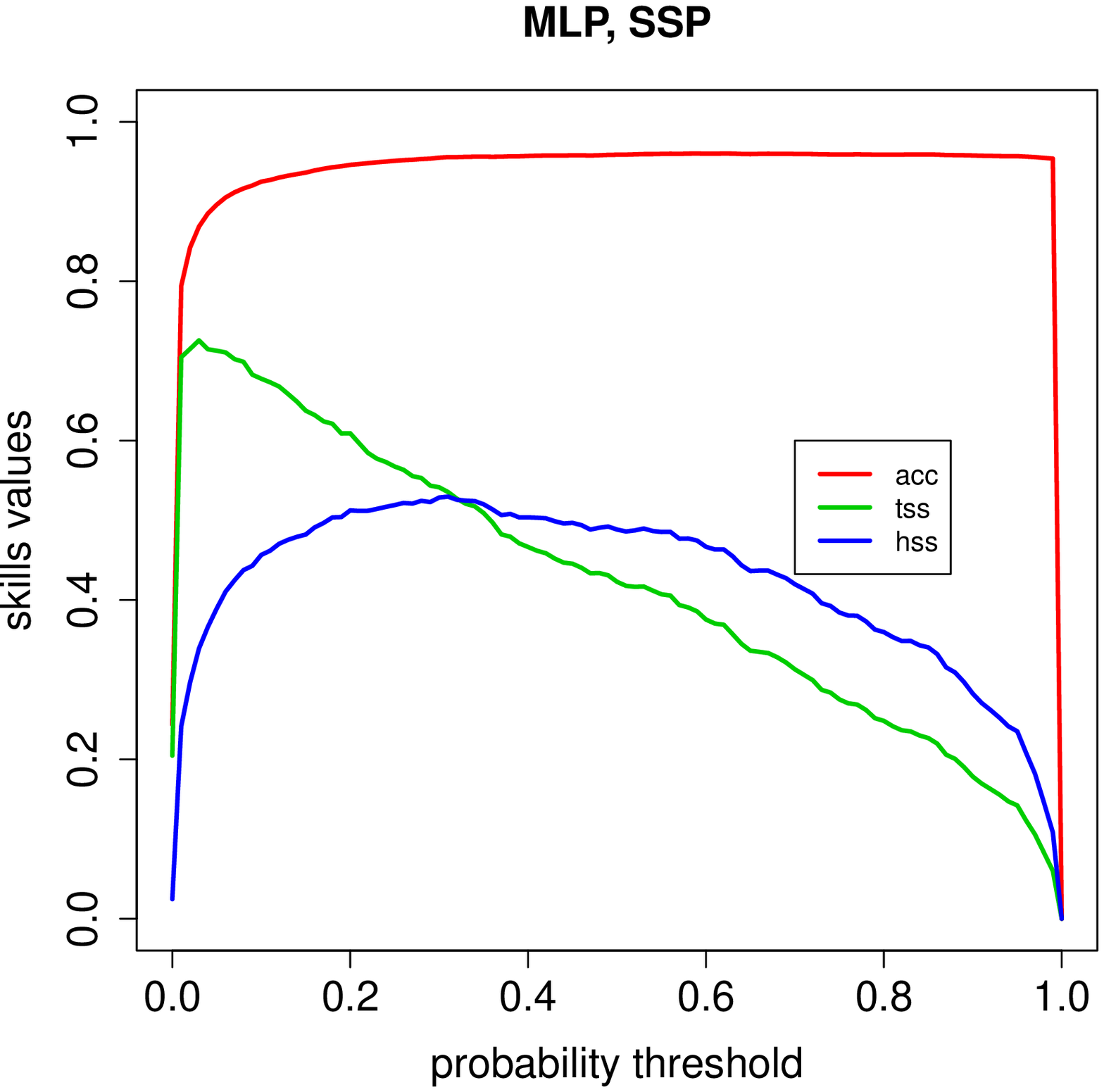}
\label{fig:a}}
\subfloat[Subfigure 2 list of figures text][MLP, ROC]{
\includegraphics[width=0.295\textwidth]{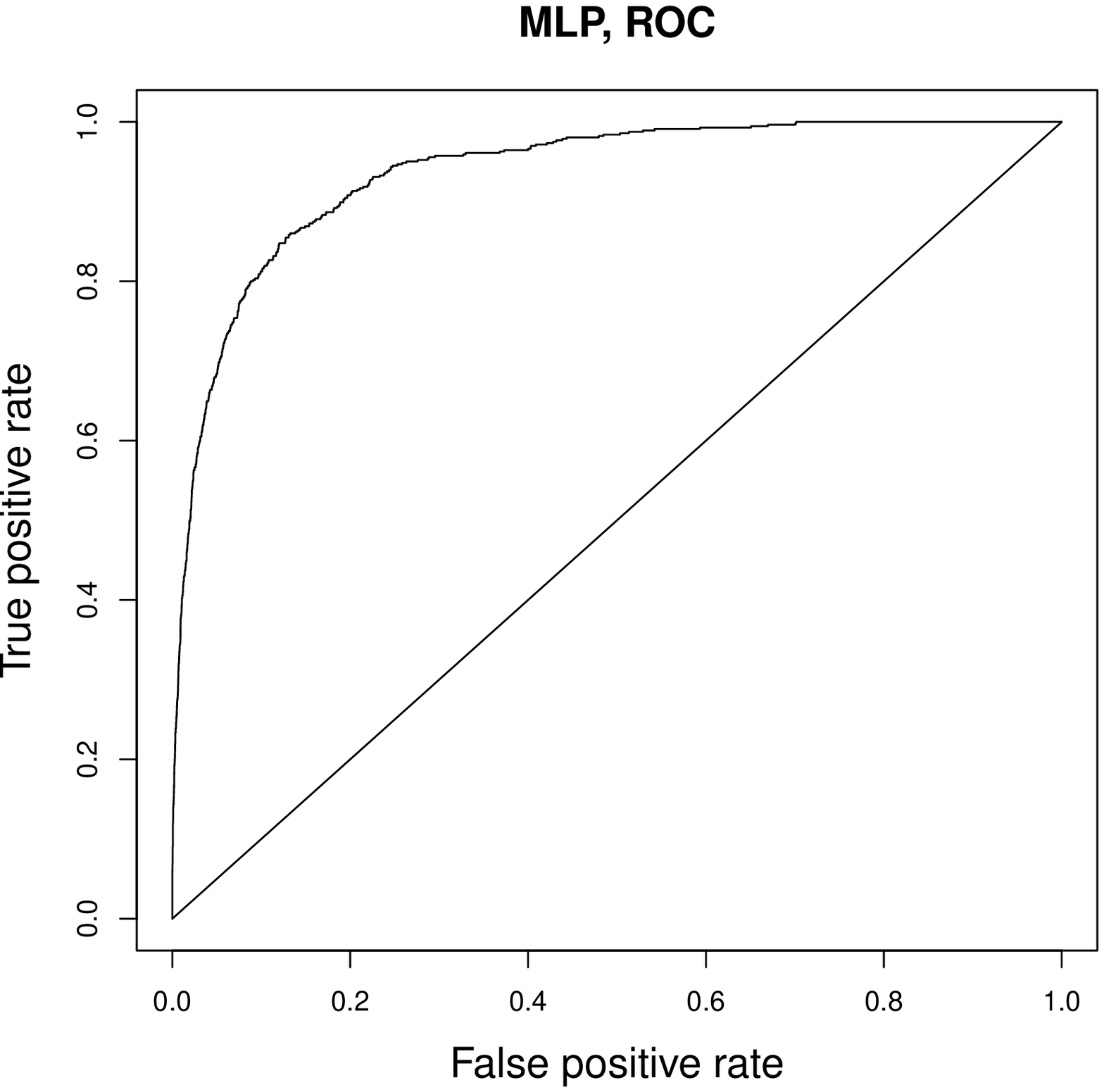}
\label{fig:b}}
\subfloat[Subfigure 3 list of figures text][MLP, RD]{
\includegraphics[width=0.295\textwidth]{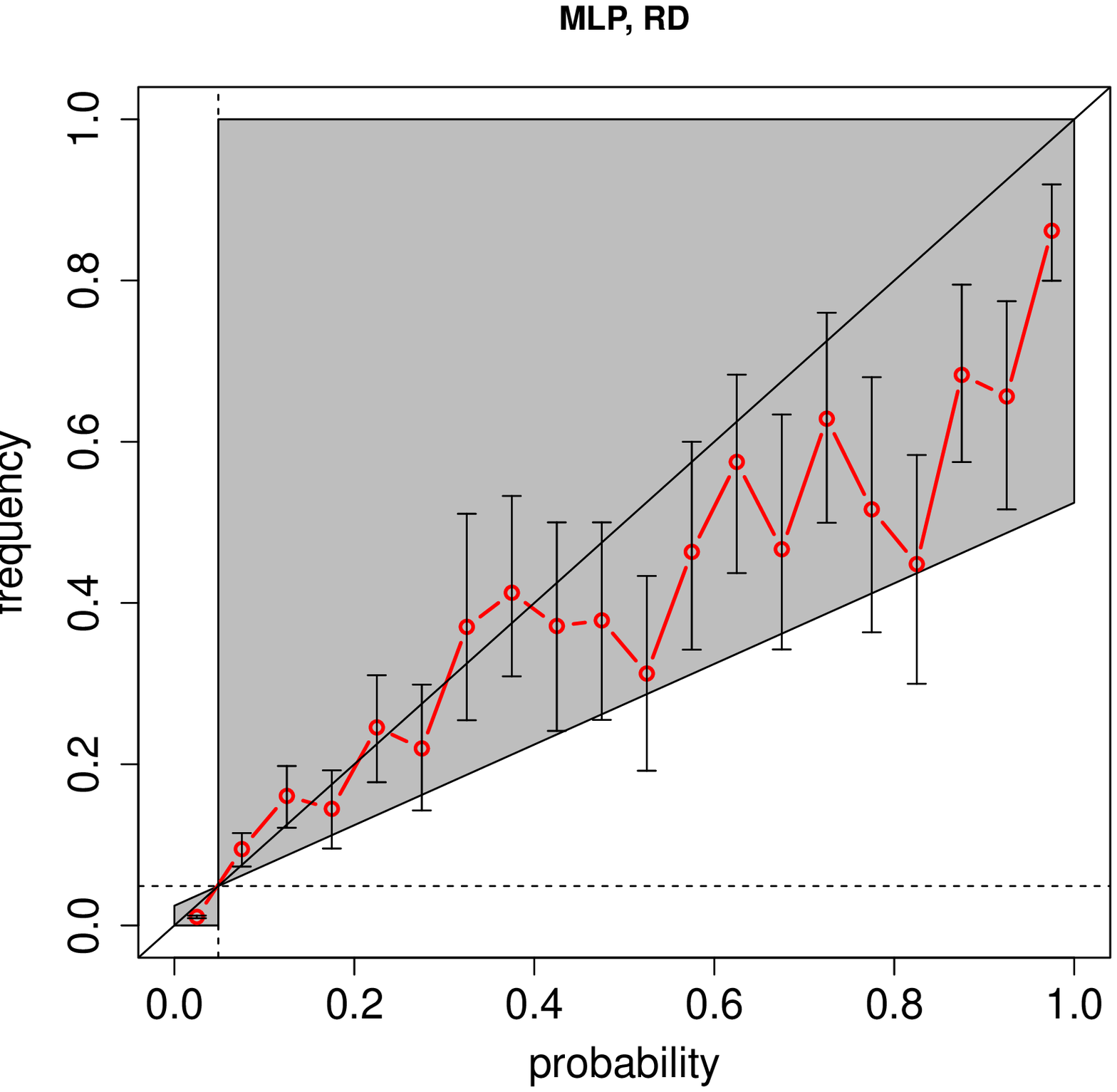}
\label{fig:c}}
\qquad
\subfloat[Subfigure 1 list of figures text][SVM, SSP]{
\includegraphics[width=0.295\textwidth]{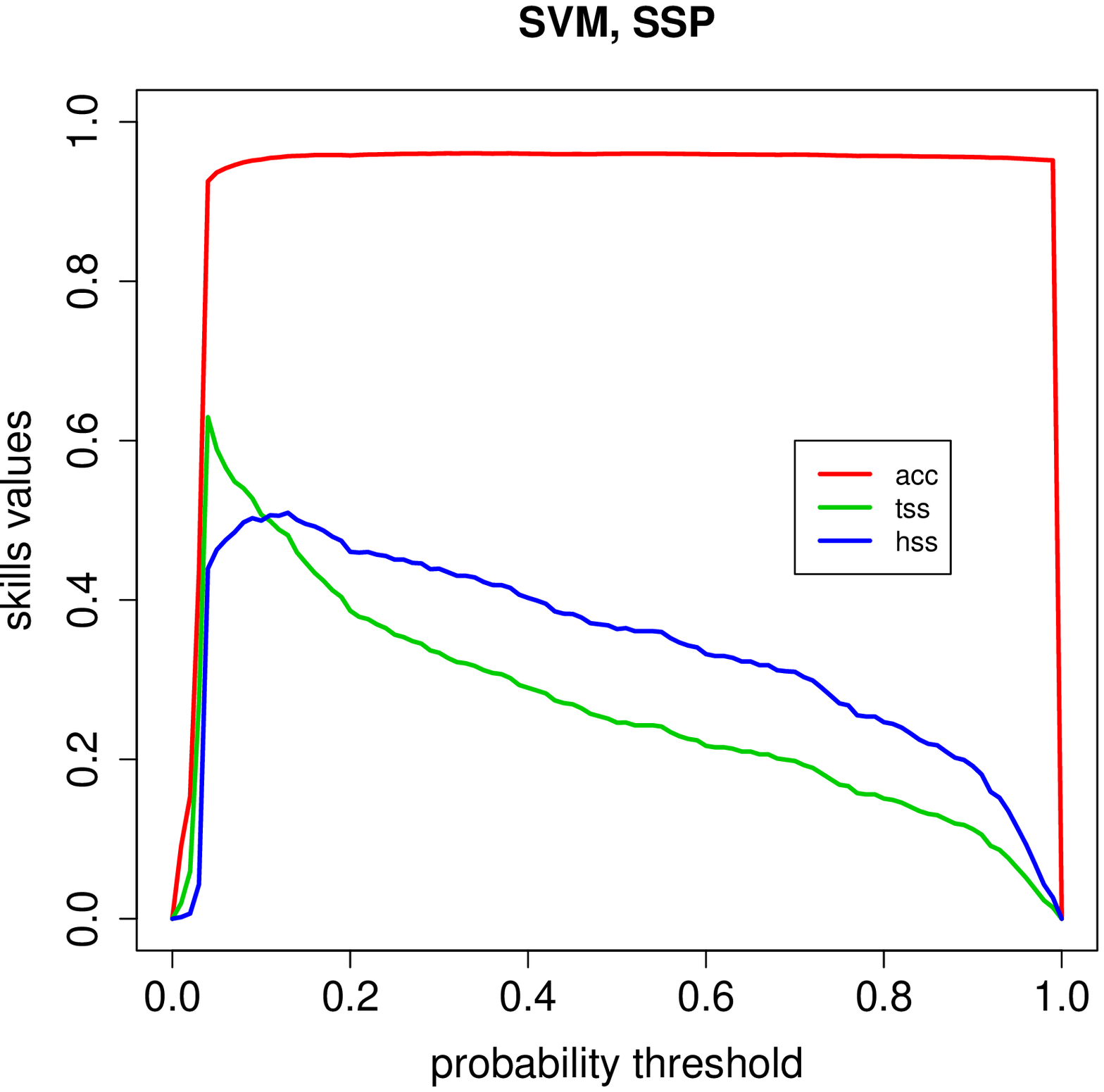}
\label{fig:d}}
\subfloat[Subfigure 2 list of figures text][SVM, ROC]{
\includegraphics[width=0.295\textwidth]{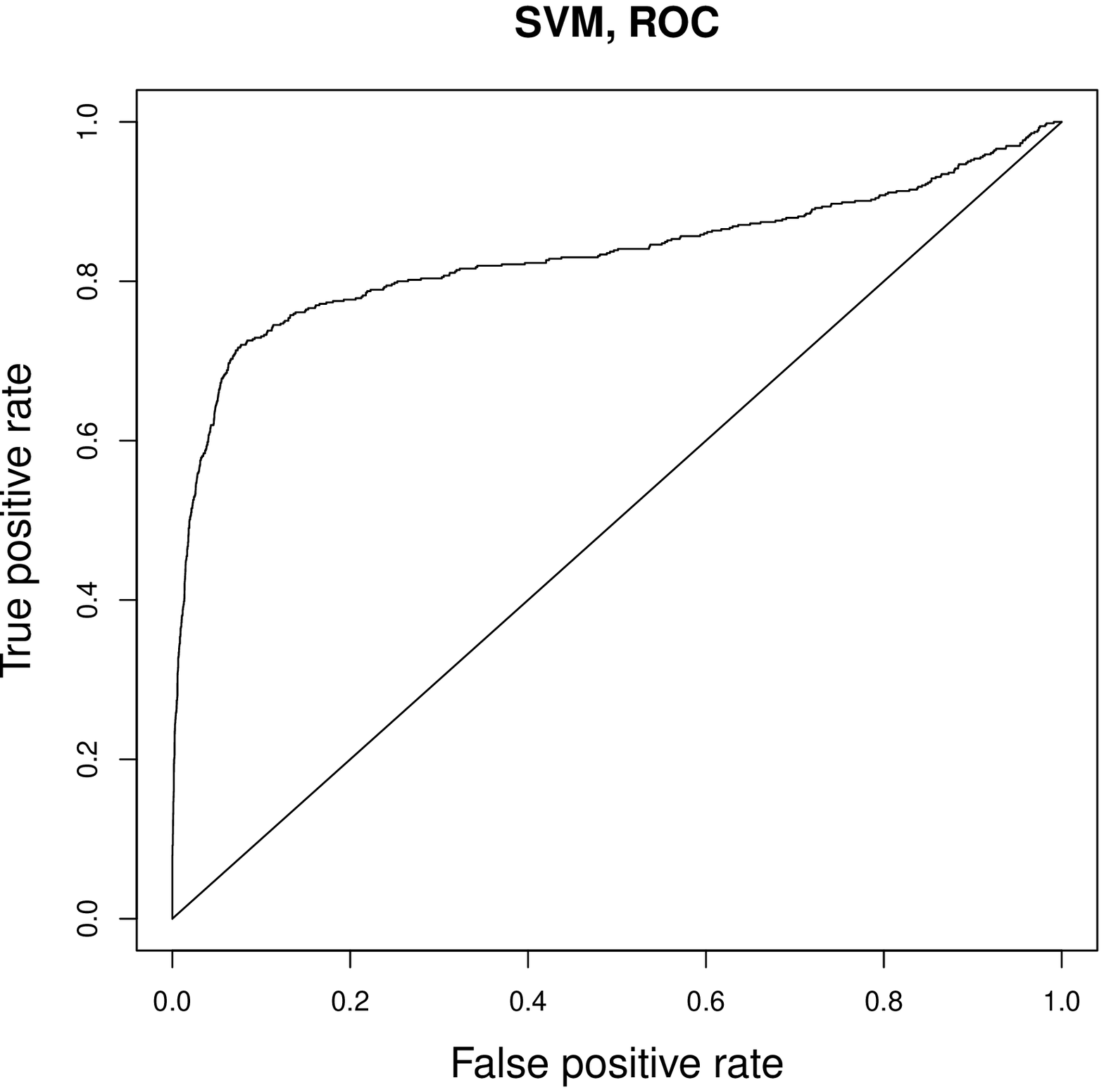}
\label{fig:e}}
\subfloat[Subfigure 3 list of figures text][SVM, RD]{
\includegraphics[width=0.295\textwidth]{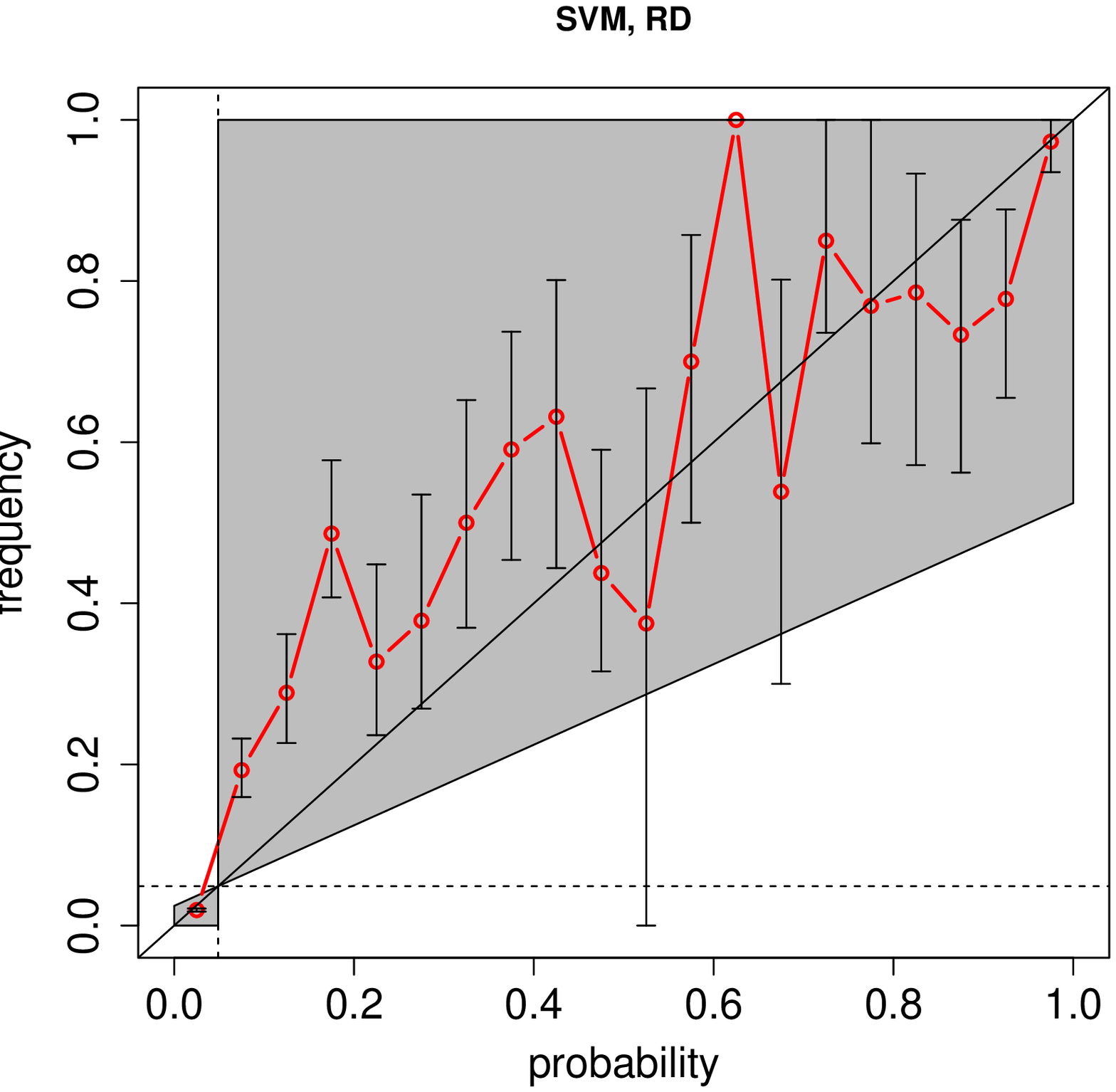}
\label{fig:f}}
\qquad
\subfloat[Subfigure 1 list of figures text][SVM weighted, SSP]{
\includegraphics[width=0.295\textwidth]{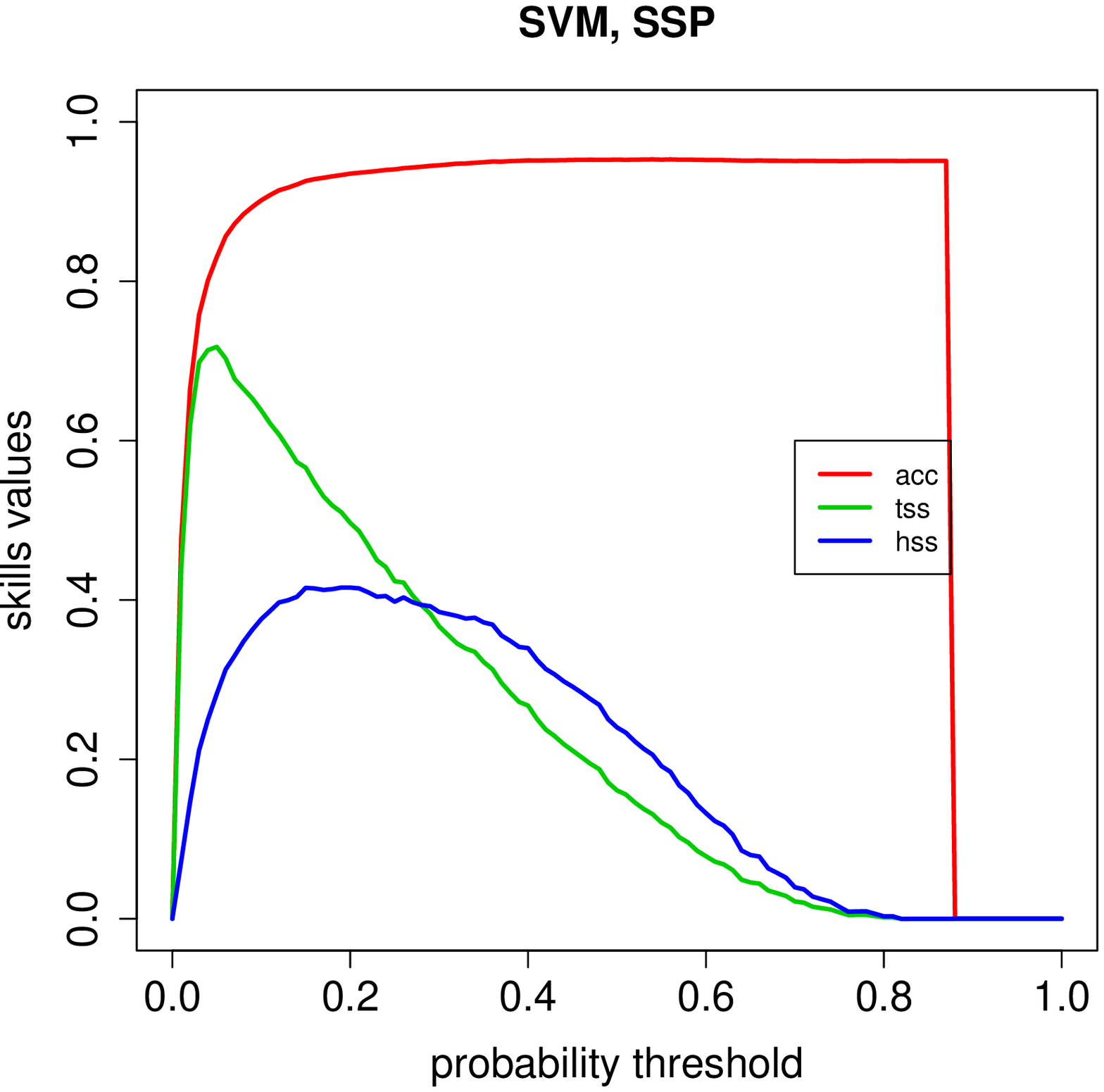}
\label{fig:d2}}
\subfloat[Subfigure 2 list of figures text][SVM weighted, ROC]{
\includegraphics[width=0.295\textwidth]{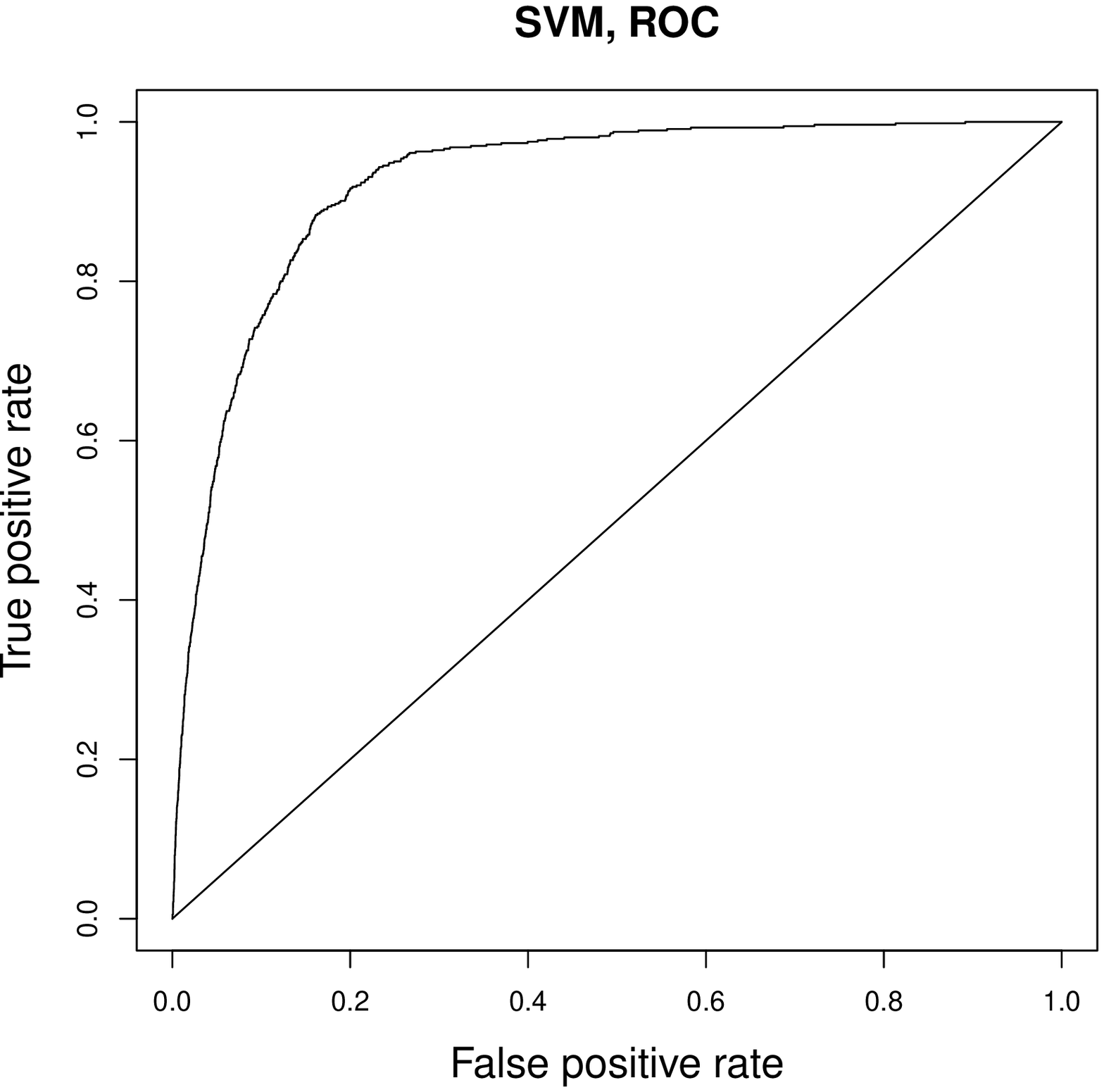}
\label{fig:e2}}
\subfloat[Subfigure 3 list of figures text][SVM weighted, RD]{
\includegraphics[width=0.295\textwidth]{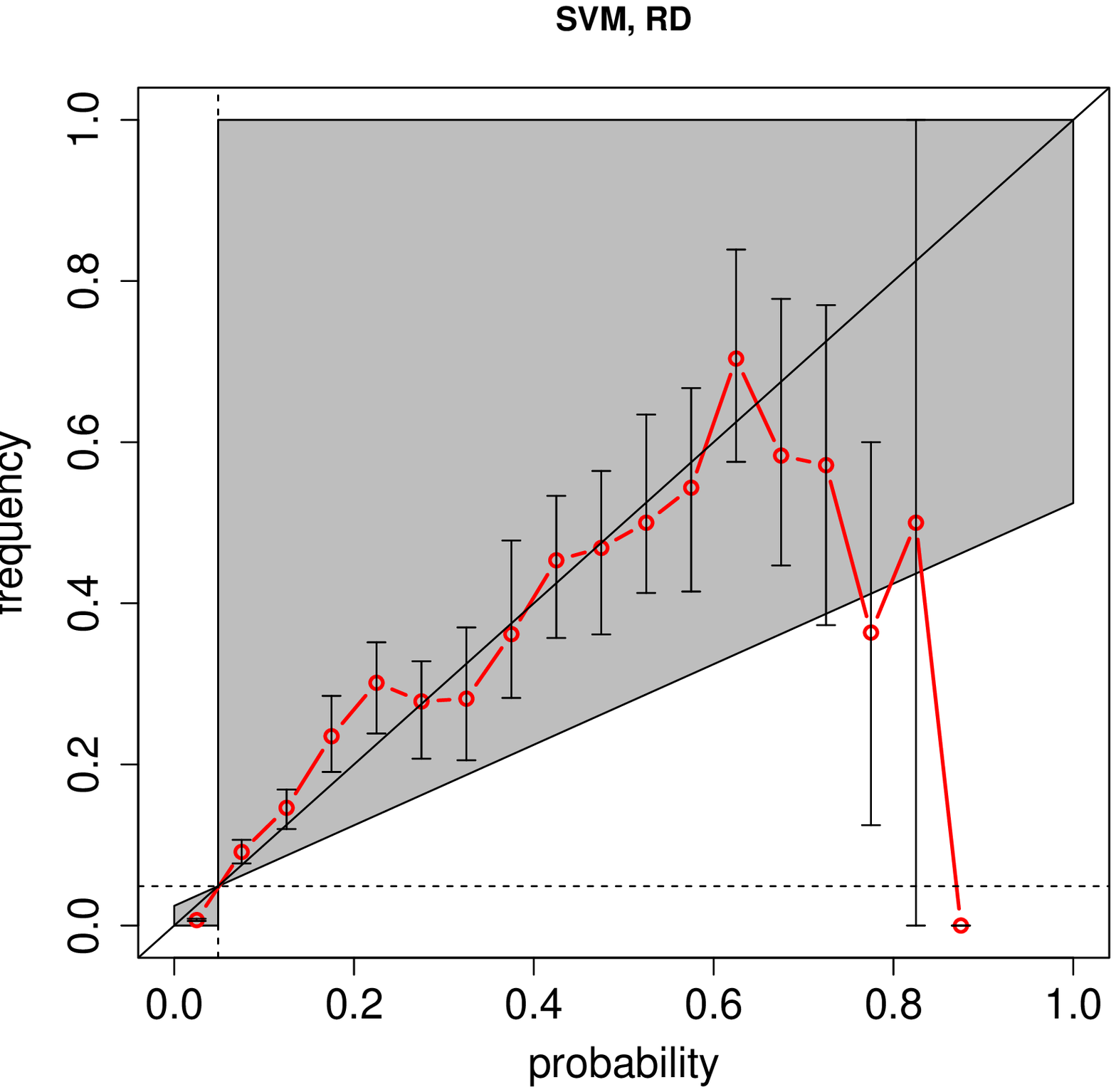}
\label{fig:f2}}
\qquad
\subfloat[Subfigure 1 list of figures text][RF, SSP]{
\includegraphics[width=0.295\textwidth]{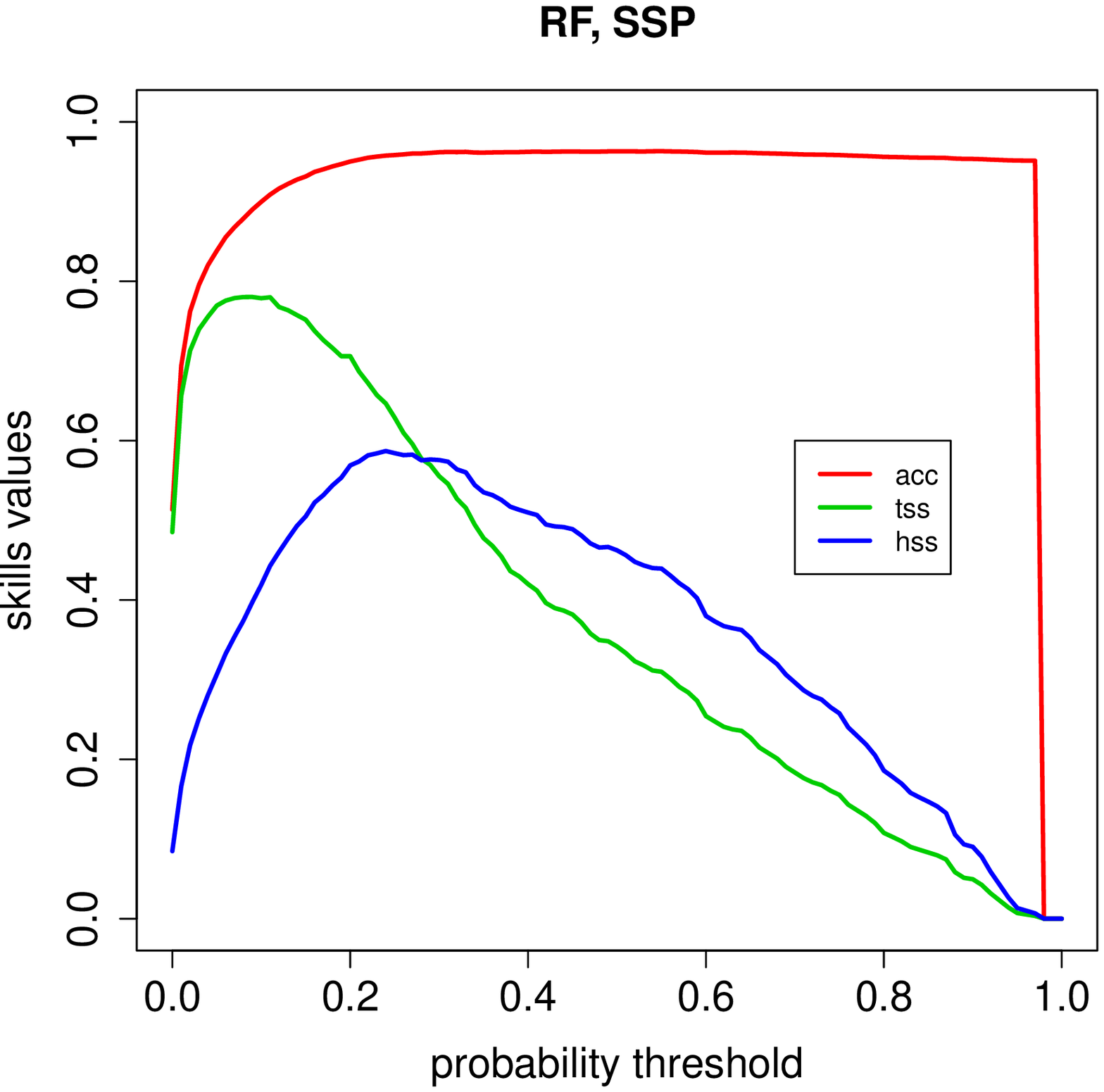}
\label{fig:g}}
\subfloat[Subfigure 2 list of figures text][RF, ROC]{
\includegraphics[width=0.295\textwidth]{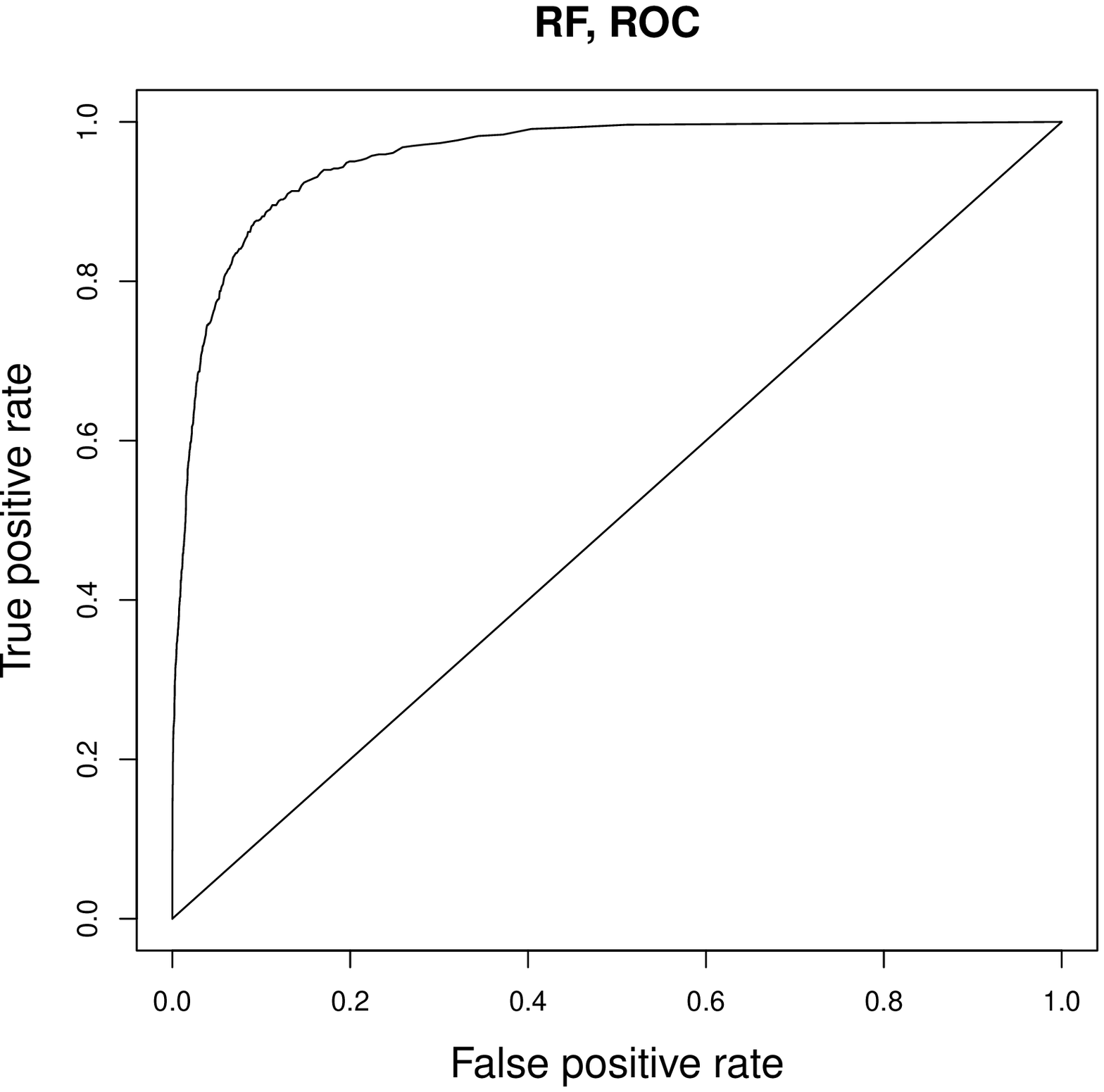}
\label{fig:h}}
\subfloat[Subfigure 3 list of figures text][RF, RD]{
\includegraphics[width=0.295\textwidth]{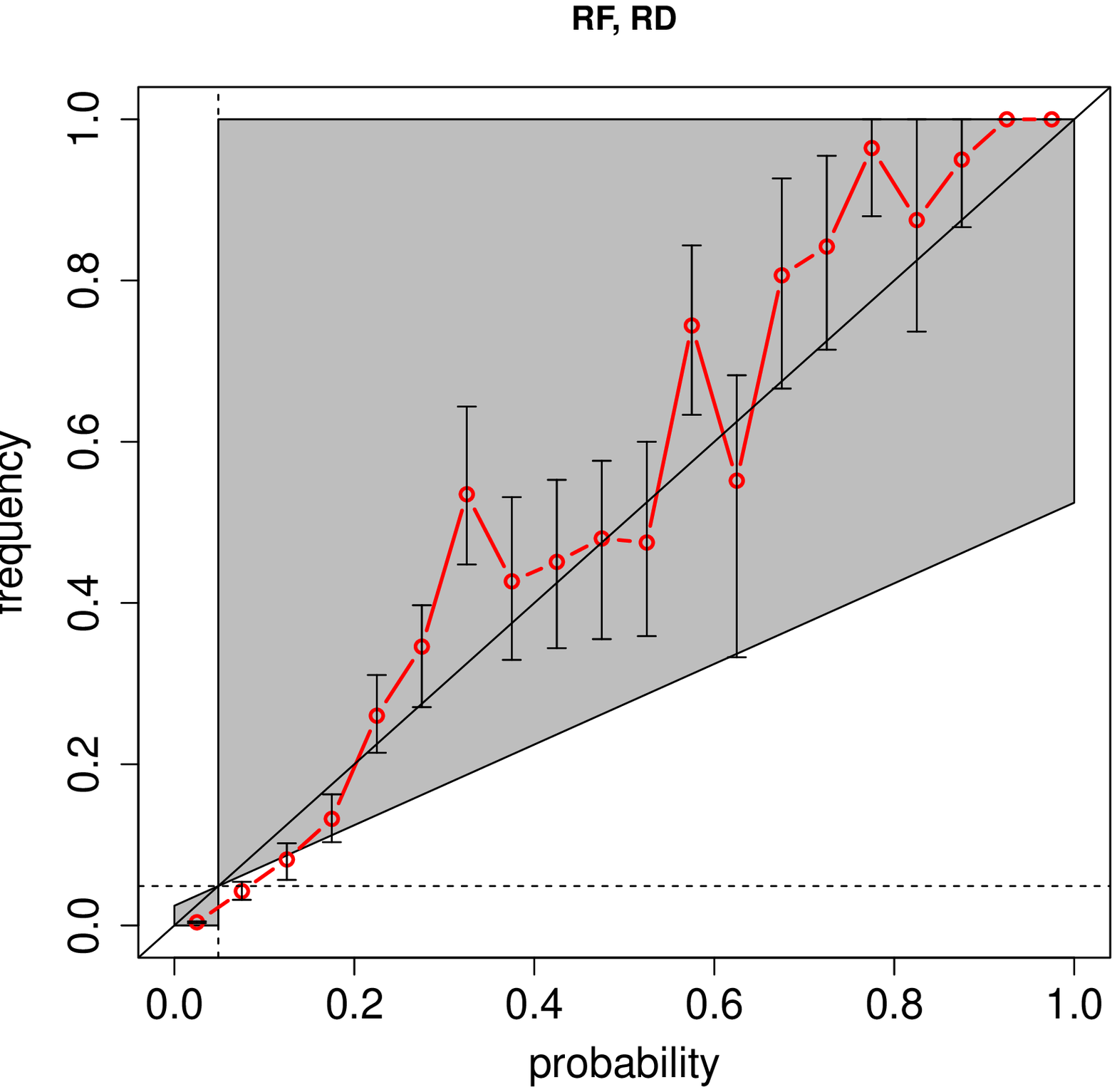}
\label{fig:i}}
\caption{ML methods comparison for $>$M1 GOES flares prediction for (from \textit{top} to \textit{bottom}) MLP, SVM, weighted SVM and RF. From \textit{left} to \textit{right} we present the corresponding SSP, ROC and RD.}
\label{fig:1}

\end{figure}

\begin{figure}[h]
\centering
\subfloat[Subfigure 1 list of figures text][LM, SSP]{
\includegraphics[width=0.295\textwidth]{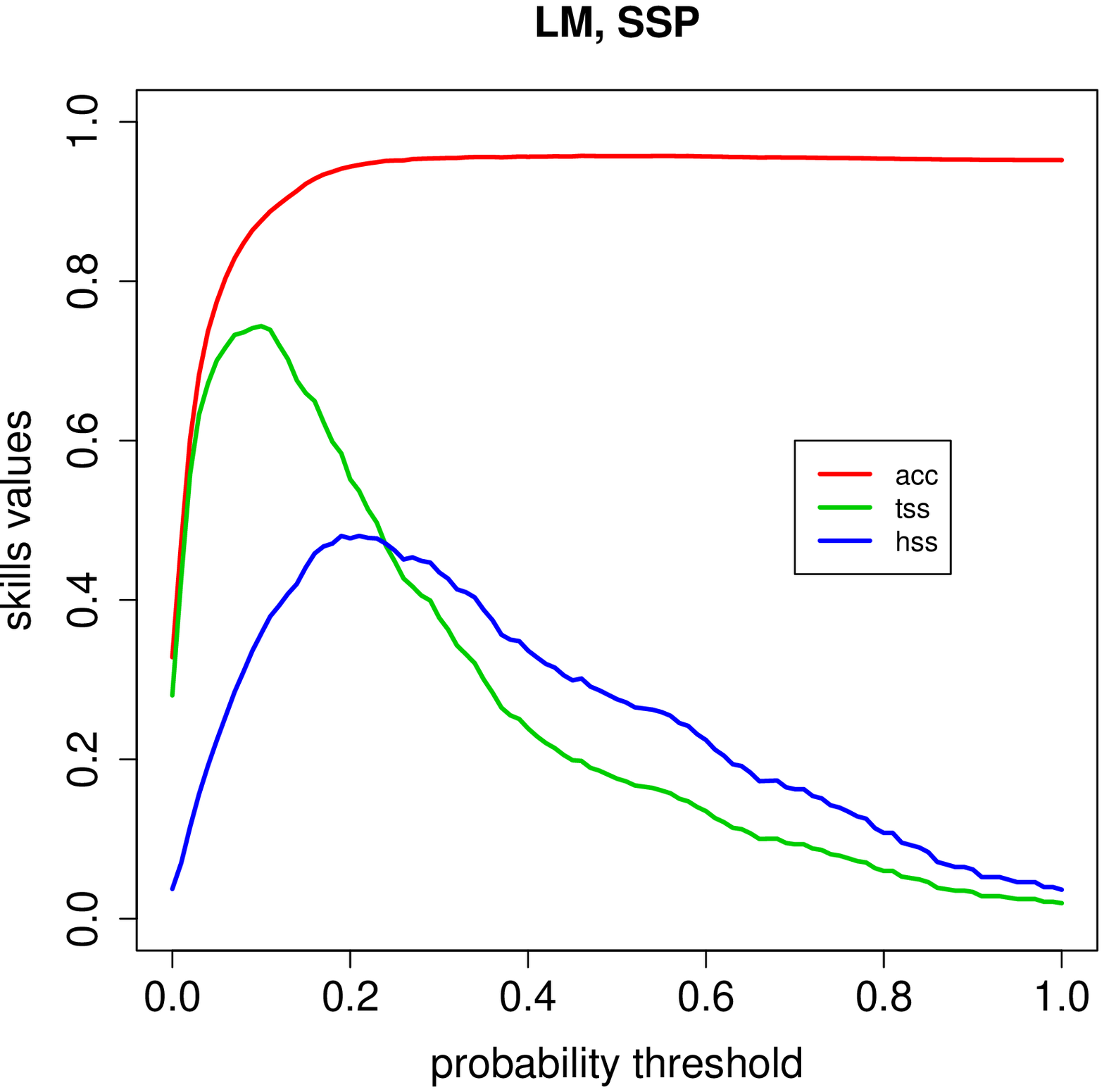}
\label{fig:aa}}
\subfloat[Subfigure 2 list of figures text][LM, ROC]{
\includegraphics[width=0.295\textwidth]{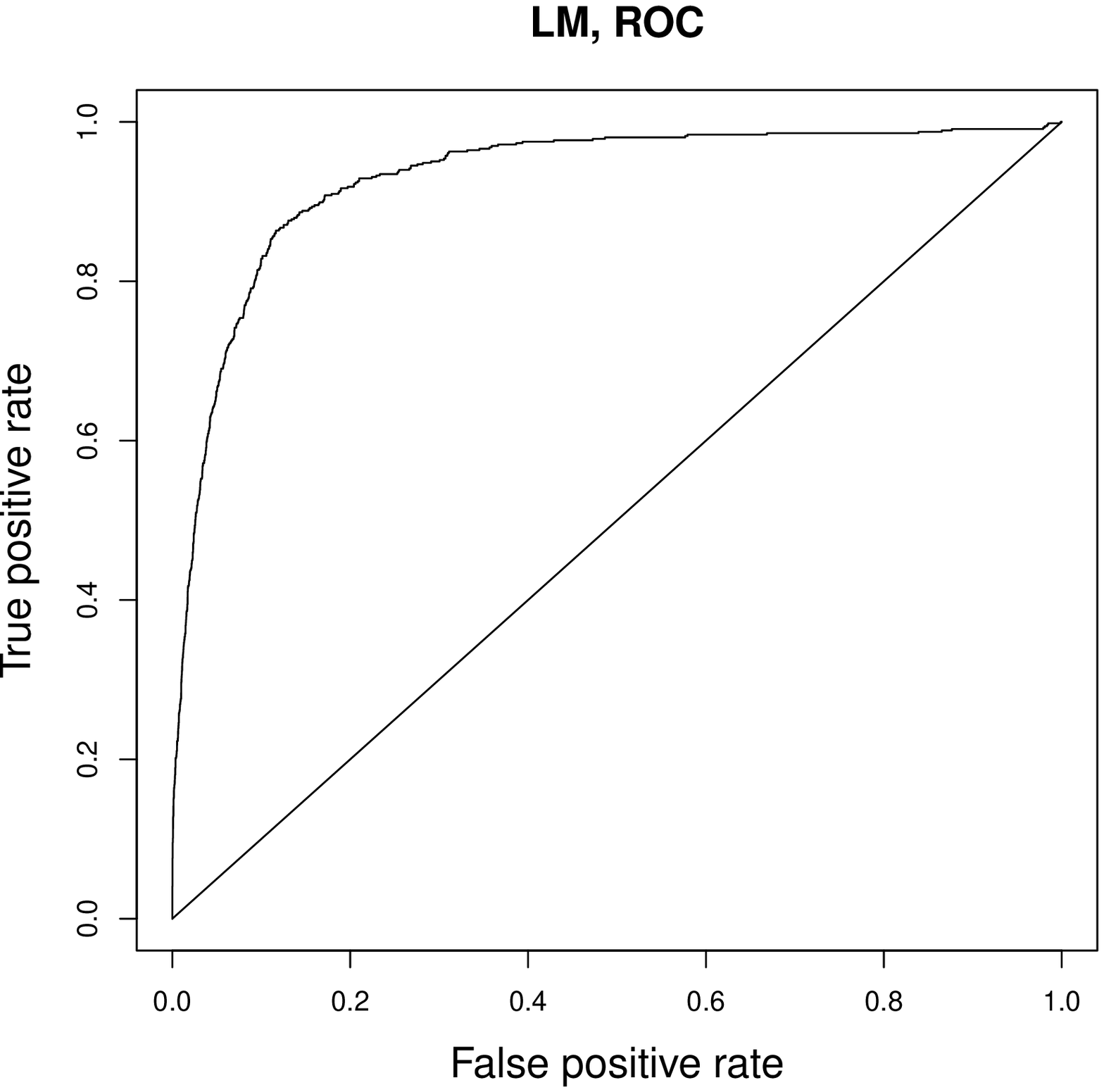}
\label{fig:bb}}
\subfloat[Subfigure 3 list of figures text][LM, RD]{
\includegraphics[width=0.295\textwidth]{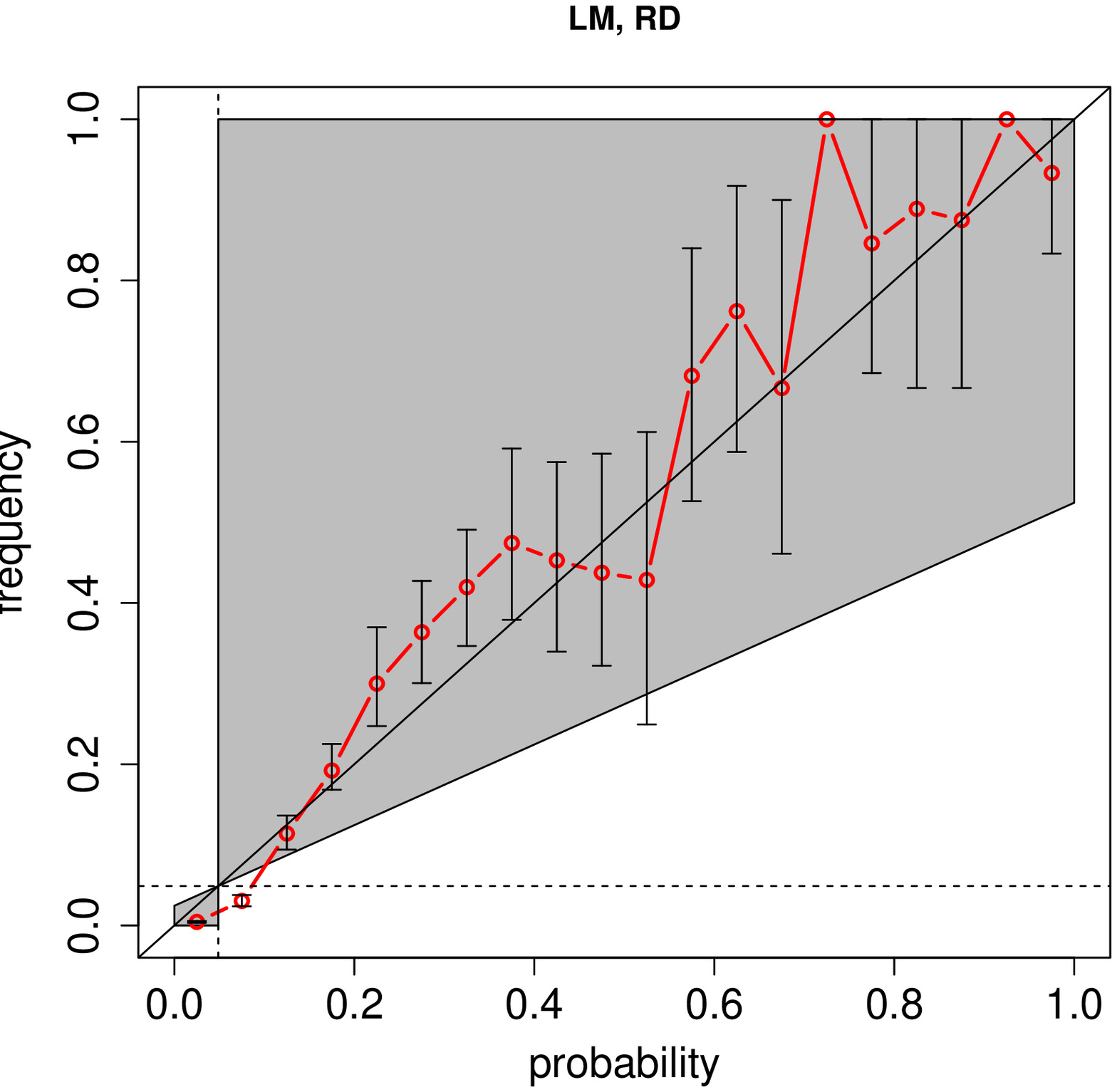}
\label{fig:cc}}
\qquad
\subfloat[Subfigure 1 list of figures text][PR, SSP]{
\includegraphics[width=0.295\textwidth]{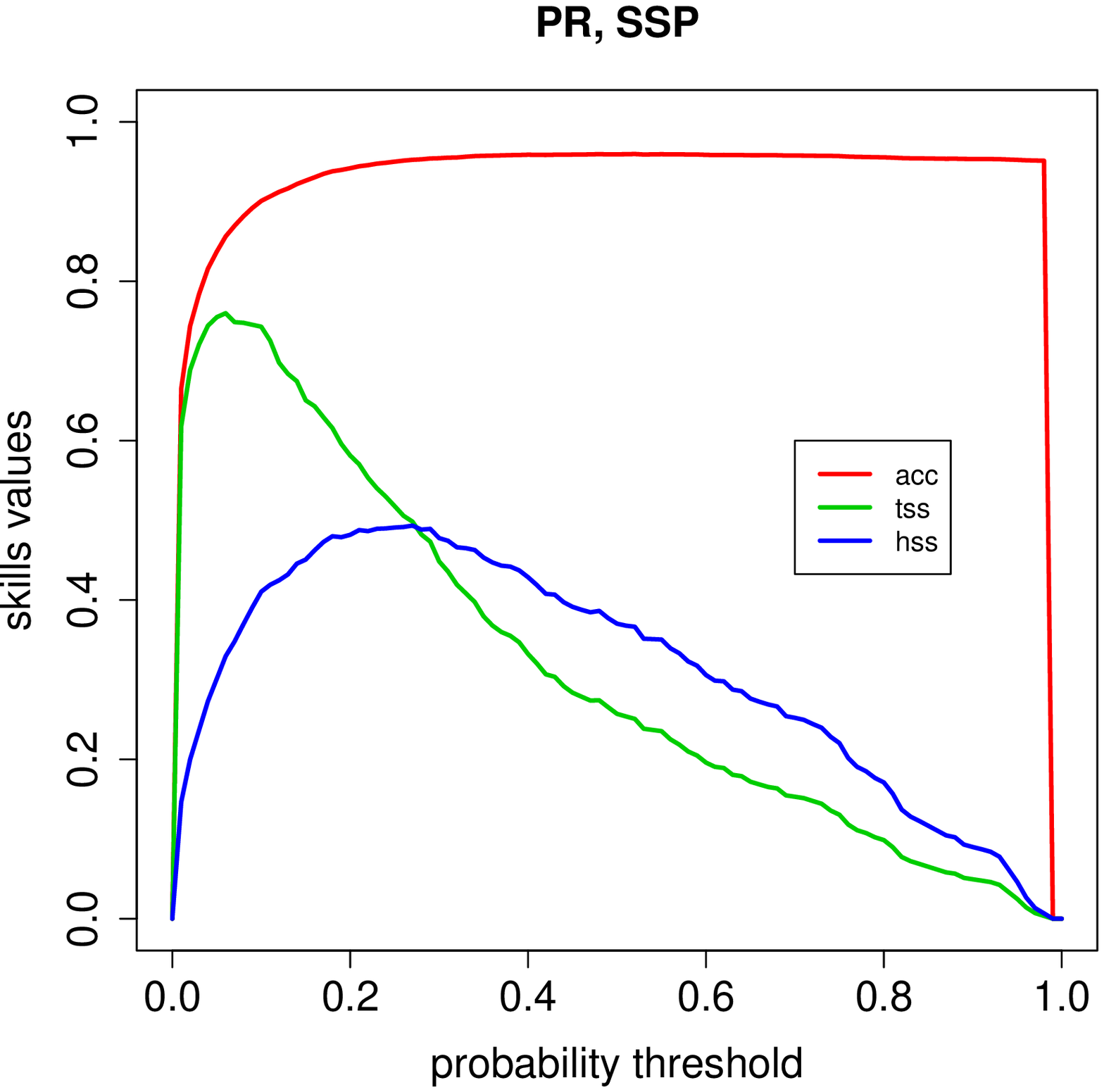}
\label{fig:dd}}
\subfloat[Subfigure 2 list of figures text][PR, ROC]{
\includegraphics[width=0.295\textwidth]{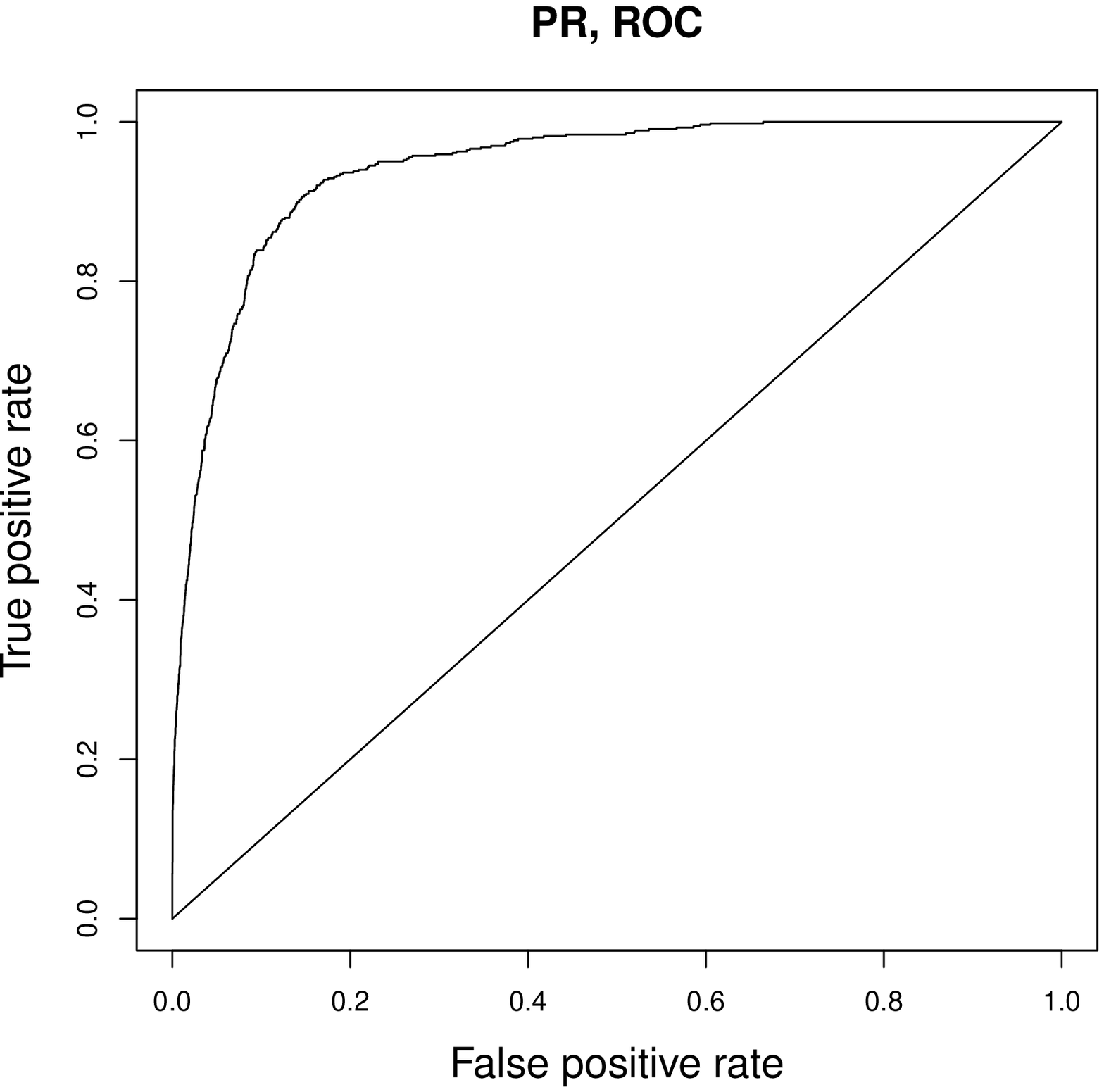}
\label{fig:ee}}
\subfloat[Subfigure 3 list of figures text][PR, RD]{
\includegraphics[width=0.295\textwidth]{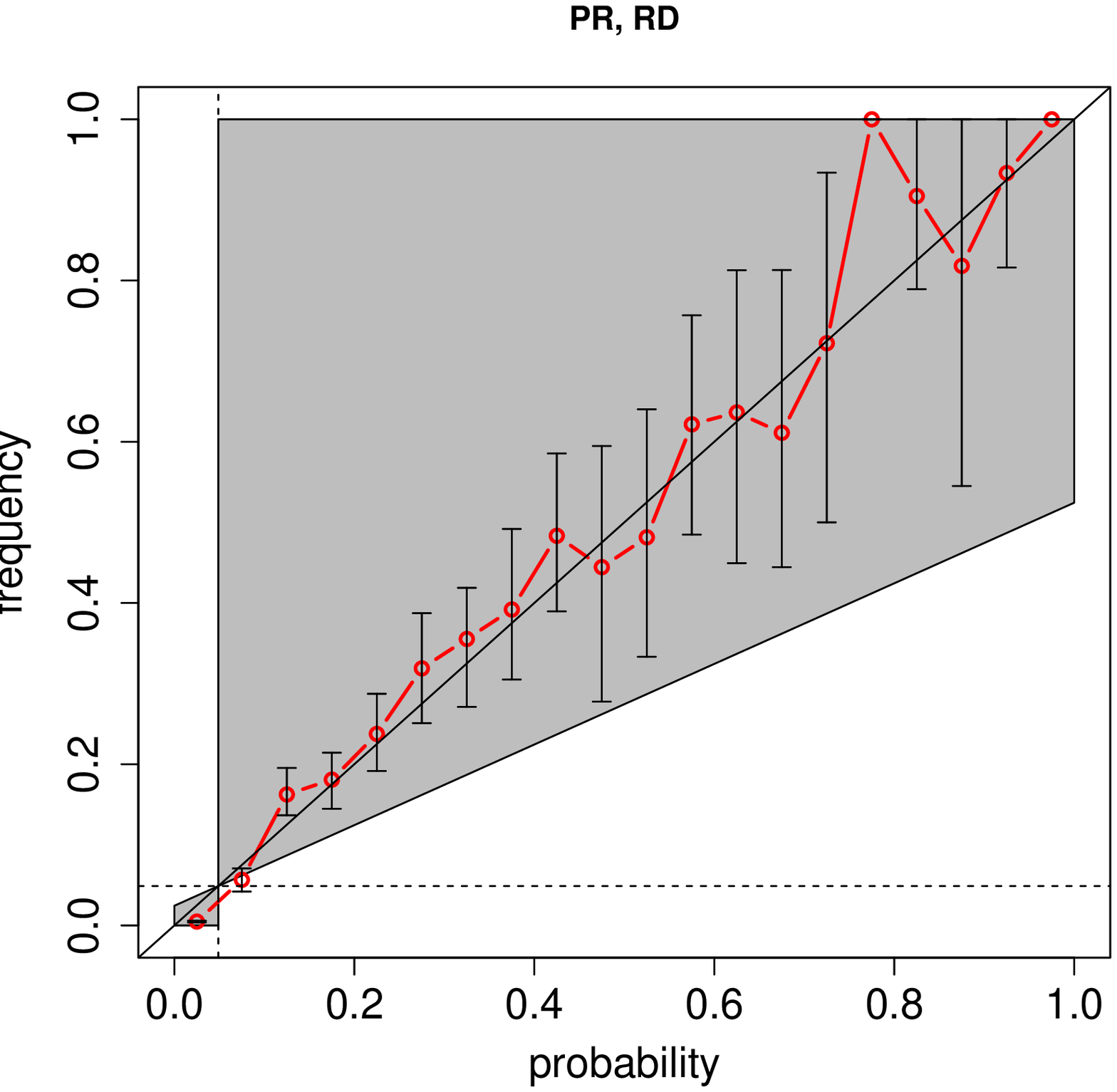}
\label{fig:ff}}
\qquad
\subfloat[Subfigure 1 list of figures text][LG, SSP]{
\includegraphics[width=0.295\textwidth]{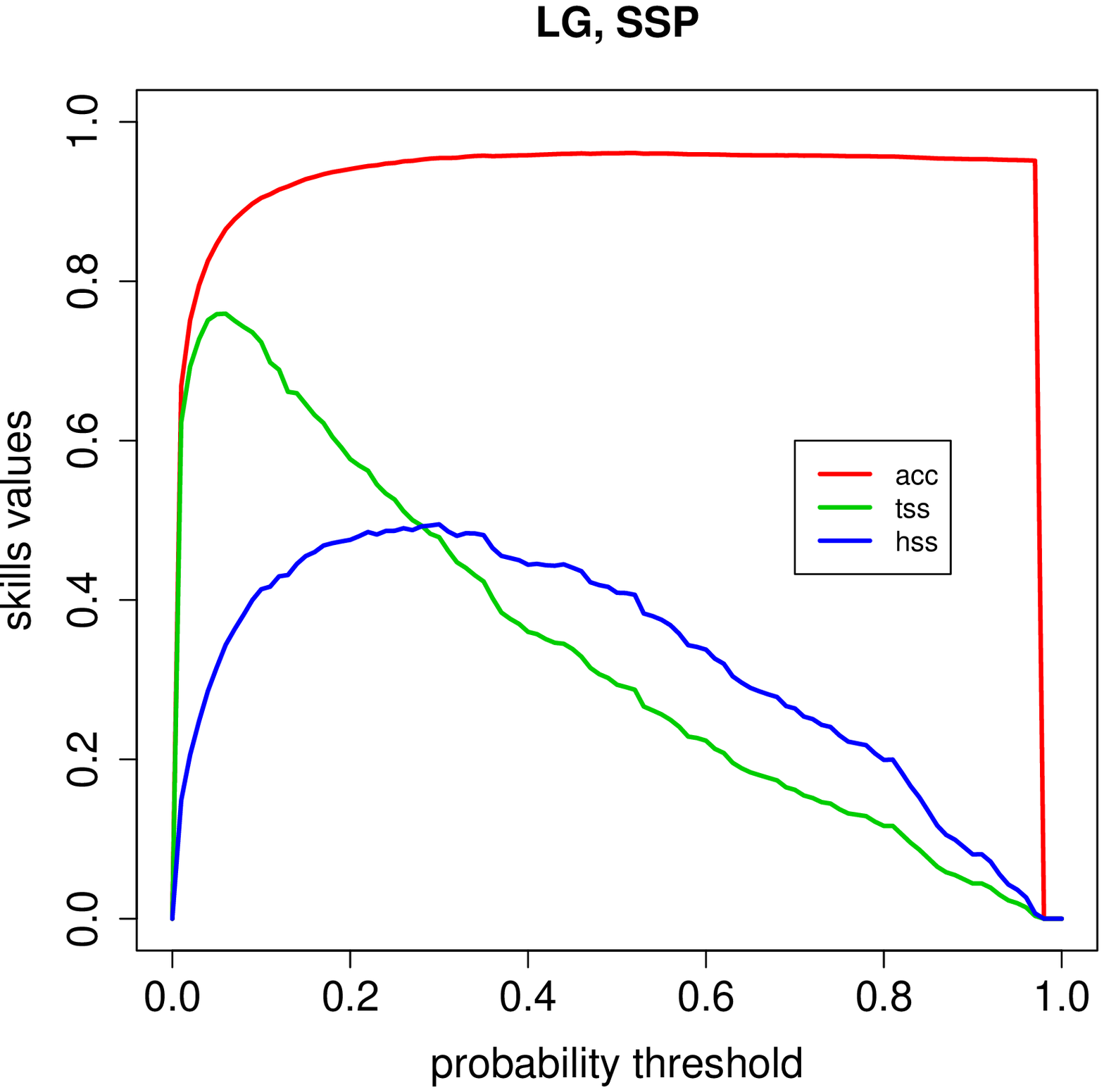}
\label{fig:gg}}
\subfloat[Subfigure 2 list of figures text][LG, ROC]{
\includegraphics[width=0.295\textwidth]{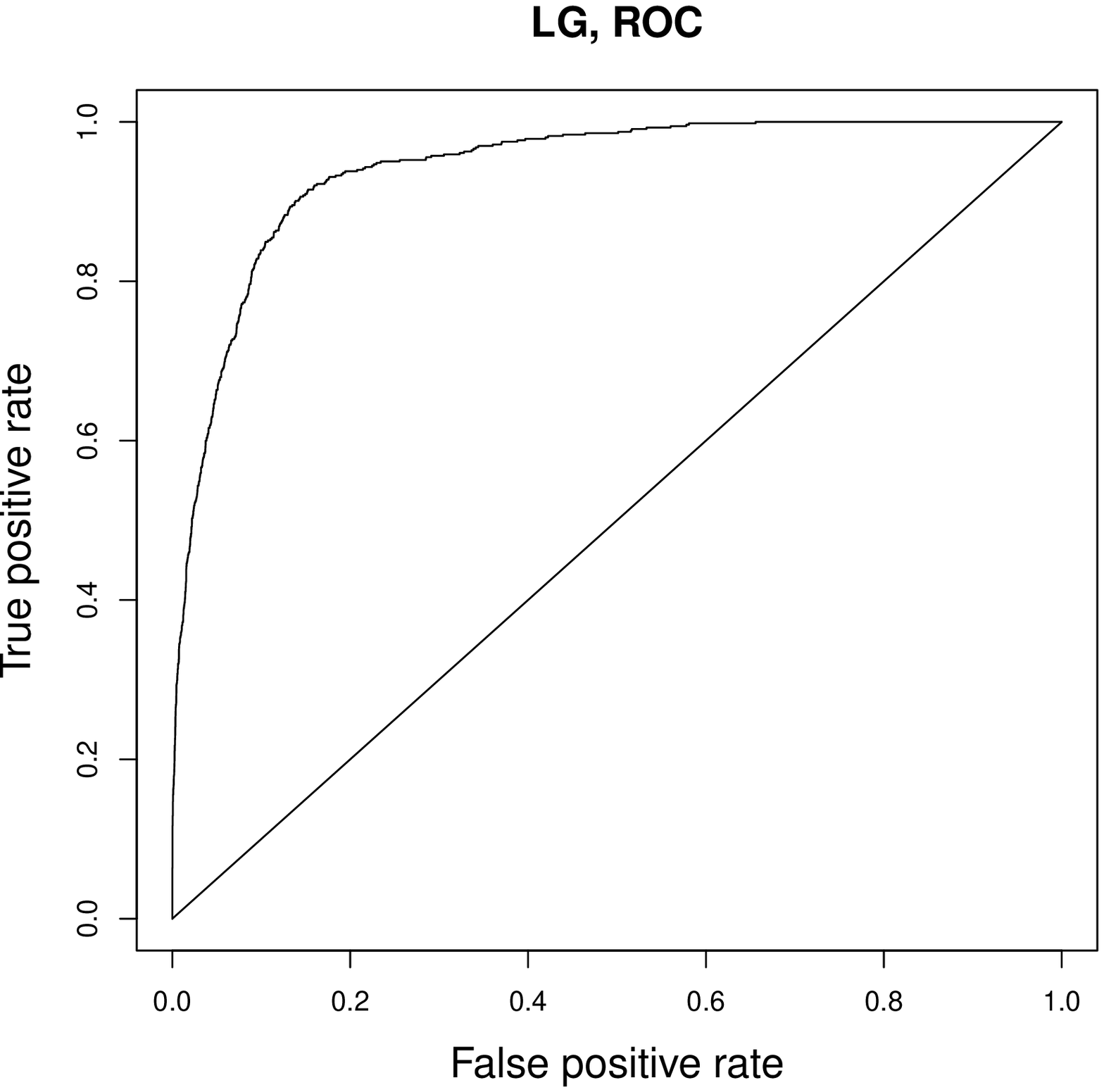}
\label{fig:hh}}
\subfloat[Subfigure 3 list of figures text][LG, RD]{
\includegraphics[width=0.295\textwidth]{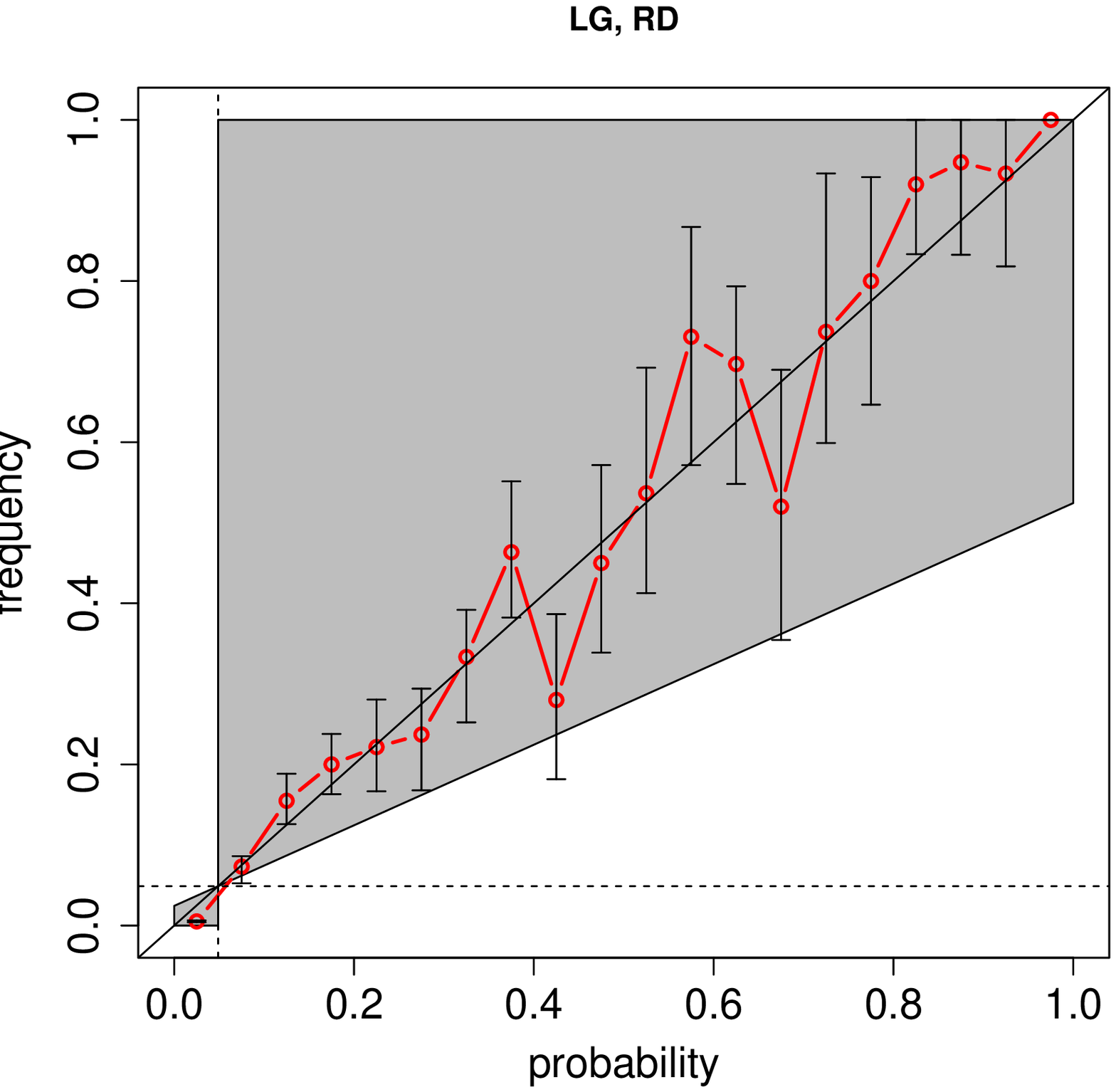}
\label{fig:ii}}
\caption{Same as Figure \ref{fig:1}, but for statistical methods: linear regression (LM; \textit{top}), probit regression (PR; \textit{middle}), and logit regression (LG; \textit{bottom}).}
\label{fig:11}

\end{figure}


\subsection{Results on $>$C1 Flare Prediction}
\subsubsection{Prediction of $>$C1 Flare Events Using Machine Learning}

\noindent We continue our computational experiments by training and performing our algorithms to the prediction of GOES $>$C1 flares. Figure \ref{fig:111} shows the forecast performances of the three tested ML methods, for $>$C1 flare prediction. In particular:

\begin{figure}[h]
\centering
\subfloat[Subfigure 1 list of figures text][MLP, SSP]{
\includegraphics[width=0.295\textwidth]{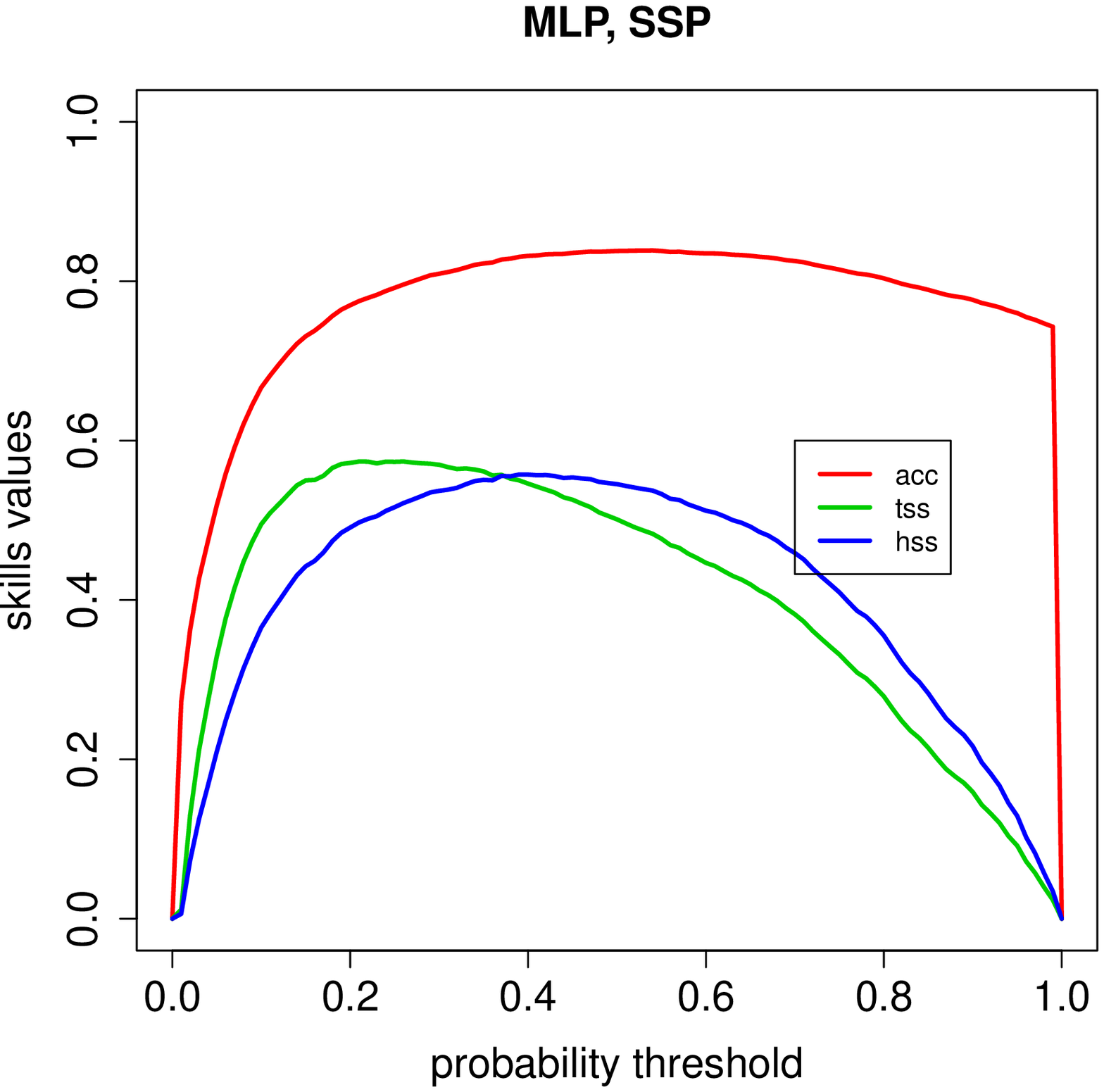}
\label{fig:aaa}}
\subfloat[Subfigure 2 list of figures text][MLP, ROC]{
\includegraphics[width=0.295\textwidth]{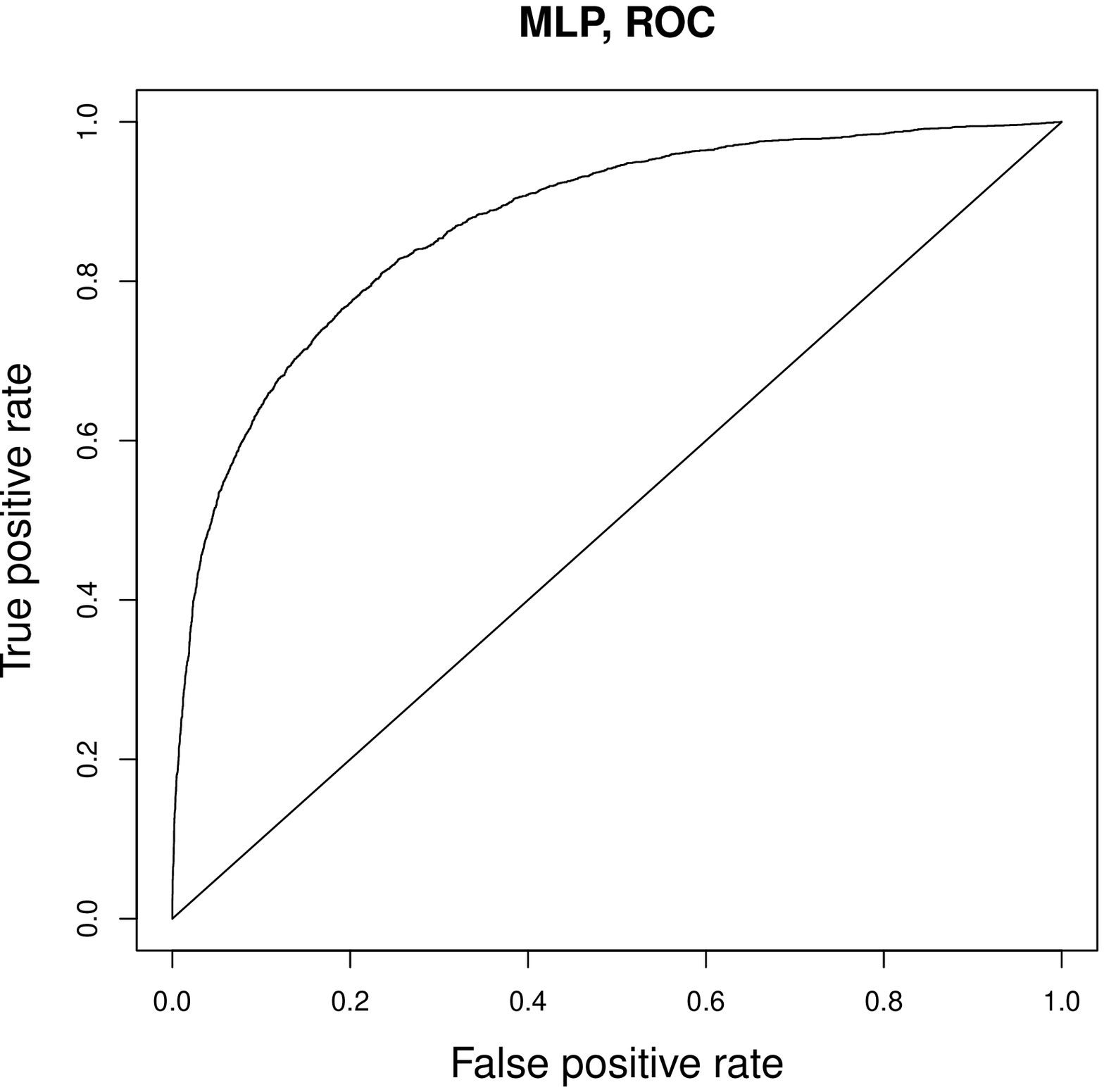}
\label{fig:bbb}}
\subfloat[Subfigure 3 list of figures text][MLP, RD]{
\includegraphics[width=0.295\textwidth]{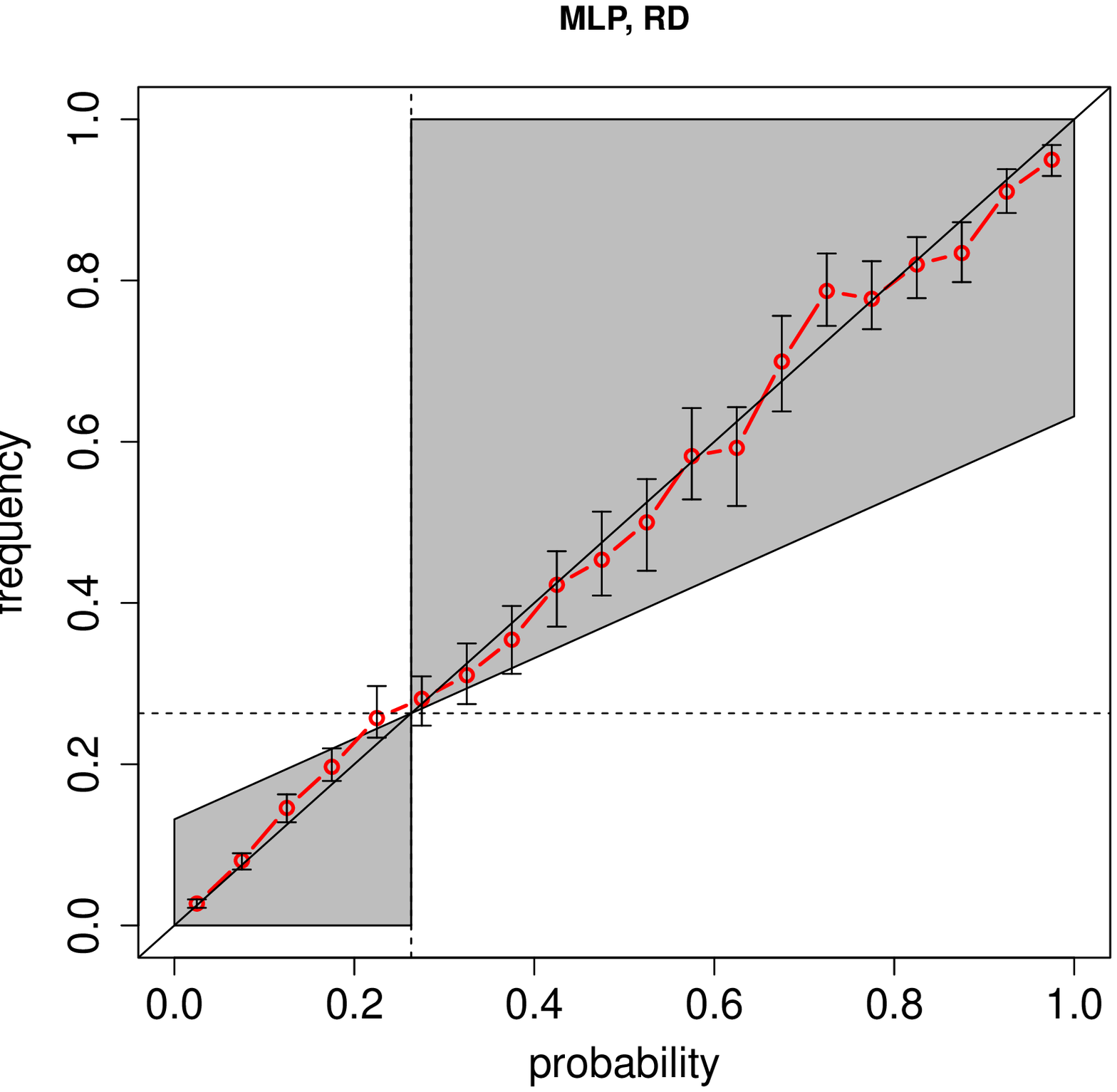}
\label{fig:ccc}}
\qquad
\subfloat[Subfigure 1 list of figures text][SVM, SSP]{
\includegraphics[width=0.295\textwidth]{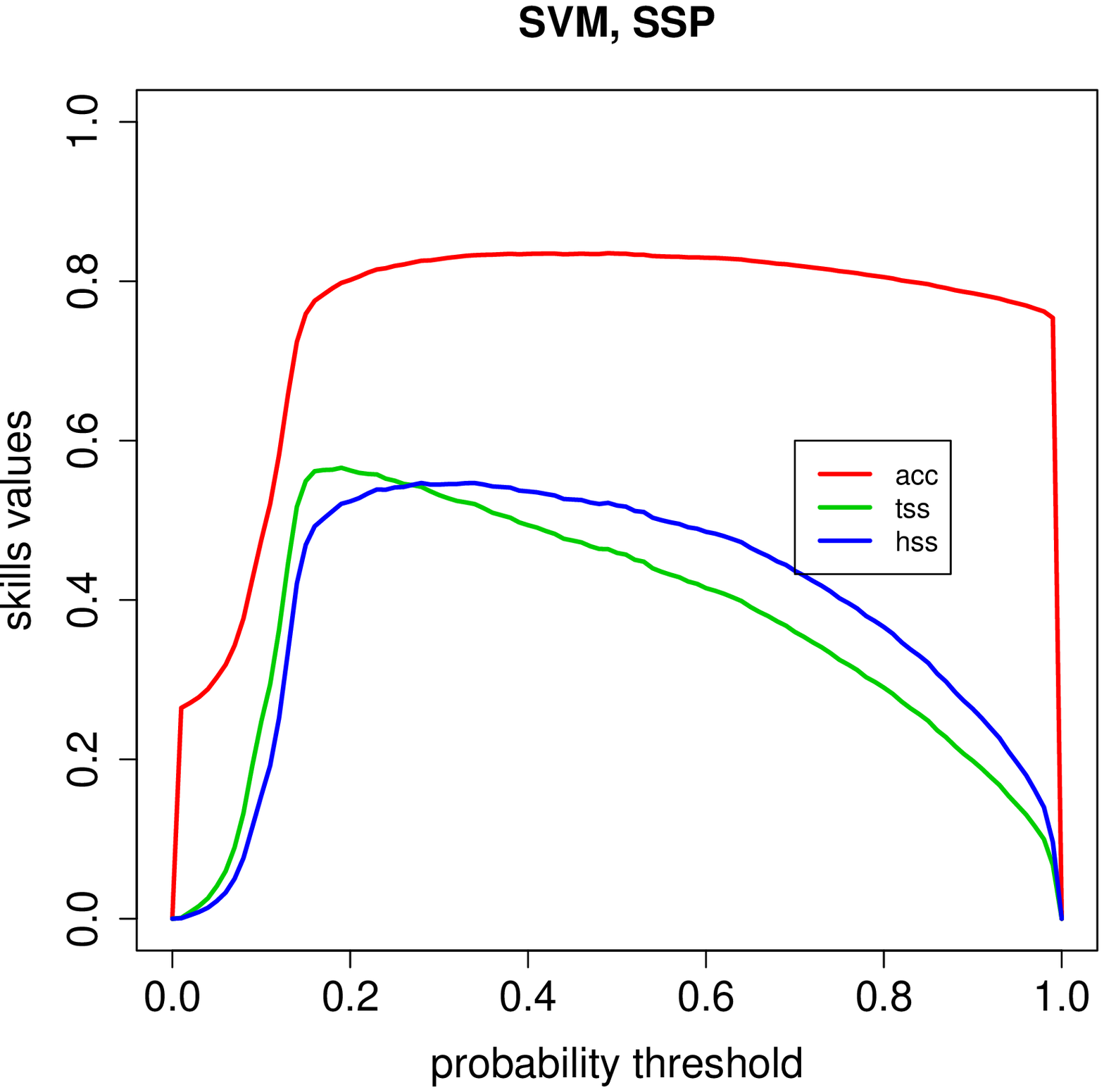}
\label{fig:ddd}}
\subfloat[Subfigure 2 list of figures text][SVM, ROC]{
\includegraphics[width=0.295\textwidth]{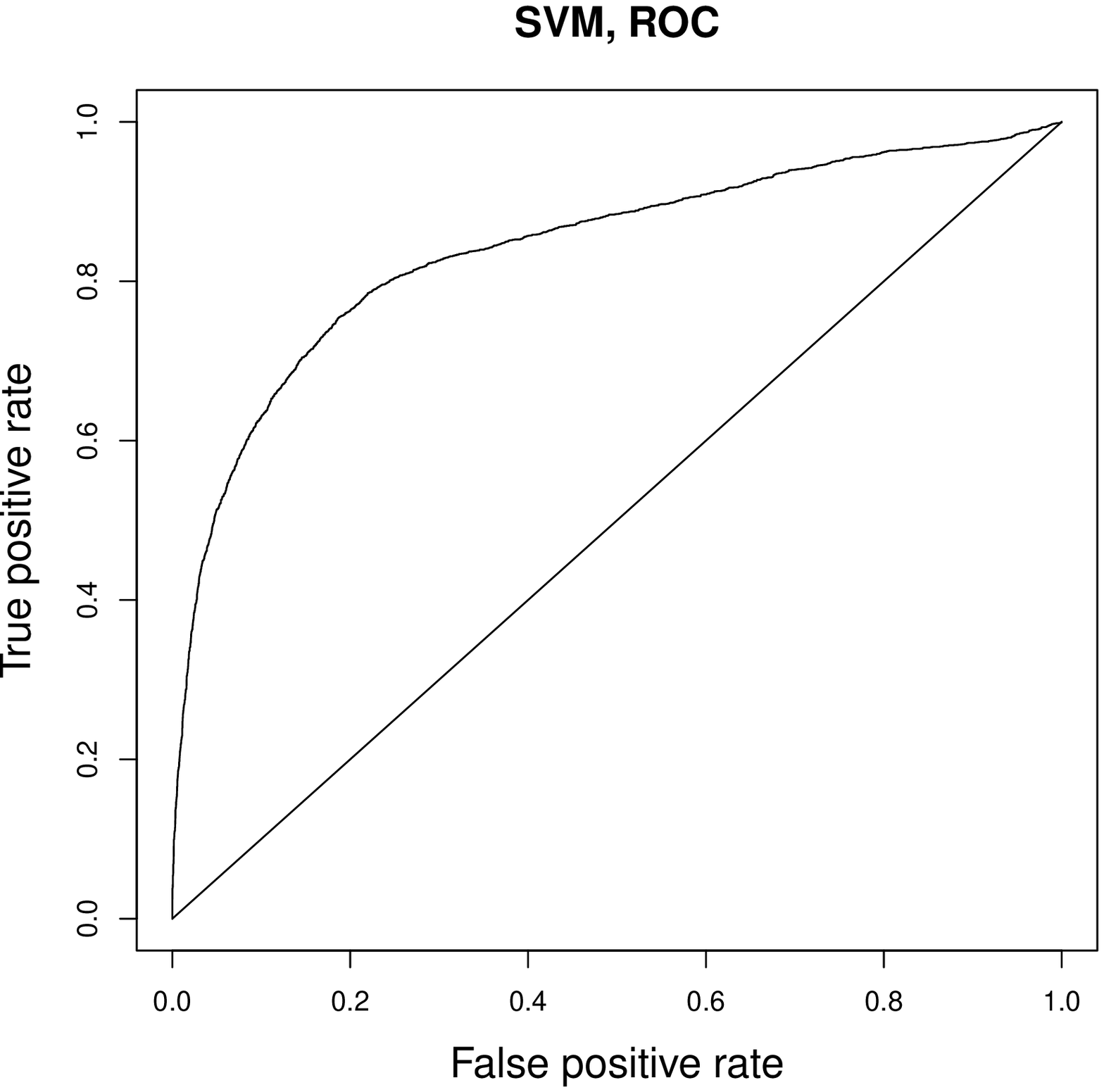}
\label{fig:eee}}
\subfloat[Subfigure 3 list of figures text][SVM, RD]{
\includegraphics[width=0.295\textwidth]{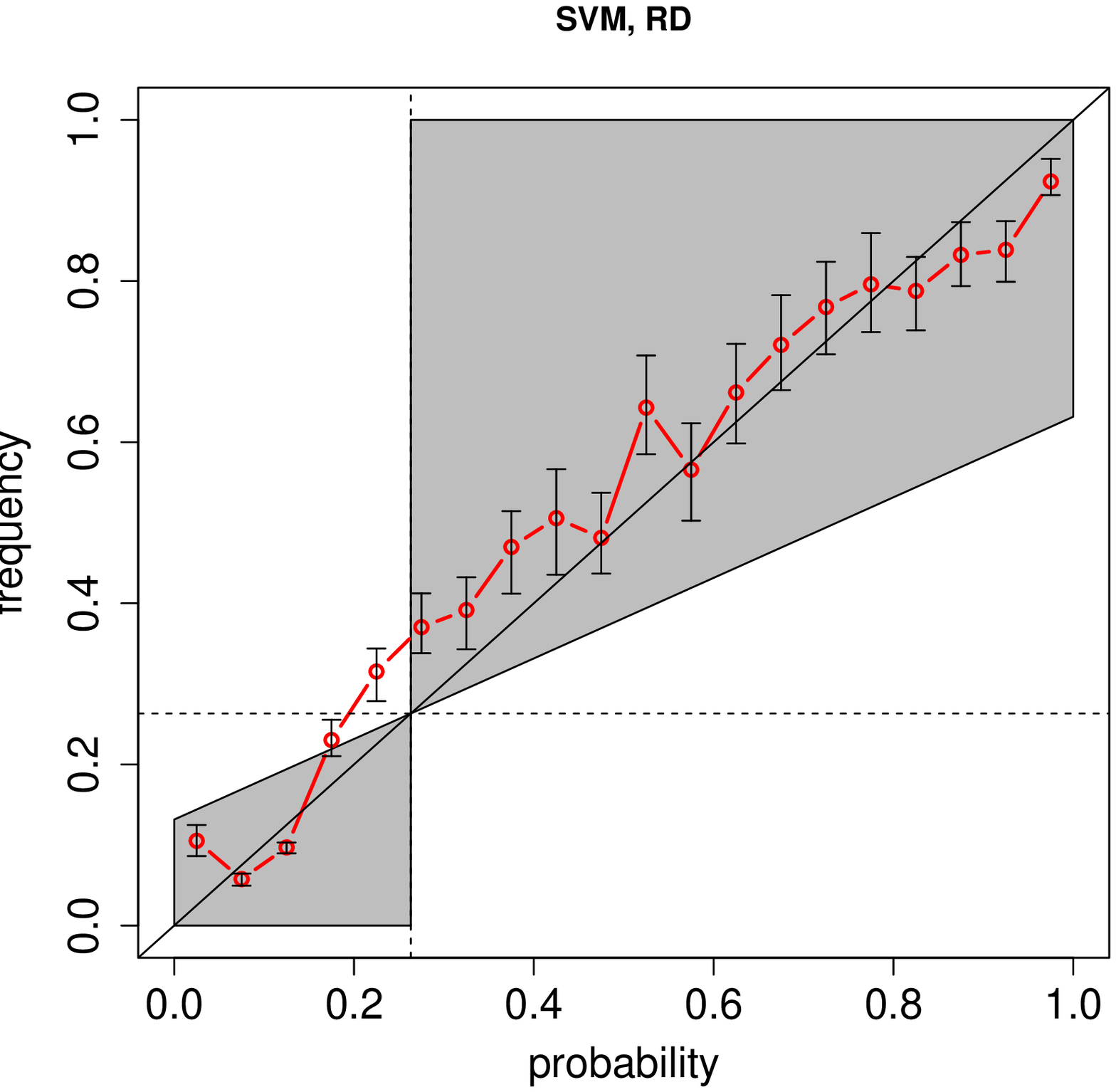}
\label{fig:fff}}
\qquad
\subfloat[Subfigure 1 list of figures text][RF, SSP]{
\includegraphics[width=0.295\textwidth]{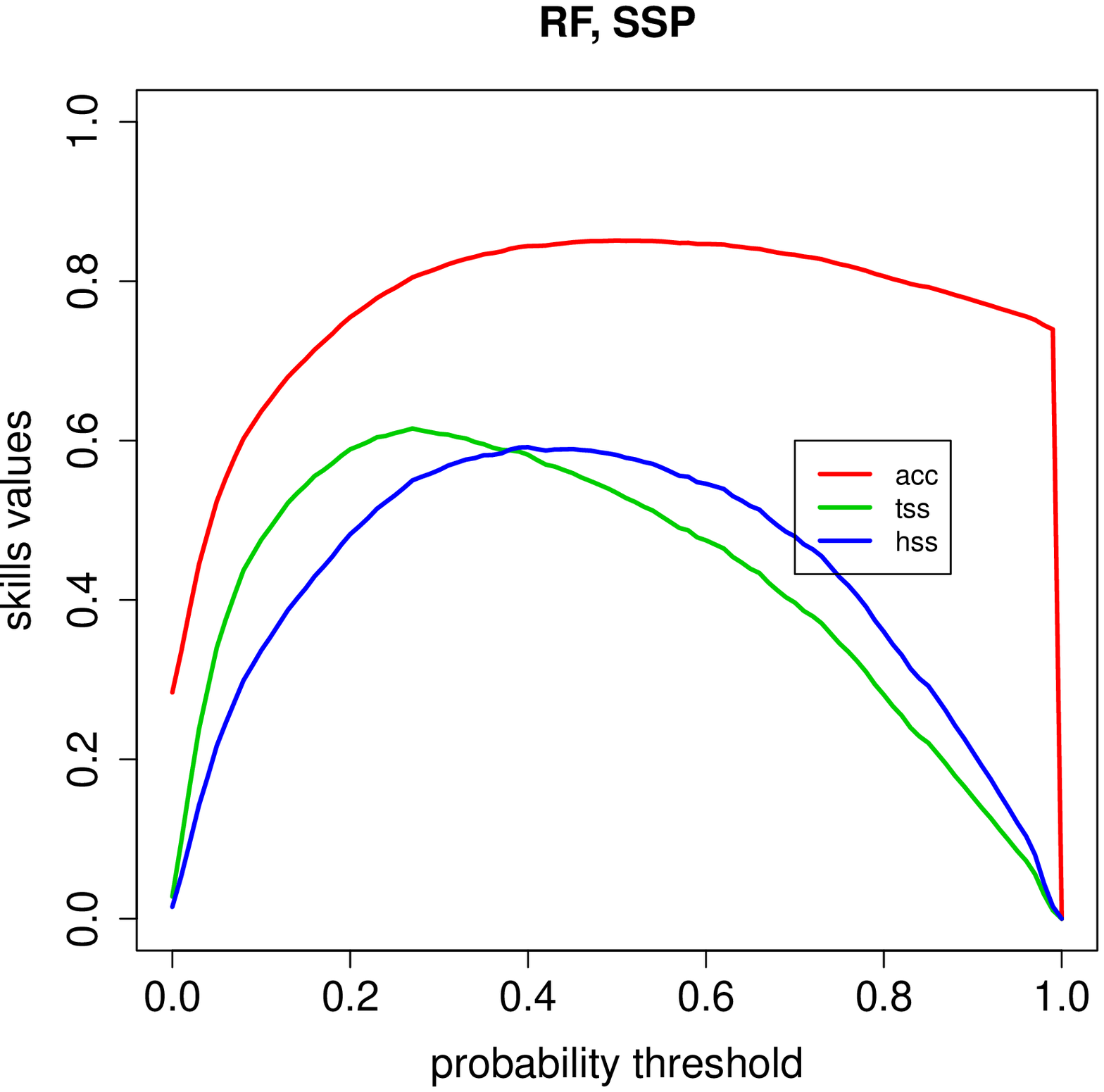}
\label{fig:ggg}}
\subfloat[Subfigure 2 list of figures text][RF, ROC]{
\includegraphics[width=0.295\textwidth]{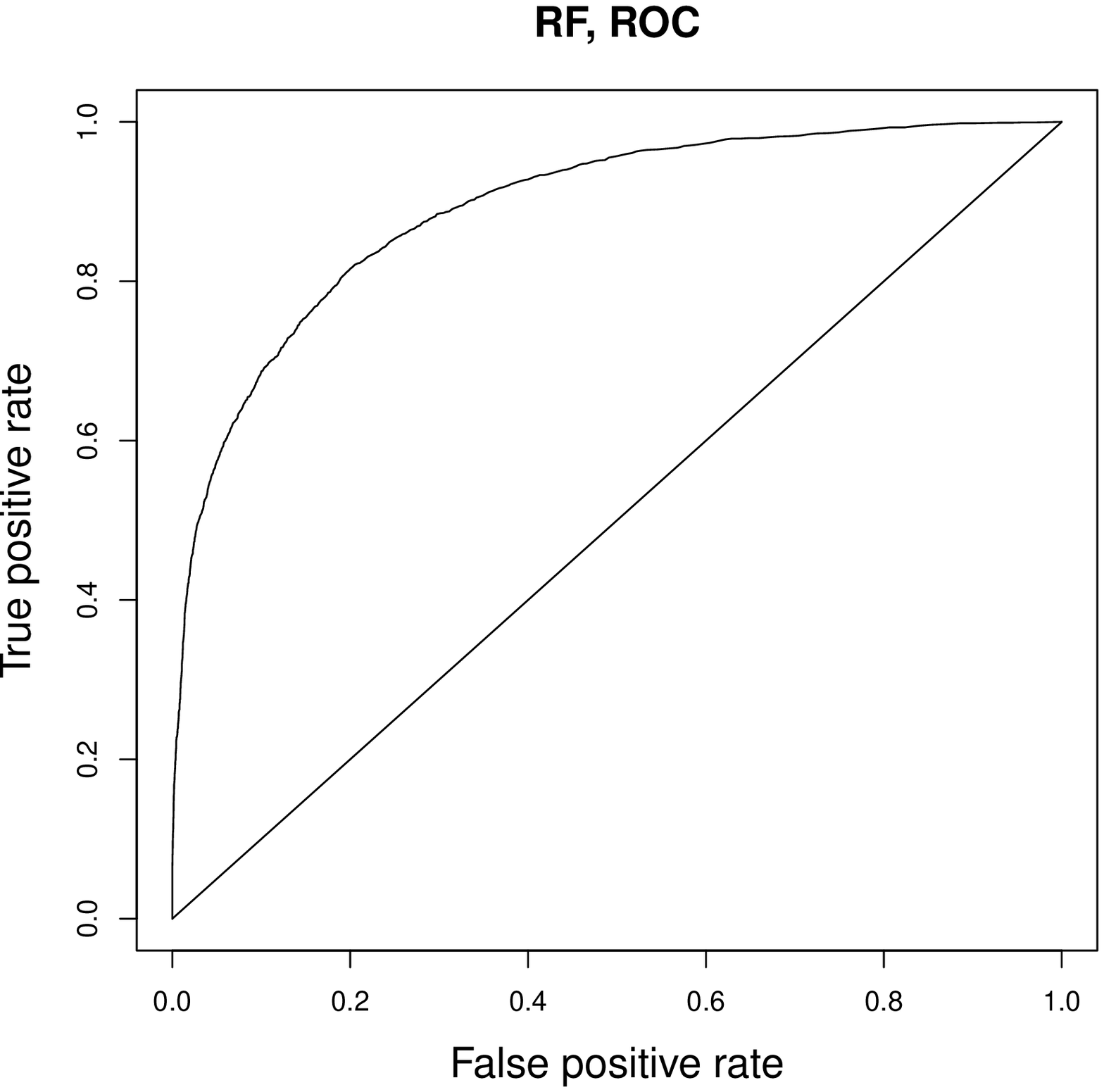}
\label{fig:hhh}}
\subfloat[Subfigure 3 list of figures text][RF, RD]{
\includegraphics[width=0.295\textwidth]{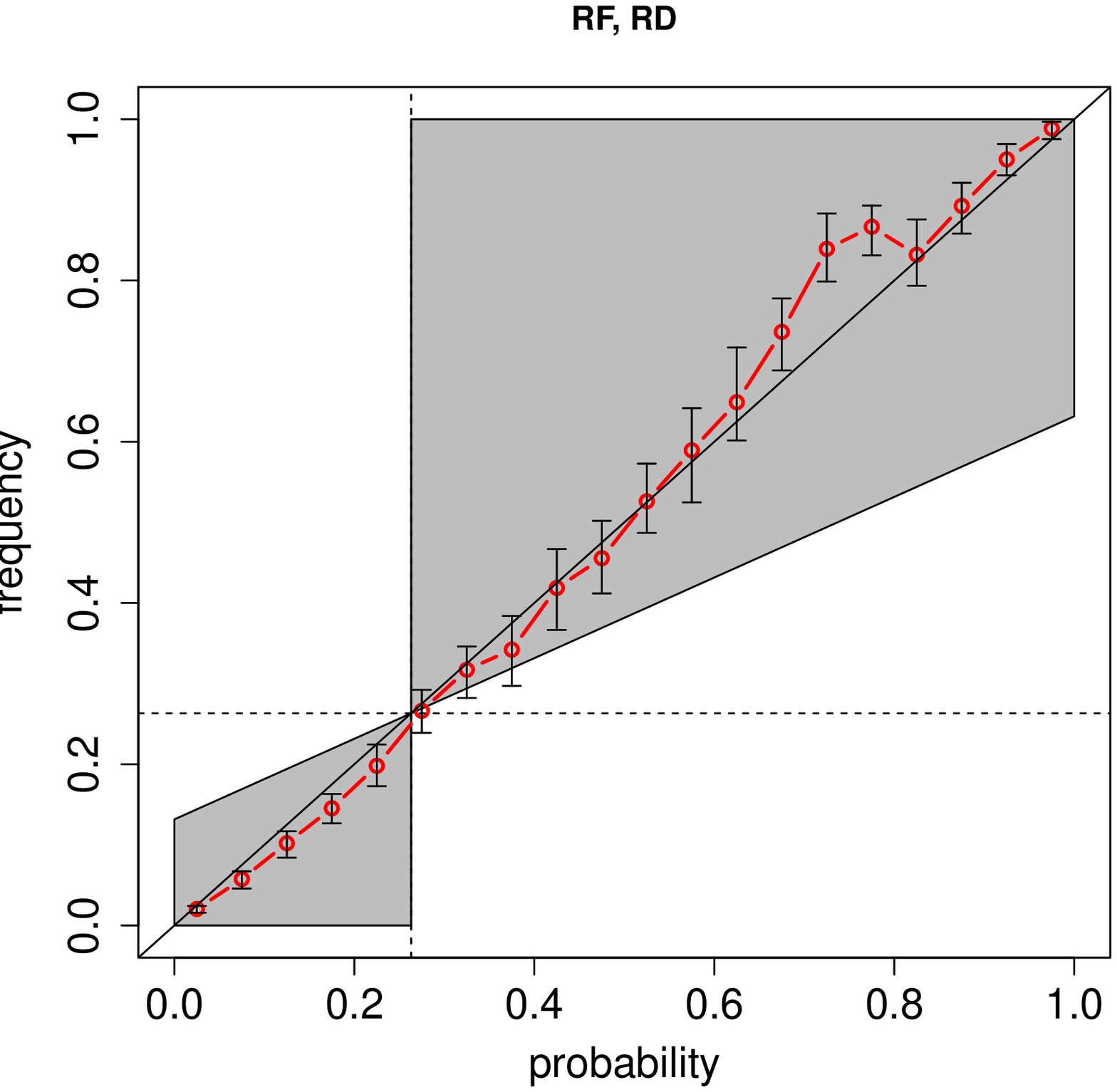}
\label{fig:iii}}
\caption{Same as Figure \ref{fig:1}, but for $>$C1 flare prediction.}
\label{fig:111}

\end{figure}

Regarding the MLP, we notice that since for the $>$C1 flares the number of hidden nodes selected is {\rm size}=4, plateaus in {\rm HSS} and {\rm TSS} are not so eminent, contrary to the case of $>$M1 flare prediction. The ROC curve seems satisfactory with maximum {\rm TSS}=0.574 and the RD is quite significant, showing no systematic over- or under-prediction.

With respect to the SVM, a purely monotonic decrease of {\rm TSS} can be seen, following an instantaneous peak. Some plateau in {\rm HSS} is also noticed, followed by a monotonic decrease. The ROC curve appears less satisfactory than in case of MLPs with maximum {\rm TSS}=0.566 and the RD shows some systematic under-prediction for most of the forecast probabilities range.

For the RFs, one notices a relatively similar behavior with MLPs, albeit with a slightly more pronounced {\rm HSS} peak. The ROC curve seems better behaved than in the previous two methods with maximum {\rm TSS}=0.615 and the RD is arguably the best achieved together with the MLP RD.

\subsubsection{Prediction of $>$C1 Flare Events Using Statistical Models}

Figure \ref{fig:1111} shows the forecast performances of the three tested statistical methods, for $>$C1 class flare prediction. In particular:

\begin{figure}[h]
\centering
\subfloat[Subfigure 1 list of figures text][LM, SSP]{
\includegraphics[width=0.295\textwidth]{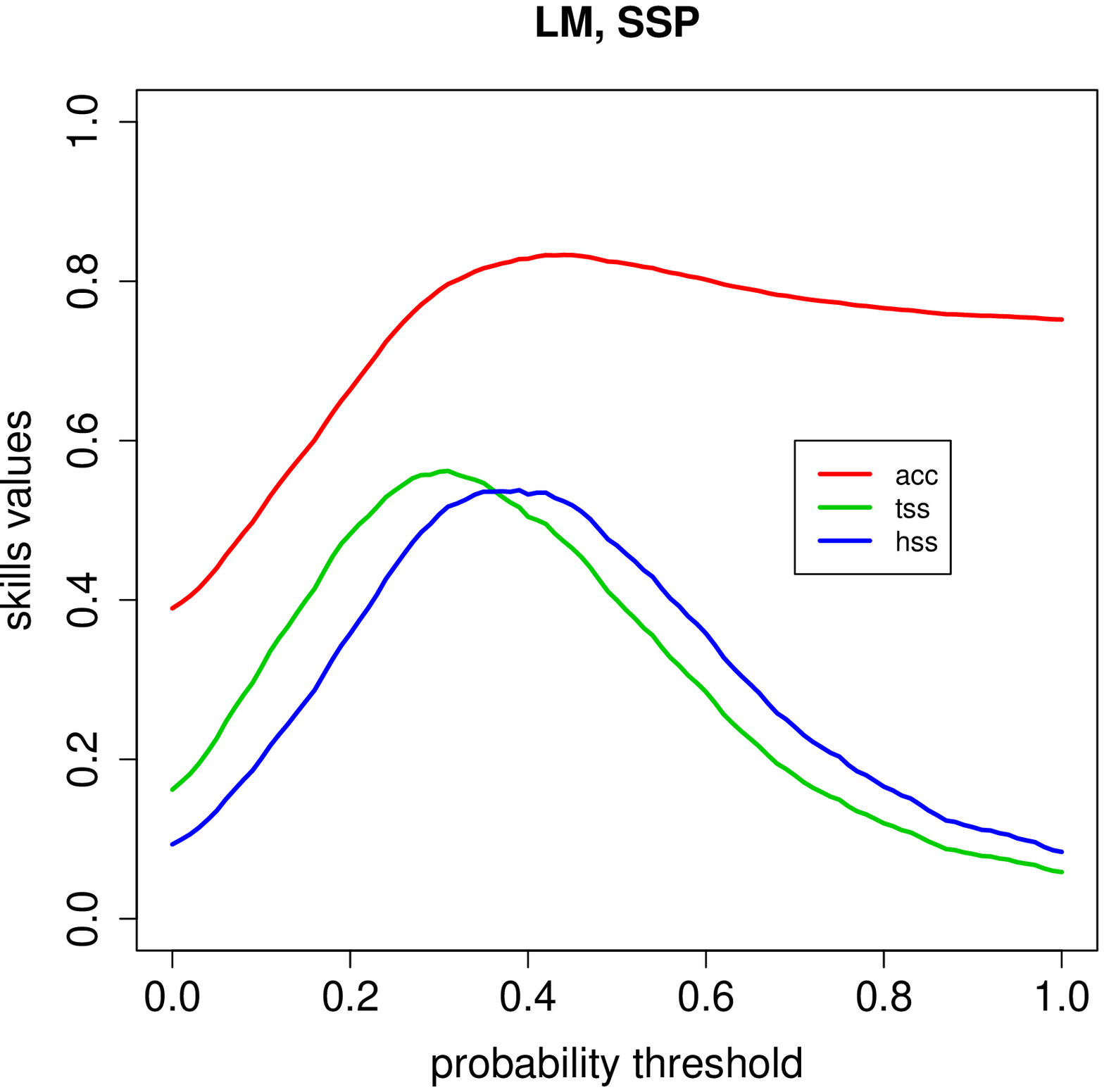}
\label{fig:aaaa}}
\subfloat[Subfigure 2 list of figures text][LM, ROC]{
\includegraphics[width=0.295\textwidth]{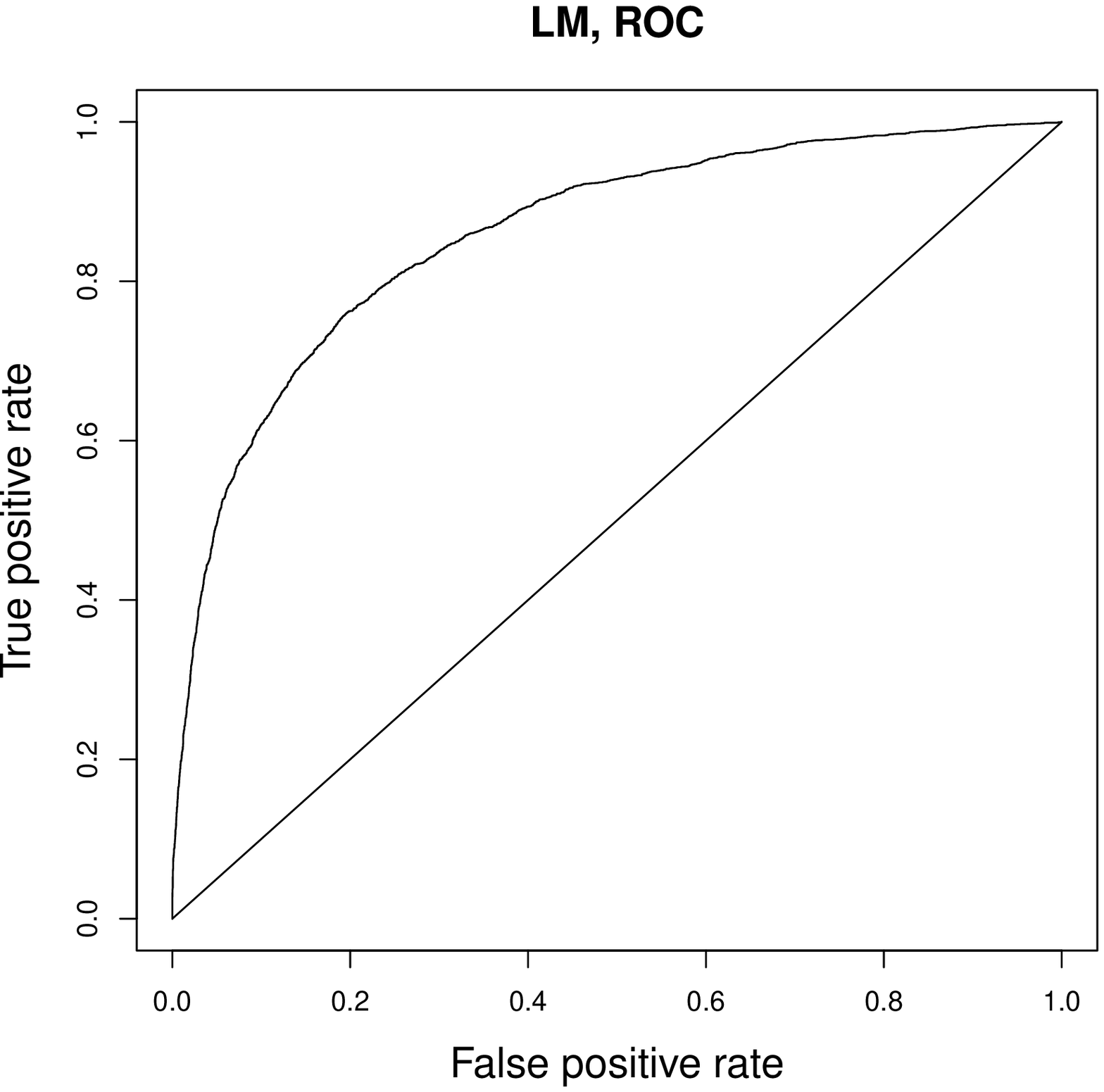}
\label{fig:bbbb}}
\subfloat[Subfigure 3 list of figures text][LM, RD]{
\includegraphics[width=0.295\textwidth]{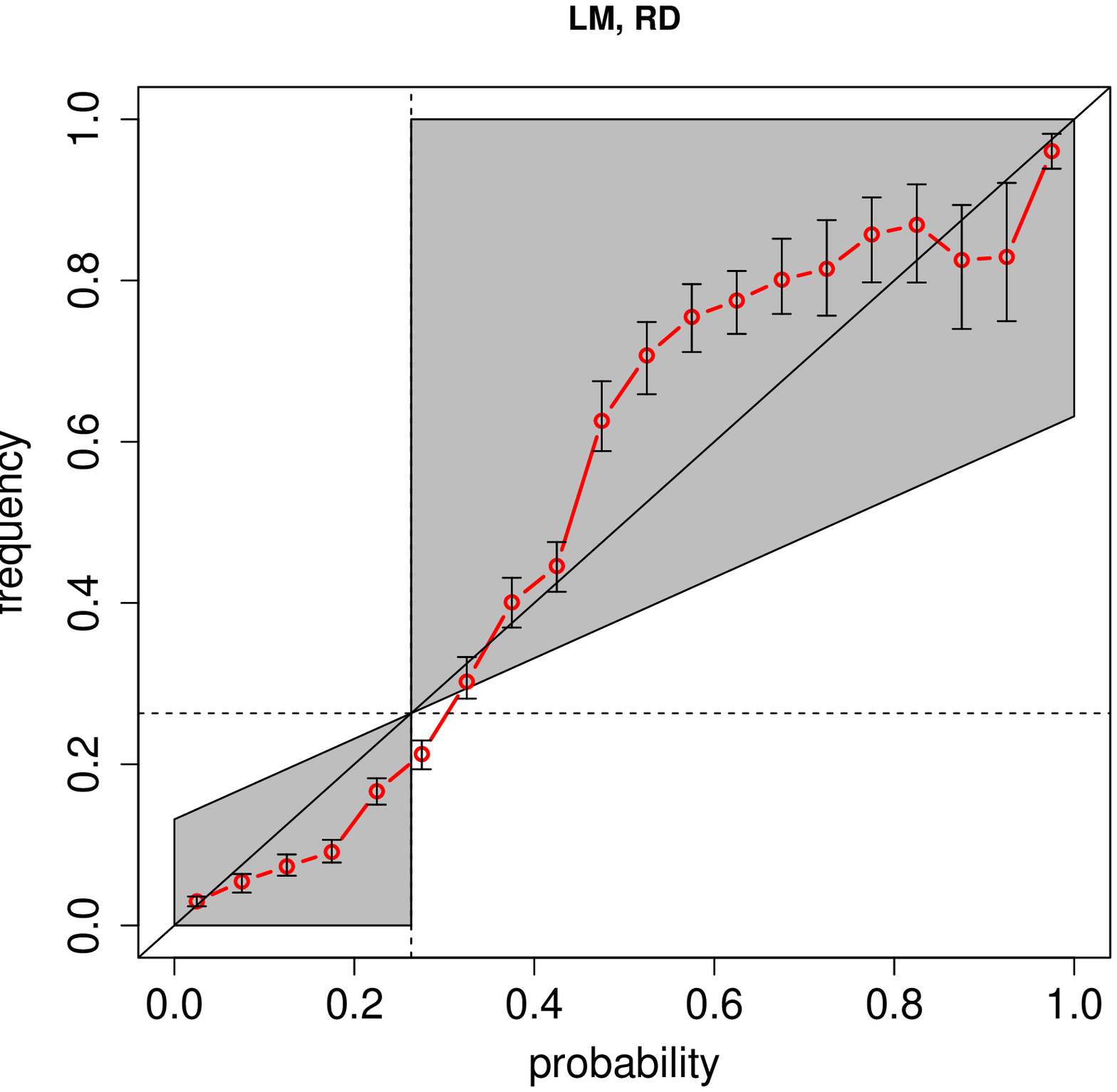}
\label{fig:cccc}}
\qquad
\subfloat[Subfigure 1 list of figures text][PR, SSP]{
\includegraphics[width=0.295\textwidth]{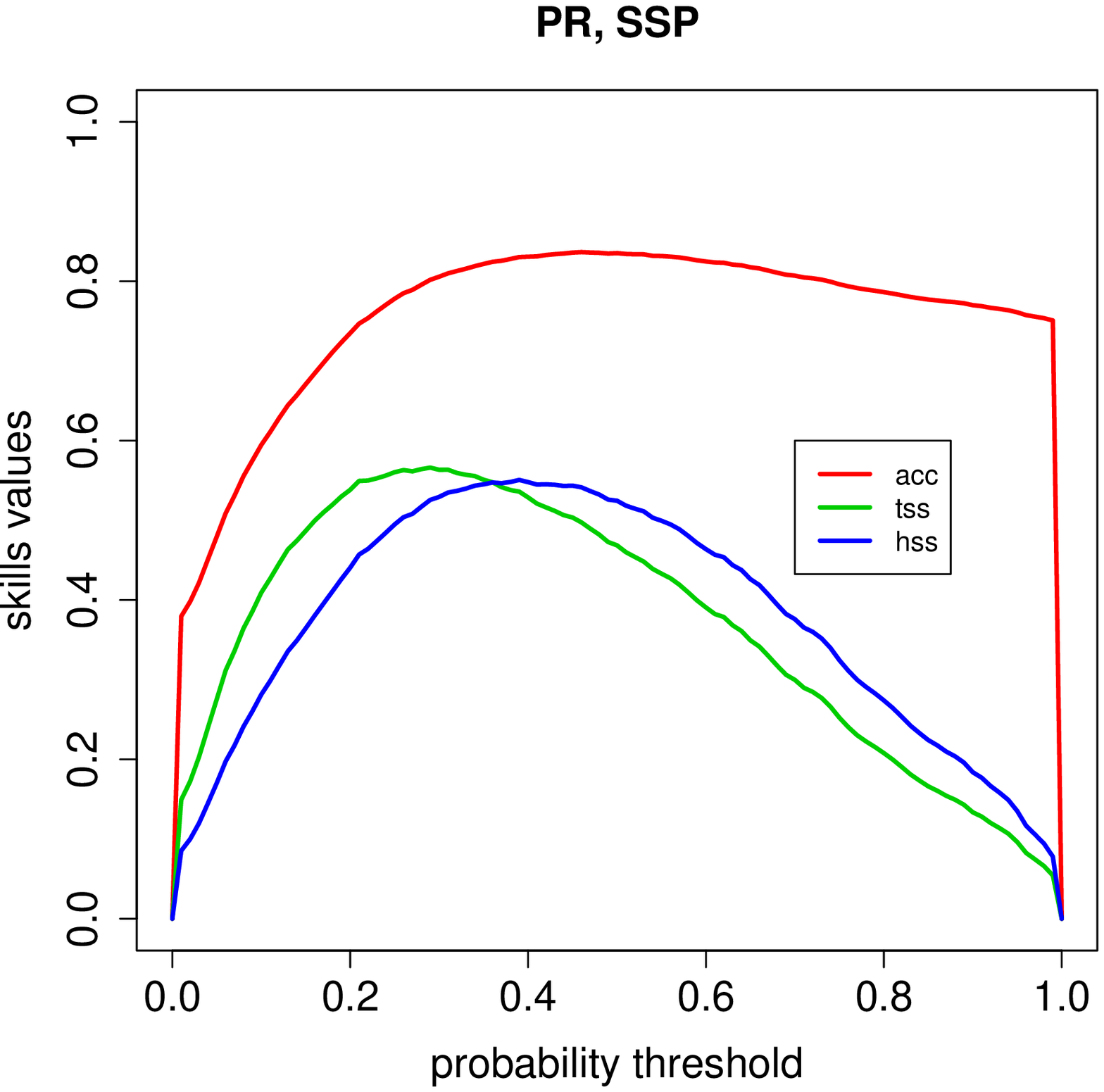}
\label{fig:dddd}}
\subfloat[Subfigure 2 list of figures text][PR, ROC]{
\includegraphics[width=0.295\textwidth]{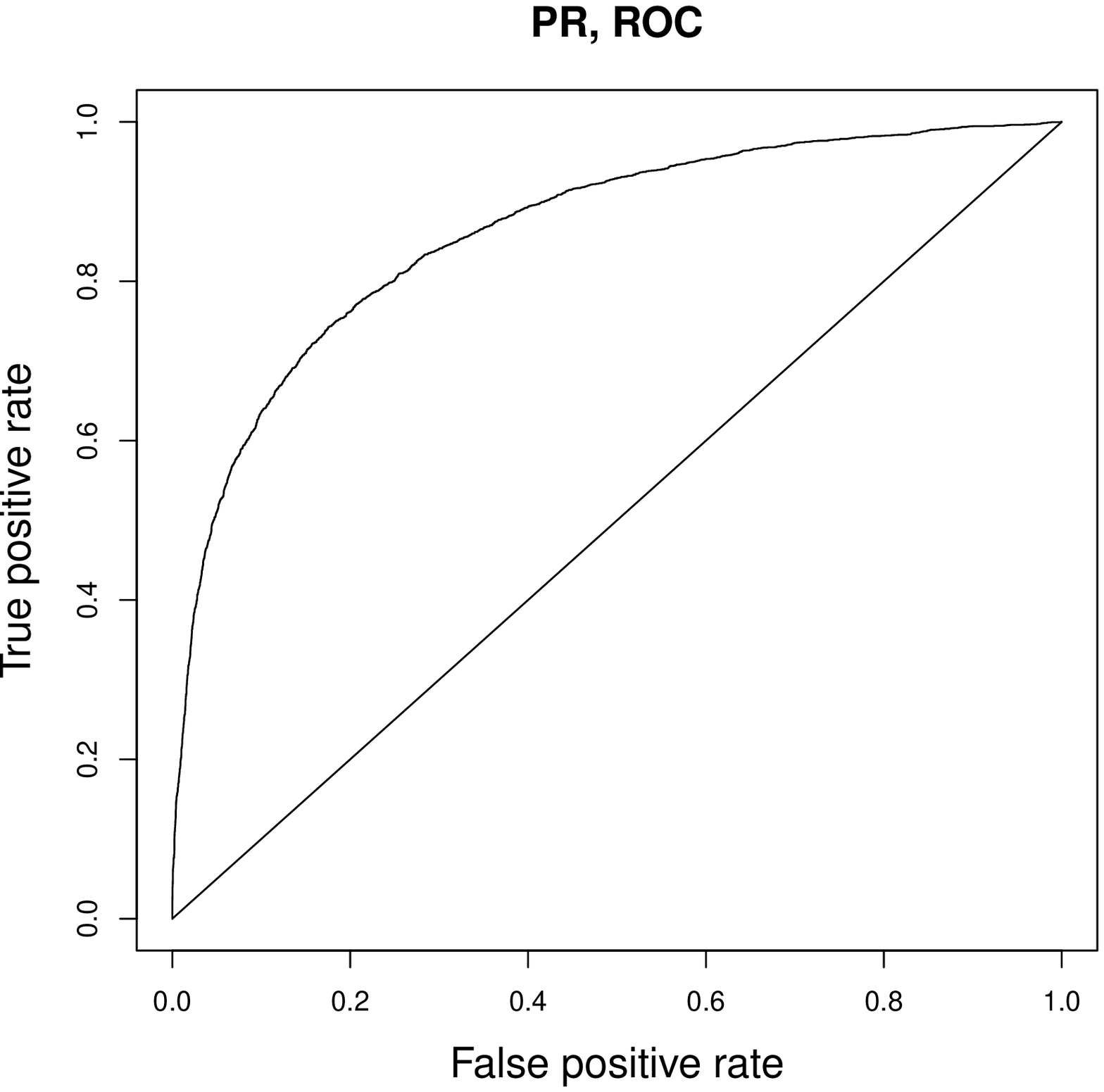}
\label{fig:eeee}}
\subfloat[Subfigure 3 list of figures text][PR, RD]{
\includegraphics[width=0.295\textwidth]{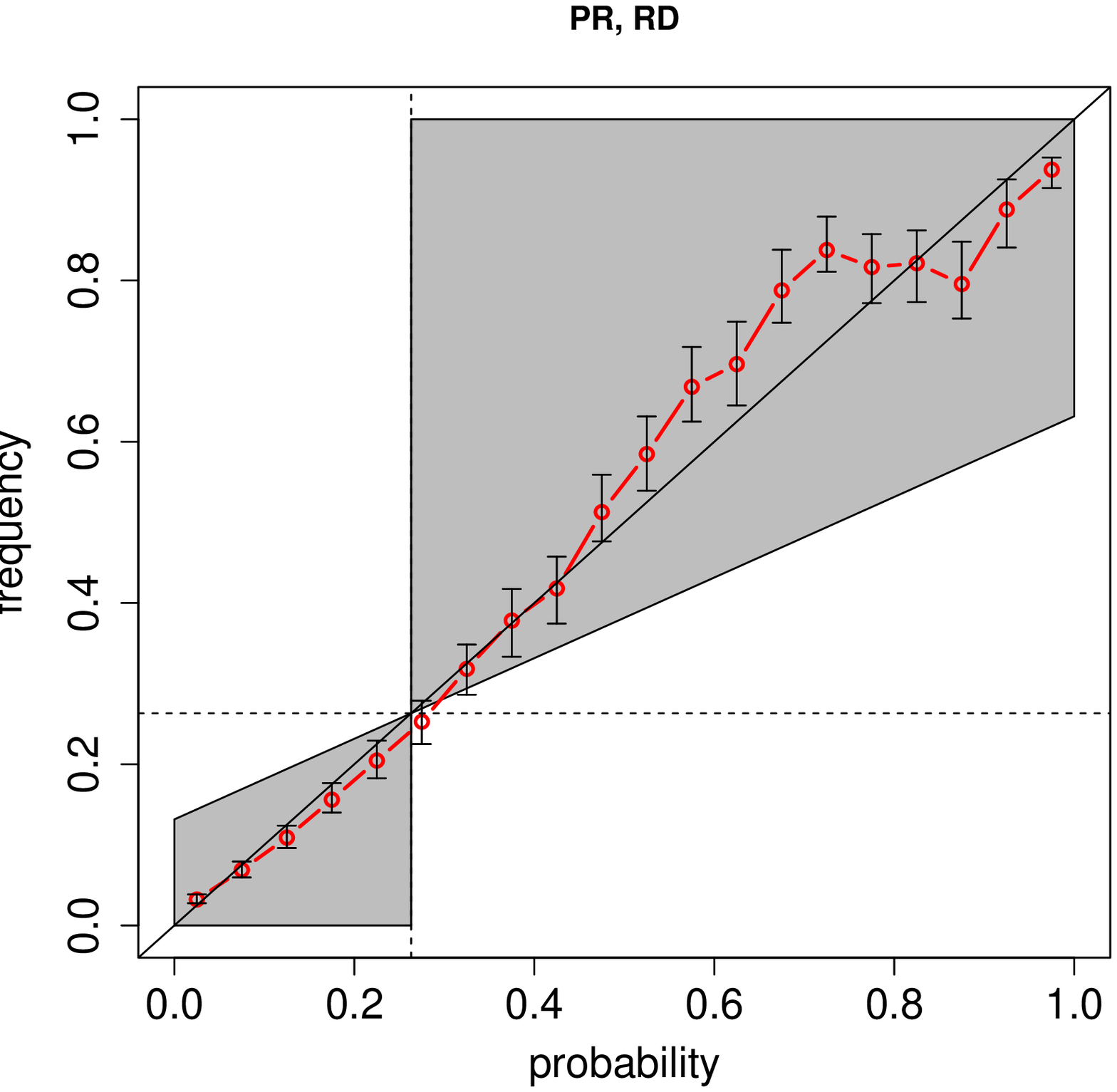}
\label{fig:ffff}}
\qquad
\subfloat[Subfigure 1 list of figures text][LG, SSP]{
\includegraphics[width=0.295\textwidth]{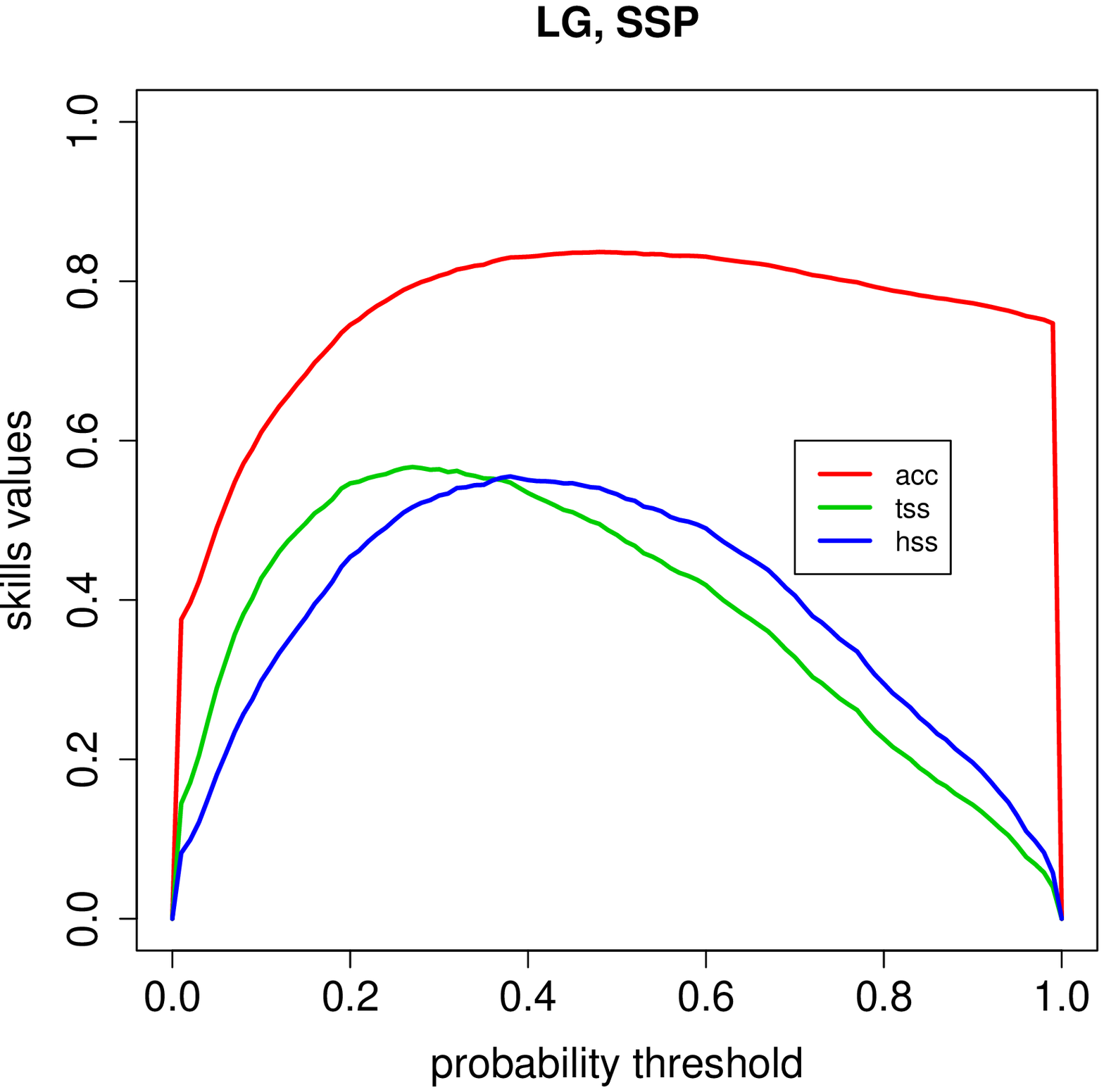}
\label{fig:gggg}}
\subfloat[Subfigure 2 list of figures text][LG, ROC]{
\includegraphics[width=0.295\textwidth]{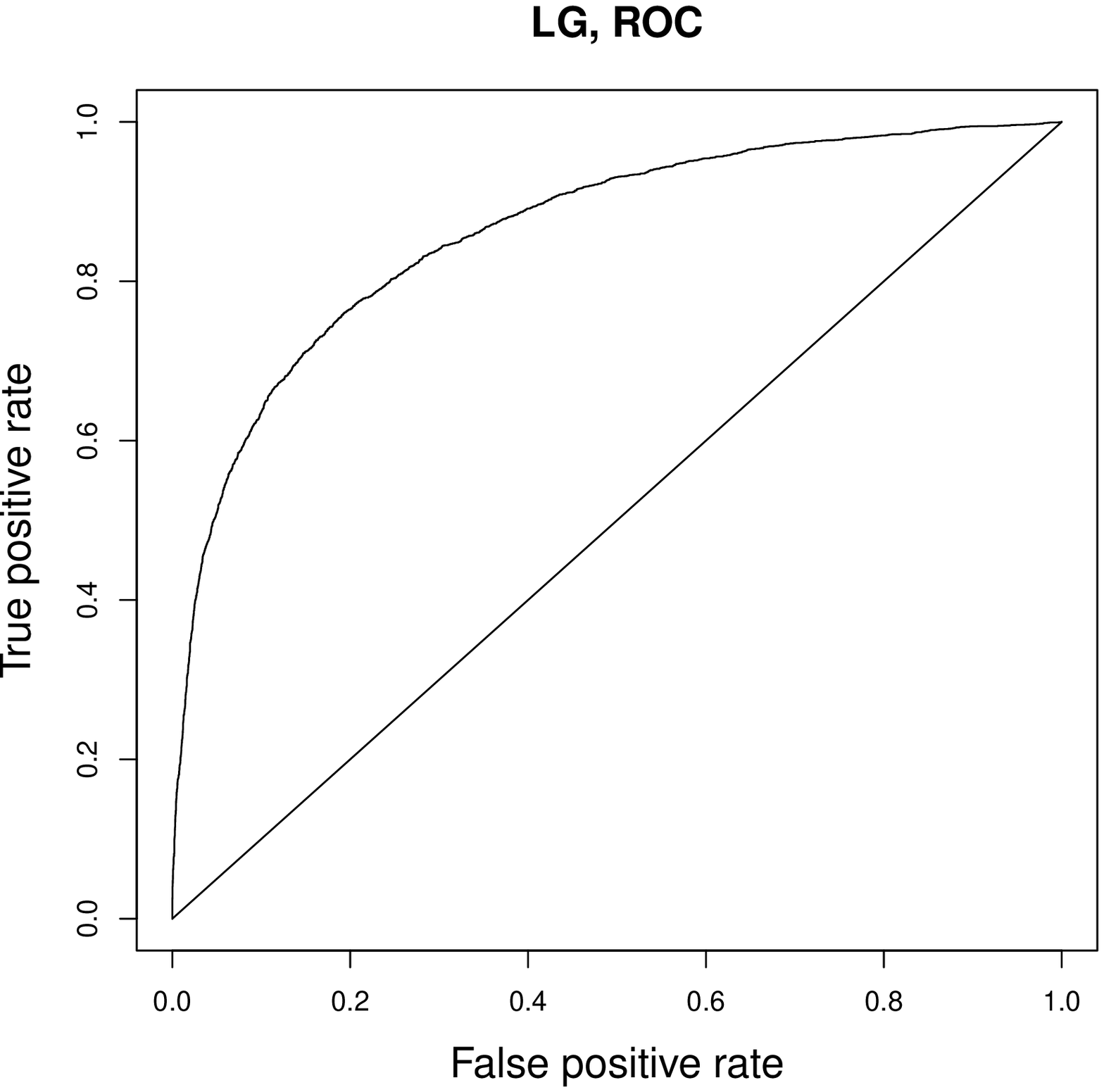}
\label{fig:hhhh}}
\subfloat[Subfigure 3 list of figures text][LG, RD]{
\includegraphics[width=0.295\textwidth]{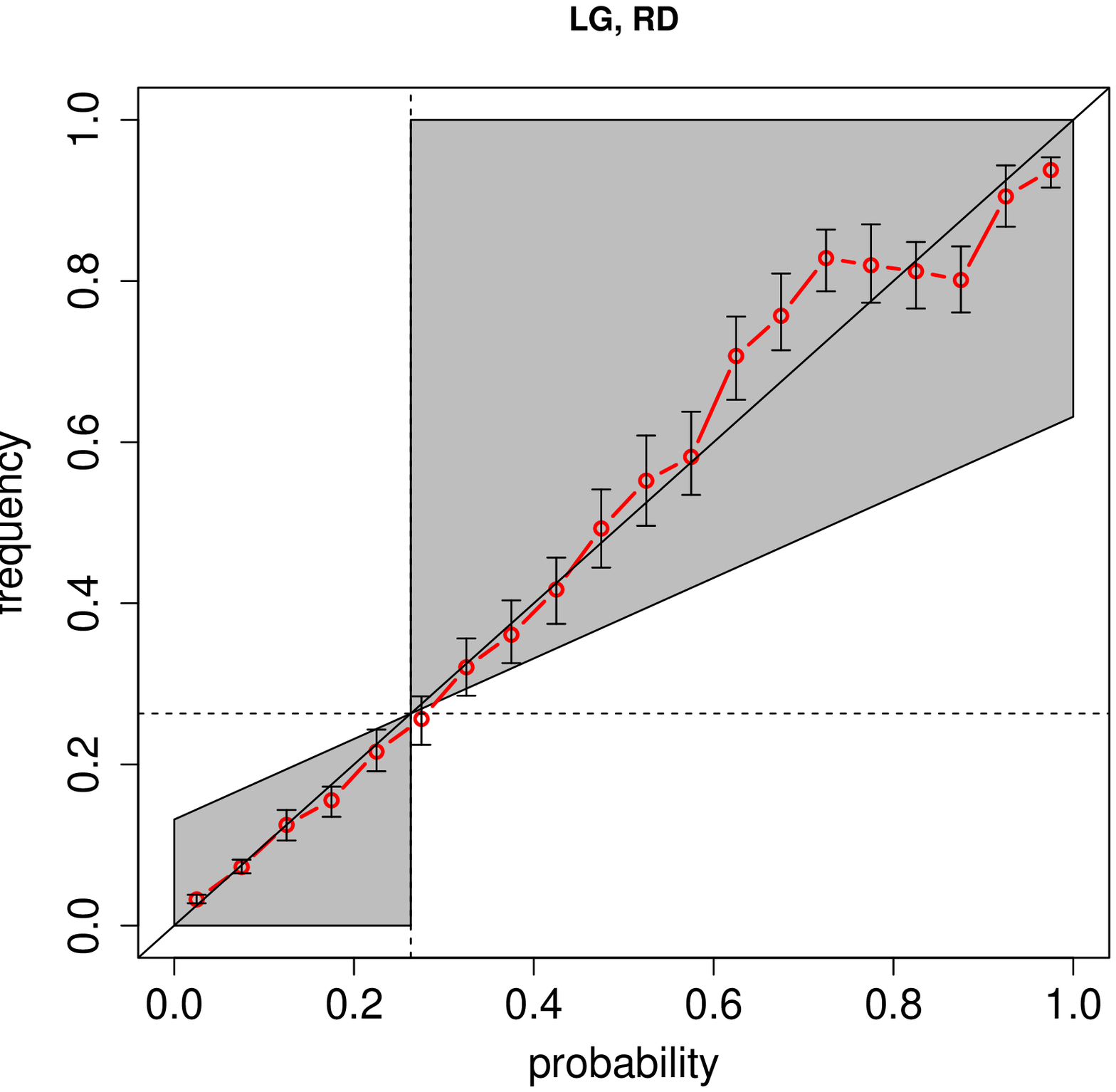}
\label{fig:iiii}}
\caption{Same as Figure \ref{fig:11}, but for $>$C1 flare prediction.}
\label{fig:1111}

\end{figure}

For the LM, we notice a decrease in the {\rm ACC} of the method and some more-or-less similar behavior in the behaviour of {\rm HSS} and {\rm TSS}. The ROC curve seems satisfactory with maximum {\rm TSS}=0.562, while the RD appears to show a systematic over-prediction below a forecast probability of 0.4 and a systematic under-prediction above a forecast probability of 0.4 (excluding probabilities $>$ 0.9).


Regarding the PR, similar behaviour with LM appears for the SSPs, while the ROC curve seems slightly better with maximum {\rm TSS}=0.566. The RD curve shows some systematic under-prediction, although generally within error bars.

Finally, for the LG, one notices a similar behaviour in the SSP, as in the case of LM and PR, but arguably a better behaved ROC curve with maximum {\rm TSS}=0.567. The RD seems to bee the best behaved, compared to those of LM and PR.


\subsubsection{Monte Carlo Simulation for $>$C1 Flares}

        \noindent In Table \ref{tab:MonteCarloC} we provide the average values of the skill scores {\rm ACC}, {\rm TSS} and {\rm HSS} for all prediction methods after the 200 replications of the Monte Carlo experiment regarding $>$C1 flares prediction. We notice from Table \ref{tab:MonteCarloC} that the maximum {\rm HSS}=0.60 is obtained with the RF method for a probability threshold of 40\%. The corresponding skill score values are {\rm ACC}=0.85$\pm$0.00, {\rm TSS}=0.59$\pm$0.01 and {\rm HSS}=0.60$\pm$0.01. The second best method in Table \ref{tab:MonteCarloC} for the same probability threshold is obtained with the LG method, with {\rm ACC}=0.83$\pm$0.00, ${\rm TSS}=0.54\pm 0.01$ and {\rm HSS}=0.56$\pm$0.01. Considering again the probability threshold where the maximum {\rm TSS} is observed, we get the optimal results for the RF method and threshold 30\% with values {\rm ACC}=0.82$\pm$0.00, {\rm TSS}=0.61$\pm$0.01 and ${\rm HSS}=0.57\pm0.01$. The second best method may be considered the MLP (or the LG in a tie) at 30\% threshold with {\rm ACC}=0.81$\pm$0.00, {\rm TSS}=0.57$\pm$0.01 and {\rm HSS}=0.53$\pm$0.01. For a range of probability thresholds (30\% -- 40\%) the method RF yields increasing values of {\rm HSS} and decreasing values of {\rm TSS}. As a result, again it is not clear which is the optimal value of the threshold probability, if we choose to simultaneously optimize both {\rm TSS} and {\rm HSS}. For example, an appealing RF forecasting model is with threshold 35\% and skill scores {\rm ACC}=0.84$\pm$0.00, {\rm TSS}=0.60$\pm$0.01 and {\rm HSS}=0.59$\pm$0.01 in Table \ref{tab:MonteCarloC}. These results are generally above those reported for $>$C1 class flares predictability, namely {\rm TSS} $\in [0.50, 0.55]$ and
{\rm HSS} $\in [0.40, 0.45]$ \citep{Al-Ghraibah:15,Boucheron:15}. In brief, we believe that our data samples, both training and testing, are comprehensive and generally unbiased.

\subsection{Assessment of Prediction Methods and Predictor Strength} \label{sec:AssessmentOfPredictions}
Following the presentation of results in Tables \ref{tab:MonteCarloM1} and \ref{tab:MonteCarloC}, we can see that both for $>$M1 and $>$C1 flare prediction, RF delivers the best skill score metrics for a wide range of probability thresholds. The second best method is MLP together with LG. In this setting we perform some additional evaluation that confirms these results.

Regarding the predictors strength, we present analytical results in Appendix A. It seems that ${\rm log}{\rm R}$ and ${\rm WL}_{{\rm SG}}$ rank in the first places both for $>$C1 and $>$M1 flare prediction, closely followed by the Ising energy and the TLMPIL.

In order to investigate the robustness of our results, we present additional results in Appendix C where we make predictions once a day (at 00:00 UT). The mean evolution (over 200 Monte Carlo iterations) of {\rm ACC}, {\rm TSS} and {\rm HSS} with respect to the probability threshold is presented. Likewise, the BS, AUC and BSS are presented. The main finding is that issuing forecasts once a day keeps similar average skill scores with issuing forecasts eight times a day, but the associated uncertainties ({\it e.g.} standard deviations) are higher in the case of daily predictions.

A final word for the comparison of ML algorithms \textit{vs.} conventional statistics models for this specific dataset and positive/negative class definitions is provided in Appendix D. There, we have included auxiliary meta-analysis of the results in Tables \ref{tab:MonteCarloM1} and \ref{tab:MonteCarloC} in order to clearly show whether the ML category of prediction algorithms does any better than the conventional statistics models in the $>$M1 and $>$C1 flare prediction cases. A multicriteria analysis using the weighted-sum (WS) method \citep{Greco:16} seems appropriate in order to aggregate the performance metrics {\rm ACC}, {\rm TSS} and {\rm HSS} of all classifiers as a function of the probability threshold ({\it e.g.} using equal weights for the aggregation). In this way, a composite index (CI), as a measure of overall utility, is computed for every algorithm and probability threshold combination. There exist $21 \times 6 = 126$ such alternatives when we use a 5\% probability threshold grid, such as the grid in Tables \ref{tab:MonteCarloM1} and \ref{tab:MonteCarloC}. The ranking, in non-increasing order, of the CI reveals the overall merit of every probabilistic classifier and also allows us to draw conclusions for groups of classifiers, such as the group of ML methods (comprising RF, SVM and MLP) and the group of conventional statistics methods (comprising LM, PR and LG). Appendix D presents this multicriteria WS analysis, revealing that overall, in $>$C1 flare prediction ML outperforms conventional statistics methods by 71\% {\it vs.} 29\% in the synthesis of the top $100(1/6)=16.6\%$ performing methods (top 21 methods out of total 126 ones). Likewise, in the $>$M1 flare prediction case, ML outperforms conventional statistics methods by 62\% {\it vs.} 38\% in the synthesis of the top $100(1/6)=16.6\%$ performing methods. So, it seems that $>$C1 flare prediction is more advantageous for ML {\it versus} statistical methods, in comparison to the $>$M1 flare case. This is due to the low performance of the SVM in $>$M1 flare prediction, which is due to the way we have implemented, for simplicity, the SVM for a highly unbalanced sample in $>$M1 flare prediction\footnote{Even by using the SVM weighted variant and recomputing the WS ranking using this variant, (\textit{e.g.} see Figure \ref{fig:1} and Table \ref{tab:AuxSkillScoresM1}), the qualitative results of the presented ranking still hold.}, using a single $C$ constant and not two different $C_1, C_2$ constants during the SVM training with Equation \ref{exp:C-SVC}.

In auxiliary runs (available upon request), we also noticed that when the sample size is very low, using ML algorithms poses no advantage over conventional statistics models. In order to have proper training, the ML algorithms need $N > 2,000$ for $K=13$, especially for the $>$M1 flare prediction.

\subsection{Statistical Tests for Random Forest {\it vs.} MLP and Calculation of AUC and Brier Skill Scores}

\noindent In Section \ref{t-tests} we present results of a $t$-test between the two best performing methods according to maximizer thresholds for either {\rm TSS} or {\rm HSS} for $>$M1 class and $>$C1 class flares cases. Section \ref{AuxSkillScores} presents additional calculations reporting on BS, BSS and AUC, used for assesing classification in the prediction.

\subsubsection{Unpaired t-tests to Compare Two Means for {\rm TSS} and {\rm HSS} of Random Forest {\it vs.} MLP} \label{t-tests}

\noindent A $t$-test compares the means of two groups. Here, we use the $t$-test to compare the mean {\rm TSS} (respectively {\rm HSS}) of the RF method {\it vs.} those of the MLP method (or in general the second best performing method). Means are considered with respect to the Monte Carlo simulations performed on the 200 replications of the previous section. The {\rm TSS}- (respectively {\rm HSS}-) values considered are those for specific probability thresholds maximizing either {\rm TSS} or {\rm HSS}. Table \ref{tab:t-tests} presents the $t$-test results regarding the best and the second best methods with respect to either {\rm TSS} or {\rm HSS} for these specific probability thresholds.

\noindent We find that RF is always ({\it i.e.} 8/8 of times) statistically better than the second best method (which is the MLP 4/8 of times), with respect to both {\rm TSS} and {\rm HSS}.

\subsubsection{Calculation of AUC and Brier Skill Scores} \label{AuxSkillScores}

\noindent Tables \ref{tab:AuxSkillScoresM1} and \ref{tab:AuxSkillScoresCclass} present the calculated mean values of BS, AUC and BSS for the $>$M1 and $>$C1 flare prediction cases, respectively.

For the $>$M1 flare case (Table \ref{tab:AuxSkillScoresM1}), results show that,  on average, the best BS and BSS results are achieved with the RF method (BS = 0.0266; BSS=0.4163). The best AUC results are achieved with the RF method (AUC = 0.9556), but also PR (AUC=0.9392) and LG (AUC=0.9391) methods. 

For the $>$C1 flare case (Table \ref{tab:AuxSkillScoresCclass}), results show that, on average, the best BS and BSS results are achieved with the RF method (BS=0.1074; BSS=0.4426). The best AUC results are also achieved with the RF method (AUC=0.8927), with other methods (except SVM) following closely. The SVM probably needs better fine-tuning, given its sensitivity on {$\gamma$} and {\rm cost} (see Section \ref{TuningMLalgorithms}).

\begin{ltable}

\caption{Monte-Carlo scenario 1, based on 200 SHARP datasets, on $>$M1 GOES flare prediction.  Numbers in boldface correspond to the most significant results of a given method (MLP: multi-layer perceptron; LM: linear regression; PR: probit regression; LG: logit regression; RF: random forest; SVM: support vector machine).}
\label{tab:MonteCarloM1}
{\tiny
\begin{tabular}{|lr|rrr|rrr|rrr|rrr|rrr|rrr|}
  \hline
 &  &  & MLP &  &  & LM &  &  & PR &  &  & LG &  &  & RF &  &  & SVM &  \\
Par & \% & ACC & TSS & HSS & ACC & TSS & HSS & ACC & TSS & HSS & ACC & TSS & HSS & ACC & TSS & HSS & ACC & TSS & HSS \\
  \hline
$val_{0}$ & 0.00 & 0.15 & 0.11 & 0.01 & 0.32 & 0.27 & 0.04 & 0.00 & 0.00 & 0.00 & 0.00 & 0.00 & 0.00 & 0.51 & 0.48 & 0.08 & 0.00 & 0.00 & 0.00 \\ 
  $val_{5}$ & 0.05 & 0.90 & 0.70 & 0.39 & 0.78 & 0.70 & 0.22 & 0.84 & 0.75 & 0.30 & 0.85 & 0.75 & 0.31 & 0.85 & 0.77 & 0.32 & 0.94 & 0.59 & 0.47 \\ 
  $val_{10}$ & 0.10 & 0.93 & 0.66 & 0.45 & \textbf{0.88} & \textbf{0.73} & \textbf{0.35} & 0.90 & 0.71 & 0.39 & 0.90 & 0.69 & 0.40 & \textbf{0.90} & \textbf{0.77} & \textbf{0.42} & 0.95 & 0.51 & 0.49 \\ 
  $val_{15}$ & 0.15 & 0.94 & 0.62 & 0.48 & 0.92 & 0.65 & 0.43 & 0.93 & 0.64 & 0.45 & 0.93 & 0.63 & 0.45 & \textbf{0.93} & \textbf{0.74} & \textbf{0.49} & 0.96 & 0.46 & 0.50 \\ 
  $val_{20}$ & 0.20 & 0.95 & 0.59 & 0.50 & 0.94 & 0.54 & 0.46 & 0.94 & 0.58 & 0.48 & 0.94 & 0.58 & 0.48 & 0.95 & 0.69 & 0.54 & 0.96 & 0.42 & 0.48 \\ 
  $val_{25}$ & 0.25 & \textbf{0.95} & \textbf{0.56} & \textbf{0.50} & 0.95 & 0.45 & 0.45 & 0.95 & 0.52 & 0.49 & 0.95 & 0.52 & 0.49 & \textbf{0.96} & \textbf{0.63} & \textbf{0.57} & 0.96 & 0.39 & 0.47 \\ 
  $val_{30}$ & 0.30 & 0.95 & 0.53 & 0.50 & 0.95 & 0.38 & 0.44 & 0.96 & 0.47 & 0.49 & 0.96 & 0.48 & 0.50 & 0.96 & 0.57 & 0.57 & 0.96 & 0.37 & 0.46 \\ 
  $val_{35}$ & 0.35 & 0.96 & 0.50 & 0.50 & 0.96 & 0.31 & 0.39 & 0.96 & 0.41 & 0.47 & 0.96 & 0.43 & 0.49 & 0.96 & 0.51 & 0.57 & 0.96 & 0.35 & 0.45 \\ 
  $val_{40}$ & 0.40 & 0.96 & 0.47 & 0.50 & 0.96 & 0.26 & 0.35 & 0.96 & 0.36 & 0.45 & 0.96 & 0.39 & 0.47 & 0.97 & 0.46 & 0.55 & 0.96 & 0.33 & 0.44 \\ 
  $val_{45}$ & 0.45 & 0.96 & 0.44 & 0.49 & 0.96 & 0.21 & 0.31 & 0.96 & 0.32 & 0.42 & 0.96 & 0.35 & 0.45 & 0.97 & 0.41 & 0.52 & 0.96 & 0.31 & 0.42 \\ 
  $val_{50}$ & 0.50 & 0.96 & 0.42 & 0.48 & 0.96 & 0.18 & 0.28 & 0.96 & 0.28 & 0.39 & 0.96 & 0.31 & 0.42 & 0.97 & 0.37 & 0.49 & 0.96 & 0.29 & 0.41 \\ 
  $val_{55}$ & 0.55 & 0.96 & 0.39 & 0.47 & 0.96 & 0.16 & 0.25 & 0.96 & 0.25 & 0.36 & 0.96 & 0.27 & 0.39 & 0.96 & 0.32 & 0.45 & 0.96 & 0.28 & 0.39 \\ 
  $val_{60}$ & 0.60 & 0.96 & 0.37 & 0.45 & 0.96 & 0.14 & 0.23 & 0.96 & 0.21 & 0.33 & 0.96 & 0.24 & 0.36 & 0.96 & 0.28 & 0.41 & 0.96 & 0.26 & 0.38 \\ 
  $val_{65}$ & 0.65 & 0.96 & 0.34 & 0.44 & 0.96 & 0.11 & 0.19 & 0.96 & 0.18 & 0.29 & 0.96 & 0.21 & 0.32 & 0.96 & 0.24 & 0.36 & 0.96 & 0.24 & 0.36 \\ 
  $val_{70}$ & 0.70 & 0.96 & 0.32 & 0.42 & 0.96 & 0.09 & 0.16 & 0.96 & 0.16 & 0.25 & 0.96 & 0.18 & 0.28 & 0.96 & 0.19 & 0.31 & 0.96 & 0.22 & 0.34 \\ 
  $val_{75}$ & 0.75 & 0.96 & 0.29 & 0.40 & 0.96 & 0.08 & 0.13 & 0.96 & 0.13 & 0.22 & 0.96 & 0.15 & 0.25 & 0.96 & 0.15 & 0.25 & 0.96 & 0.21 & 0.32 \\ 
  $val_{80}$ & 0.80 & 0.96 & 0.27 & 0.37 & 0.95 & 0.06 & 0.12 & 0.96 & 0.11 & 0.18 & 0.96 & 0.12 & 0.21 & 0.96 & 0.11 & 0.19 & 0.96 & 0.18 & 0.29 \\ 
  $val_{85}$ & 0.85 & 0.96 & 0.24 & 0.34 & 0.95 & 0.05 & 0.09 & 0.96 & 0.08 & 0.14 & 0.96 & 0.09 & 0.16 & 0.96 & 0.08 & 0.14 & 0.96 & 0.16 & 0.26 \\ 
  $val_{90}$ & 0.90 & 0.96 & 0.20 & 0.31 & 0.95 & 0.03 & 0.06 & 0.95 & 0.05 & 0.10 & 0.95 & 0.06 & 0.10 & 0.95 & 0.04 & 0.08 & 0.96 & 0.13 & 0.22 \\ 
  $val_{95}$ & 0.95 & 0.96 & 0.15 & 0.25 & 0.95 & 0.02 & 0.04 & 0.95 & 0.03 & 0.06 & 0.95 & 0.03 & 0.05 & 0.92 & 0.01 & 0.02 & 0.96 & 0.08 & 0.14 \\ 
  $val_{100}$ & 1.00 & 0.00 & 0.00 & 0.00 & 0.95 & 0.02 & 0.03 & 0.00 & 0.00 & 0.00 & 0.00 & 0.00 & 0.00 & 0.00 & 0.00 & 0.00 & 0.00 & 0.00 & 0.00 \\ 
   \hline
\end{tabular}
}

\end{ltable}


\begin{ltable}

\caption{Monte-Carlo scenario 2, based on 200 SHARP datasets, on $>$C1 GOES flare prediction.  Numbers in boldface correspond to the most significant results of a given method (MLP: multi-layer perceptron; LM: linear regression; PR: probit regression; LG: logit regression; RF: random forest; SVM: support vector machine).}
\label{tab:MonteCarloC}
{\tiny
\begin{tabular}{|lr|rrr|rrr|rrr|rrr|rrr|rrr|}
  \hline
 &  &  & MLP &  &  & LM &  &  & PR &  &  & LG &  &  & RF &  &  & SVM &  \\
Par & \% & ACC & TSS & HSS & ACC & TSS & HSS & ACC & TSS & HSS & ACC & TSS & HSS & ACC & TSS & HSS & ACC & TSS & HSS \\
\hline
$val_{0}$ & 0.00 & 0.00 & 0.00 & 0.00 & 0.39 & 0.16 & 0.09 & 0.00 & 0.00 & 0.00 & 0.00 & 0.00 & 0.00 & 0.28 & 0.03 & 0.01 & 0.00 & 0.00 & 0.00 \\ 
  $val_{5}$ & 0.05 & 0.52 & 0.33 & 0.21 & 0.44 & 0.23 & 0.14 & 0.47 & 0.27 & 0.17 & 0.48 & 0.28 & 0.17 & 0.52 & 0.34 & 0.21 & 0.29 & 0.03 & 0.02 \\ 
  $val_{10}$ & 0.10 & 0.66 & 0.49 & 0.35 & 0.51 & 0.32 & 0.20 & 0.58 & 0.40 & 0.27 & 0.60 & 0.42 & 0.29 & 0.64 & 0.48 & 0.34 & 0.44 & 0.22 & 0.13 \\ 
  $val_{15}$ & 0.15 & 0.72 & 0.55 & 0.43 & 0.58 & 0.40 & 0.27 & 0.66 & 0.49 & 0.36 & 0.68 & 0.50 & 0.38 & 0.71 & 0.55 & 0.42 & 0.75 & 0.54 & 0.45 \\ 
  $val_{20}$ & 0.20 & 0.76 & 0.57 & 0.48 & 0.66 & 0.48 & 0.35 & 0.73 & 0.55 & 0.44 & 0.74 & 0.55 & 0.45 & 0.76 & 0.59 & 0.48 & 0.80 & 0.57 & 0.53 \\ 
  $val_{25}$ & 0.25 & 0.79 & 0.57 & 0.51 & 0.74 & 0.55 & 0.45 & 0.78 & 0.56 & 0.49 & 0.78 & 0.57 & 0.50 & 0.79 & 0.61 & 0.53 & 0.82 & 0.56 & 0.55 \\ 
  $val_{30}$ & 0.30 & \textbf{0.81} & \textbf{0.57} & \textbf{0.53} & 0.79 & 0.57 & 0.51 & 0.80 & 0.57 & 0.53 & \textbf{0.81} & \textbf{0.57} & \textbf{0.53} & \textbf{0.82} & \textbf{0.61} & \textbf{0.57} & 0.83 & 0.54 & 0.55 \\ 
  $val_{35}$ & 0.35 & 0.82 & 0.56 & 0.55 & 0.82 & 0.55 & 0.54 & 0.82 & 0.56 & 0.55 & 0.82 & 0.56 & 0.55 & \textbf{0.84} & \textbf{0.60} & \textbf{0.59} & 0.84 & 0.52 & 0.55 \\ 
  $val_{40}$ & 0.40 & 0.83 & 0.55 & 0.55 & 0.83 & 0.52 & 0.54 & 0.83 & 0.53 & 0.55 & \textbf{0.83} & \textbf{0.54} & \textbf{0.56} & \textbf{0.85} & \textbf{0.59} & \textbf{0.60} & 0.84 & 0.50 & 0.54 \\ 
  $val_{45}$ & 0.45 & 0.84 & 0.53 & 0.55 & 0.83 & 0.47 & 0.52 & 0.84 & 0.50 & 0.55 & 0.84 & 0.51 & 0.55 & 0.85 & 0.56 & 0.59 & 0.84 & 0.48 & 0.53 \\ 
  $val_{50}$ & 0.50 & 0.84 & 0.50 & 0.55 & 0.83 & 0.40 & 0.47 & 0.84 & 0.47 & 0.53 & 0.84 & 0.48 & 0.53 & 0.85 & 0.54 & 0.59 & 0.84 & 0.46 & 0.52 \\ 
  $val_{55}$ & 0.55 & 0.84 & 0.48 & 0.53 & 0.81 & 0.34 & 0.41 & 0.83 & 0.43 & 0.50 & 0.84 & 0.45 & 0.52 & 0.85 & 0.51 & 0.57 & 0.83 & 0.44 & 0.51 \\ 
  $val_{60}$ & 0.60 & 0.84 & 0.45 & 0.51 & 0.80 & 0.28 & 0.35 & 0.82 & 0.38 & 0.46 & 0.83 & 0.41 & 0.48 & 0.85 & 0.48 & 0.55 & 0.83 & 0.42 & 0.49 \\ 
  $val_{65}$ & 0.65 & 0.83 & 0.41 & 0.49 & 0.79 & 0.22 & 0.29 & 0.82 & 0.34 & 0.42 & 0.82 & 0.37 & 0.44 & 0.84 & 0.44 & 0.52 & 0.83 & 0.38 & 0.46 \\ 
  $val_{70}$ & 0.70 & 0.82 & 0.37 & 0.45 & 0.78 & 0.18 & 0.24 & 0.81 & 0.29 & 0.37 & 0.81 & 0.32 & 0.40 & 0.83 & 0.40 & 0.48 & 0.82 & 0.35 & 0.43 \\ 
  $val_{75}$ & 0.75 & 0.82 & 0.33 & 0.41 & 0.77 & 0.15 & 0.20 & 0.80 & 0.25 & 0.32 & 0.80 & 0.27 & 0.35 & 0.82 & 0.34 & 0.43 & 0.81 & 0.32 & 0.39 \\ 
  $val_{80}$ & 0.80 & 0.81 & 0.28 & 0.36 & 0.77 & 0.12 & 0.17 & 0.79 & 0.20 & 0.27 & 0.79 & 0.22 & 0.29 & 0.81 & 0.28 & 0.36 & 0.81 & 0.28 & 0.36 \\ 
  $val_{85}$ & 0.85 & 0.79 & 0.22 & 0.29 & 0.76 & 0.10 & 0.14 & 0.78 & 0.16 & 0.22 & 0.78 & 0.18 & 0.24 & 0.79 & 0.21 & 0.29 & 0.80 & 0.24 & 0.31 \\ 
  $val_{90}$ & 0.90 & 0.78 & 0.16 & 0.21 & 0.76 & 0.08 & 0.11 & 0.77 & 0.13 & 0.18 & 0.77 & 0.14 & 0.19 & 0.78 & 0.15 & 0.21 & 0.79 & 0.19 & 0.25 \\ 
  $val_{95}$ & 0.95 & 0.75 & 0.08 & 0.11 & 0.76 & 0.07 & 0.10 & 0.76 & 0.09 & 0.13 & 0.76 & 0.09 & 0.13 & 0.76 & 0.08 & 0.12 & 0.77 & 0.14 & 0.19 \\ 
  $val_{100}$ & 1.00 & 0.00 & 0.00 & 0.00 & 0.75 & 0.06 & 0.08 & 0.00 & 0.00 & 0.00 & 0.00 & 0.00 & 0.00 & 0.00 & 0.00 & 0.00 & 0.00 & 0.00 & 0.00 \\ 
   \hline
\end{tabular}
}

\end{ltable}



\begin{table}[ht]
\caption{Unpaired $t$-tests to compare the means of {\rm TSS} and {\rm HSS} metrics (out-of-sample) for the best and the second best methods in $>$M1 and $>$C1 flare forecasting.}
\label{tab:t-tests}
\begin{tabular}{cccccc}
\hline
\multicolumn{6}{c}{$>$M1 class flares prediction}\\
 \hline
 No. & Metric & Threshold (\%) & Best & Second Best & $p$-value \\
 1 & TSS & 10 & RF & LM & $<10^{-4}$ \\ 
 2 & HSS & 10 & RF & LM & $<10^{-4}$ \\ 
 3 & TSS & 25 & RF & MLP & $<10^{-4}$ \\ 
 4 & HSS & 25 & RF & MLP & $<10^{-4}$ \\ 
  \hline
  \multicolumn{6}{c}{$>$C1 class flares prediction}\\
  \hline
 No. & Metric & Threshold (\%) & Best & Second Best & $p$-value \\
 5 & TSS & 30 & RF & MLP & $<10^{-4}$  \\ 
 6 & HSS & 30 & RF & MLP & $<10^{-4}$  \\ 
 7 & TSS & 40 & RF & LG & $<10^{-4}$  \\ 
 8 & HSS & 40 & RF & LG & $<10^{-4}$  \\ 
  \hline
\end{tabular}
\end{table}

\begin{table}[ht]
\caption{Mean values for BS, BSS and AUC for all tested models on the prediction of $>$M1 flares. Means are obtained after 200 Monte-Carlo replications. Parentheses underneath values denote standard deviations. Notice that smaller values indicate better performance for BS, whereas higher values indicate better performance for AUC and BSS.}
\label{tab:AuxSkillScoresM1}
\begin{tabular}{ccccccc}
\hline
\multicolumn{6}{c}{BS}\\
 \hline
 MLP & LM & PR & LG & RF & SVM & ${\rm SVM}_{\rm weighted}$ \\
 0.0324 & 0.0331 & 0.0305 & 0.0302 & 0.0266 & 0.0327 & 0.0357 \\ 
 (0.0013) & (0.0008) & (0.0008) & (0.0008) & (0.0008) & (0.0012) & (0.0011) \\ 
  \hline
  \multicolumn{6}{c}{AUC}\\
  \hline
 MLP & LM & PR & LG & RF & SVM & ${\rm SVM}_{\rm weighted}$ \\
 0.9301 & 0.9278 & 0.9392 & 0.9391 & 0.9556 & 0.8320 & 0.9175 \\ 
 (0.0067) & (0.0043) & (0.0033) & (0.0033) & (0.0035) & (0.0168) & (0.0059) \\
  \hline
  \multicolumn{6}{c}{BSS}\\
  \hline
 MLP & LM & PR & LG & RF & SVM & ${\rm SVM}_{\rm weighted}$ \\
 0.2903 & 0.2745 & 0.3323 & 0.3375 & 0.4163 & 0.2829 & 0.2181 \\ 
 (0.0267) & (0.0117) & (0.0128) & (0.0134) & (0.0126) & (0.0159) & (0.0154) \\ 
   \hline
\end{tabular}
\end{table}

\begin{table}[ht]
\caption{Same as Table \ref{tab:AuxSkillScoresM1}, but for the prediction of $>$C1 flares.}
\label{tab:AuxSkillScoresCclass}
\begin{tabular}{cccccc}
\hline
\multicolumn{6}{c}{BS}\\
  \hline
 MLP & LM & PR & LG & RF & SVM \\
 0.1167 & 0.1292 & 0.1201 & 0.1191 & 0.1074 & 0.1226 \\ 
 (0.0014) & (0.0012) & (0.0012) & (0.0012) & (0.0012) & (0.0015) \\ 
   \hline
   \multicolumn{6}{c}{AUC}\\
   \hline
 MLP & LM & PR & LG & RF & SVM \\
 0.8731 & 0.8638 & 0.8665 & 0.8669 & 0.8927 & 0.8466 \\ 
 (0.0029) & (0.0029) & (0.0027) & (0.0027) & (0.0026) & (0.0033) \\ 
  \hline
  \multicolumn{6}{c}{BSS}\\
  \hline
 MLP & LM & PR & LG & RF & SVM \\
 0.3940 & 0.3293 & 0.3767 & 0.3818 & 0.4426 & 0.3636 \\ 
 (0.0069) & (0.0052) & (0.0055) & (0.0058) & (0.0056) & (0.0063) \\ 
   \hline
\end{tabular}
\end{table}

\newpage

\subsection{Related Published Work and Comparison To Our Results}

\noindent \cite{Ahmed:13} presented prediction results for $>$C1 class flares using cross-validation with 60\% training and 40\% testing subsets, with 10 iterations in operational and segmented mode. Since our analysis focuses in operational mode, the golden standard for near-real-time operational systems such as FLARECAST, we present here their results on the operational mode for the period Apr. 1996 - Dec. 2010:
${\rm POD}=0.455$ \&
${\rm POFD}=0.010$
thus
${\rm TSS}={\rm POD}-{\rm POFD}=0.445$
and
${\rm HSS}=0.539$. Hence, Ahmed et al. reported (using a variant of a neural network, and threshold 50\%) results for flares $>$C1: ${\rm TSS}=0.445$ and ${\rm HSS}=0.539$.

\cite{Li:08} presented results using a SVM coupled with k-nearest neighbor (KNN) for flare prediction $>$M1 in a way that, unfortunately, cannot be used to recover {\rm TSS} and {\rm HSS} values. Instead, they report
Equal = {\rm TN} + {\rm TP},
High = {\rm FP},
Low = {\rm FN}.
The accuracy achieved is only {\rm ACC}=57.02\% for SVM and {\rm ACC}=63.91\% for SVM-KNN for the testing year 2002.

\cite{Song:09} presented results using an ordinal logistic regression model classifying the C-, M- and X-class flares with response values 1, 2 and 3, respectively. The B-class flares (or no flares) category received class 0 (baseline). Their sample contains 34 X-class flares, 68 M-class flares, 65 C-class flares, and 63 B-class or no-flare cases. A clear drawback of this sample is that it is not taken using a random number generator but seems to be hand-picked aiming at studying the considered 230 events during the period 1998-2005. As a result, the sample is biased in that the occurrence rates of the various flare classes are not representative of an actual solar cycle. Perhaps not surprisingly, these authors presented high {\rm TSS} and {\rm HSS} values that, given the sample, might be taken with a conservative outlook. From the results of Model 4 in that study \citep[{\it i.e.} Table 8 of][]{Song:09}, we are able to infer that for C class flares, Song \textit{et al.} computed values {\rm TSS}=0.65 and {\rm HSS}=0.623 (C1-C9 flares). Moreover, we maintain an impression that these numbers are obtained in-sample for the dataset with 230 events in \cite{Song:09}.

\cite{Yu:09} used a sliding window approach to account for the evolution of three magnetic flare predictors with importance index above 10 (for the definition of the flare importance index, see \cite{Yu:09}). The time period is 1996 to 2004, with a cadence of 96 minutes. The authors use the C4.5 decision tree algorithm and the Learning Vector Quantization (LVQ) Neural network, both implemented in WEKA \citep{witten:16,hall:09}. The authors use a 10-fold cross-validation approach with 90\% training and 10\% testing sets from the original sample. The sliding window size was 45 observations. Their results showed that the sliding window versions of C4.5 and LVQ neural network algorithms improved the results obtained with the same algorithms for sliding window size equal to 0. Since the authors present only the {\rm TP} rate and the {\rm TN} rate results, we are not able to recover their {\rm HSS} value. Their recovered {\rm TSS} is {\rm TSS}=0.651 for the C4.5 algorithm with a sliding window of 45 observations and {\rm TSS}=0.667 for the LVQ also with a sliding window of 45 observations.

\cite{Yuan:10} used the same dataset as in \cite{Song:09} and proposed a cascading approach using,  first, an ordinal logistic regression model to produce probabilities for GOES flare classes B, C, M and X (associated with response levels 0,1,2 and 3, respectively) and, second, feeding the probability values to an SVM in order to obtain the final class membership. Their results, according to \cite{Yuan:10} improve the prediction especially for X-class flares (response level = 3 in the ordinal logistic regression) but, still, are not exceptionally high. For example, for level = 1, therefore for C-class flares, we were able to recover the following {\rm TSS} values for the used methods:
\noindent Logistic Regression: {\rm TSS}=0.22,
SVM: {\rm TSS}=0.08,
Logistic Regression + SVM: {\rm TSS}=0.09,
\noindent These rather fair results, as can be seen from the contingency tables presented in \cite{Yuan:10}, may be due to the selection of a probability threshold value at 50\% for levels 0, 1 and 3 in the ordinal logistic regression model and at 25\% for the level 3 (X-class flares) in the same model. Choosing a threshold equal to 50\% maximizes {\rm ACC} but not {\rm TSS} / {\rm HSS}, as can be seen both here and in \cite{Bloomfield:12}.

\cite{Colak:09} developed an online solar flare forecasting system called ASAP. Their prediction algorithm is a combination of two neural networks with the Sum-of-Squared Error (SSE) objective function, where the first neural network predicts whether a flare of all types (C, M or X) will occur and, if the prediction is yes, the second neural network predicts whether a C-, M-, or X-class flare will occur. The ASAP system was developed in C++ and has been validated with data from 1999 to 2002 (around the peak of Solar Cycle 23). The predictors were the sunspot area and characteristics from the McIntosh classification of sunspots (Zpc scheme). They obtained {\rm HSS}=49.3\% (C-class flares) and {\rm HSS}=47\% (M-class flares) for a forecast window of 24h.

\cite{Wang:08} developed a MLP neural network using three input variables for the prediction of solar flares of class $>$M1. The predictors were the maximum horizontal gradient $|{\rm grad}_h(B_z)|$, the length $L$ of the neutral line and the number of singular points $\eta$. A limitation of the study is that only flaring active regions (at GOES C1 and above) are sampled and considered. The forecast window is 48h. The authors presented prediction results for the period 1996-2002 (training set: Apr. 1996 to Dec. 2001, testing set: Jan 2002 to Dec 2002). The results were presented as plots of the X-ray flux associated with the predicted/observed flares for the test year 2002, so comparison with the authors' skill scores is not possible. This work reported {\rm ACC}=69\% for the test year.

\cite{Bobra:15} applied a SVM to a sample of 5,000 non-flaring and 303 flaring (at the GOES $>$M1 level) AR. Those $N=5,303$ AR with $N=5,000$ negative examples and $P=303$ positive examples (ratio $N/P=16.5$), were sampled from the $\approx$1.5 million patches of the SHARP product \citep{Bobra:14} between years 2010 and 2014. The authors selected 285 M-class flares and 18 X-class flares observed between 2010 May and 2014 May. By comparison, our study herein relies on a representative sample of flaring/non-flaring AR in the period 2012-2016 and for flares $>$M1, with a ratio $N/P=19.9$ ($P=1108$ and $N=22,026$). By inspecting Table 3 of \cite{Bobra:15} we see that the authors report results: {\rm ACC}=0.924$\pm$0.007, {\rm TSS}=0.761$\pm$0.039 and ${\rm HSS_2}$=0.517$\pm$0.035 while our results are {\rm ACC}=0.93$\pm$0.00, {\rm TSS}=0.74$\pm$0.02 and {\rm HSS}=0.49$\pm$0.01 (their definition of ${\rm HSS_2}$ is the same as the {\rm HSS} definition in Section \ref{sec:metricconcepts}). Thus, our results with random forests are competitive with those of \cite{Bobra:15}. We note that we use a 50/50 rule for splitting training/testing sets, while \cite{Bobra:15} use a 70/30 rule. Also, N/P in \cite{Bobra:15} is 16.5 while in our case N/P is 19.9. Finally, we use no fine-tuning in the parameters of the Random Forest, while \cite{Bobra:15} carefully tune the C, $\gamma$ and $C_1 / C_2$ of their Equation 2, 5 and 6, respectively. Regardless, we believe that \cite{Bobra:15} represent the state-of-the art in solar flare forecasting so far.

\cite{Boucheron:15} applied support vector regression (SVR) to 38 predictors characterizing the magnetic field of solar AR in order to predict: i) the flare size and ii) the time-to-flare using SVR modeling. The forecast window used varies between 2 and 24 hours with a step of 2 hours (12 cases of forecast windows). By using the size regression with appropriate thresholds \citep[different to the usual probability thresholds, for example, in][]{Bloomfield:12}, the authors achieved prediction results for $>$C1 flares with {\rm TSS}=0.55 and {\rm HSS}=0.46, while reporting that using the same data, \cite{Al-Ghraibah:15} achieved ${\rm TSS} \approx 0.50$ and ${\rm HSS} \approx 0.40$, respectively, for the prediction of $>$C1 class flares.

\cite{Al-Ghraibah:15} applied relevance vector machines (RVM), a technique that is a generalization of SVM, to a set of 38 magnetic properties characterizing 2124 AR in a total of 122,060 images across different time points for all AR. They predicted $>$C1 flares using either the full set of properties or suitable subsets thereof. The magnetic properties are of three types: i) snapshots in space and time, ii) evolution in time and iii) structures of multiple size scales. \cite{Al-Ghraibah:15} reported results (\textit{e.g.}, see their Table 5 and Figure 6) in the range ${\rm TSS} \approx 0.51$ and ${\rm HSS} \approx 0.39$, which is a baseline result for the literature when no temporal information is included in the predictor set ({\it i.e.} static images are used).

\section{Conclusions} 
      \label{S-Conclusion} 
\noindent We present a new approach for the efficient prediction of $>$M1 and $>$C1 solar flares: classic and modern machine learning (ML) methods, such as multi-layer perceptrons (MLP), support vector machines (SVM) and random forests (RF) were used in order to build the prediction models. The predictor variables were based on the SDO/HMI SHARP data product, available since 2012.

The sample was representative of the solar activity during a five-year period of Solar Cycle 24 (2012 -- 2016), with all calendar days within this period included in the sample. The cadence of properties, or predictors, within the chosen days was 3 hours.

We show that the RF methodology could be our prediction method of choice, both for the prediction of $>$M1 flares (with a relative frequency of 4.8\%, or 1108 events) and for the prediction of $>$C1 flares (with a relative frequency of 26.1\%, or 6029 events). In terms of categorical skill scores, a probability threshold of 15\% for $>$M1 flares gives rise to mean (after 200 replications) RF skill scores of the order {\rm TSS}=$0.74 \pm 0.02$ and {\rm HSS}=$0.49 \pm 0.01$, while a probability threshold of 35\% for $>$C1 flares gives rise to mean {\rm TSS}=$0.60 \pm 0.01$ and {\rm HSS}=$0.59 \pm 0.01$. The respective accuracy values are {\rm ACC}=0.93 and {\rm ACC}=0.84. In terms of probabilistic skill scores, the ranking of the ML techniques with respect to their BSS against climatology is RF (0.42), MLP (0.29) and SVM (0.28) for $>$M1 flares and RF (0.44), MLP (0.39) and SVM (0.36) for $>$C1 flares.

We further indicate that for $>$M1 flare prediction, SVM and MLP need additional tuning of their hyperparameters (Section \ref{TuningMLalgorithms}) in order to produce comparable results with RF. Moreover, several statistical methods (linear regression, probit, logit) produced acceptable forecast results when compared with the ML methods. By increasing the number of hidden nodes, the MLP networks provide flatter skill scores profiles ({\it i.e.} {\rm ACC}, {\rm TSS}, {\rm HSS} as a function of the threshold probability), but the peak values of the corresponding curves are smaller than those achieved by MLP networks with fewer hidden nodes. Regarding the $>$C1 flares, all forecast methods work acceptably, although the best method is, again, RF. A Monte Carlo experiment showed that results are robust with respect to different realizations of the training/testing pair, with different random seeds. Monte Carlo modeling also manages to decrease the amplitudes of the applicable standard deviations of skill scores. Typically standard deviations are larger for the $>$M1 flare case compared to that of $>$C1 flares. This is to be attributed to the different occurrence frequency of flares in the two cases.

RF is a relatively new approach to solar flare prediction. Nonetheless, it may be preferable over other widely used ML algorithms, at least for the data sets exploited so far, giving competitive results without much tuning of the RF hyperparameters. This generates hope for future meaningful developments in the formidable solar flare prediction problem, at the same time aligning with excellent performance for RF reported in several classification benchmarks \citep{Fernandez-Delgado:14}. This important statement made, it appears that even with the application of RF, solar flare prediction in the foreseeable future will likely continue to be probabilistic ({\it i.e.} 0.0 -- 1.0, continuous), rather than binary ({\it i.e.} 0 or 1).

In terms of the predictors importance, Schrijver's $R$ is found to be among the most statistically significant predictors together with ${\rm WL}_{\rm SG}$. Also, the Ising energy and the TLMPIL are considered as important, ranking slightly below the previous two predictors. This stems from the importance calculations according to the Fisher score and random forest importance for the $>$C1 and $>$M1 flare cases in Appendix A. This result is also in line with the common knowledge that flares occur mostly when strong and highly sheared MPILs are formed. Other MPIL-highlighting predictors, such as the effective connected magnetic field strength, $B_{\rm eff}$ \citep{Georgoulis07} remain to be tested, in conjunction with $R$ and ${\rm WL}_{\rm SG}$ as their cadence was lower than 3\;h at the time this study was performed. 

An interesting finding for the RF technique (Appendix B) is obtained by the predictors' ranking information according to their importance, as measured by the Fisher score. Namely, when we create prediction models with a varying number of the most important predictors included, the RF prediction performance (in terms of {\rm TSS} and {\rm HSS}) continues to improve monotonically with the number of included parameters. On the contrary, the MLP and SVM algorithms achieve only slight improvements in prediction results (again in terms of {\rm TSS} and {\rm HSS}) by adding more than, say, the six most important predictors. This interesting finding may further improve forecasting when more viable predictors become available.

For future FLARECAST-supported research we plan to enlarge our analysis sample by reducing the property cadence from 3\;h to 1\;h or even less (the limit is the inherent cadence of SDO/HMI SHARP data, namely 12\;min). Another direction of future research is to investigate the robustness of our results for samples created with a larger cadence of 12\;h (24\;h) coupled with a forecast window of 12\;h (24\;h), respectively. Furthermore, we plan to exploit the substantial time-series aspect of our data using recurrent neural networks, possibly trained with evolutionary algorithms. The present work, along with a series of similar concluded or still ongoing studies are considered for possible integration in the final FLARECAST online system and forecasting tool, to be deployed by early 2018.



\begin{acks}
\noindent We would like to thank the anonymous referee for very helpful comments that greatly improved the initial manuscript. This research has been supported by the EU Horizon 2020 Research and Innovation Action under grant agreement No.640216 for the ``Flare Likelihood And Region Eruption foreCASTing'' ({\sc{FLARECAST}}) project. Data were provided by the MEDOC data and operations centre (CNES /
CNRS / Univ. Paris-Sud), http://medoc.ias.u-psud.fr/ and the GOES team.
\end{acks}

\vspace{0.5cm}

\noindent \textbf{\scriptsize{Disclosure of Potential Conflicts of Interest.}} The authors declare that they have no conflicts of interest. 




\numberwithin{table}{section}
\numberwithin{figure}{section}
\numberwithin{equation}{section}

\setcounter{table}{0}
\setcounter{figure}{0}
\renewcommand{\thetable}{\Alph{section}\arabic{table}}

\appendix   

\section{Importance of predictors for flare prediction}
\begin{figure}[h]
\centering
\subfloat[Subfigure 1 list of figures text][Fisher score, $>$C1 flares]{
\includegraphics[width=0.475\textwidth]{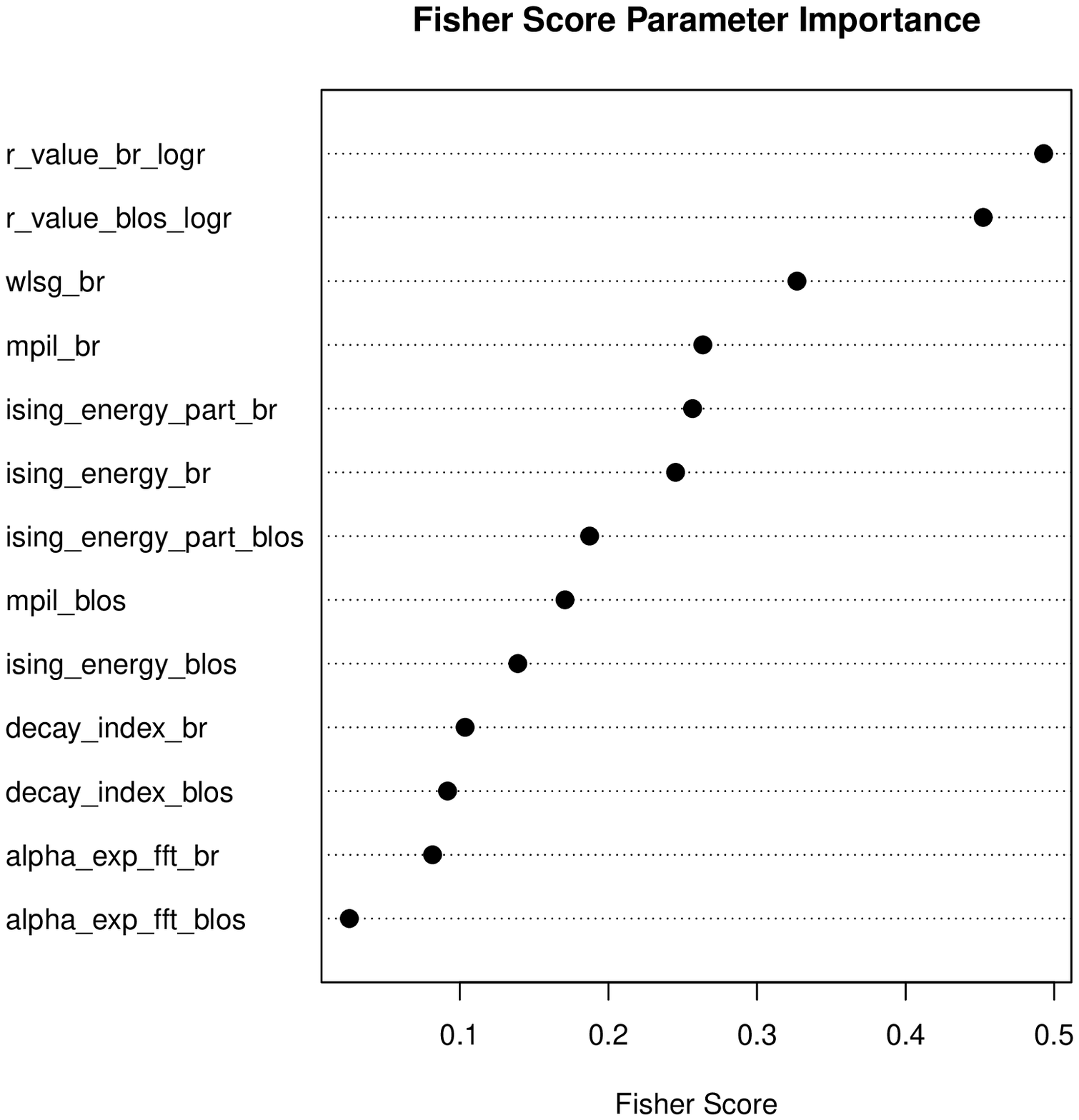}
\label{fig:aaaa}}
\subfloat[Subfigure 2 list of figures text][Gini importance, $>$C1 flares]{
\includegraphics[width=0.475\textwidth]{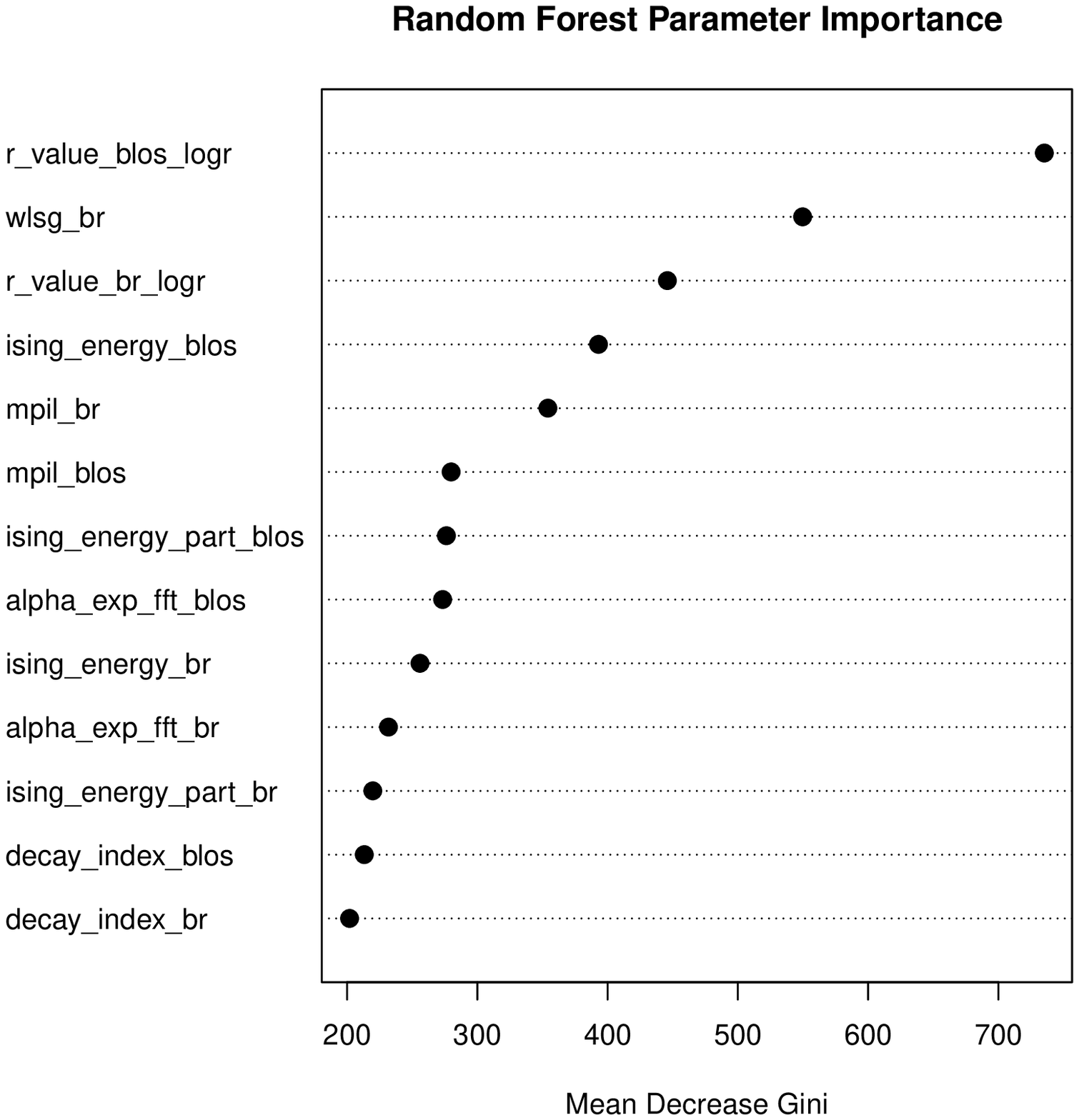}
\label{fig:bbbb}}
\qquad
\subfloat[Subfigure 3 list of figures text][Correlation, $>$C1 flares]{
\includegraphics[width=0.875\textwidth]{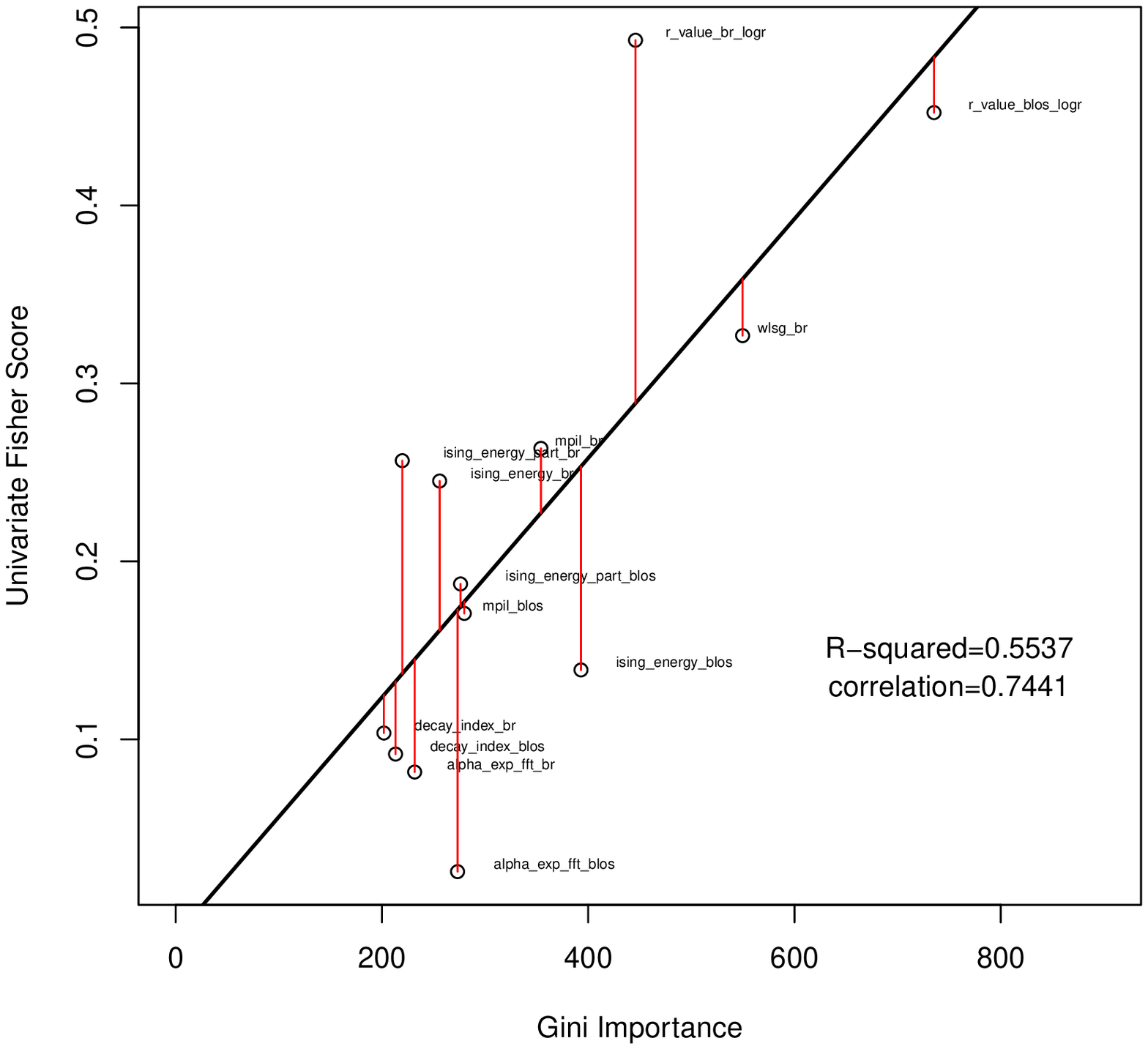}
\label{fig:cccc}}
\caption{Importance of several predictors while predicting $>$C1 flares.}
\label{fig:A1}

\end{figure}

\begin{figure}[h]
\centering
\subfloat[Subfigure 1 list of figures text][Fisher score, $>$M1 flares]{
\includegraphics[width=0.475\textwidth]{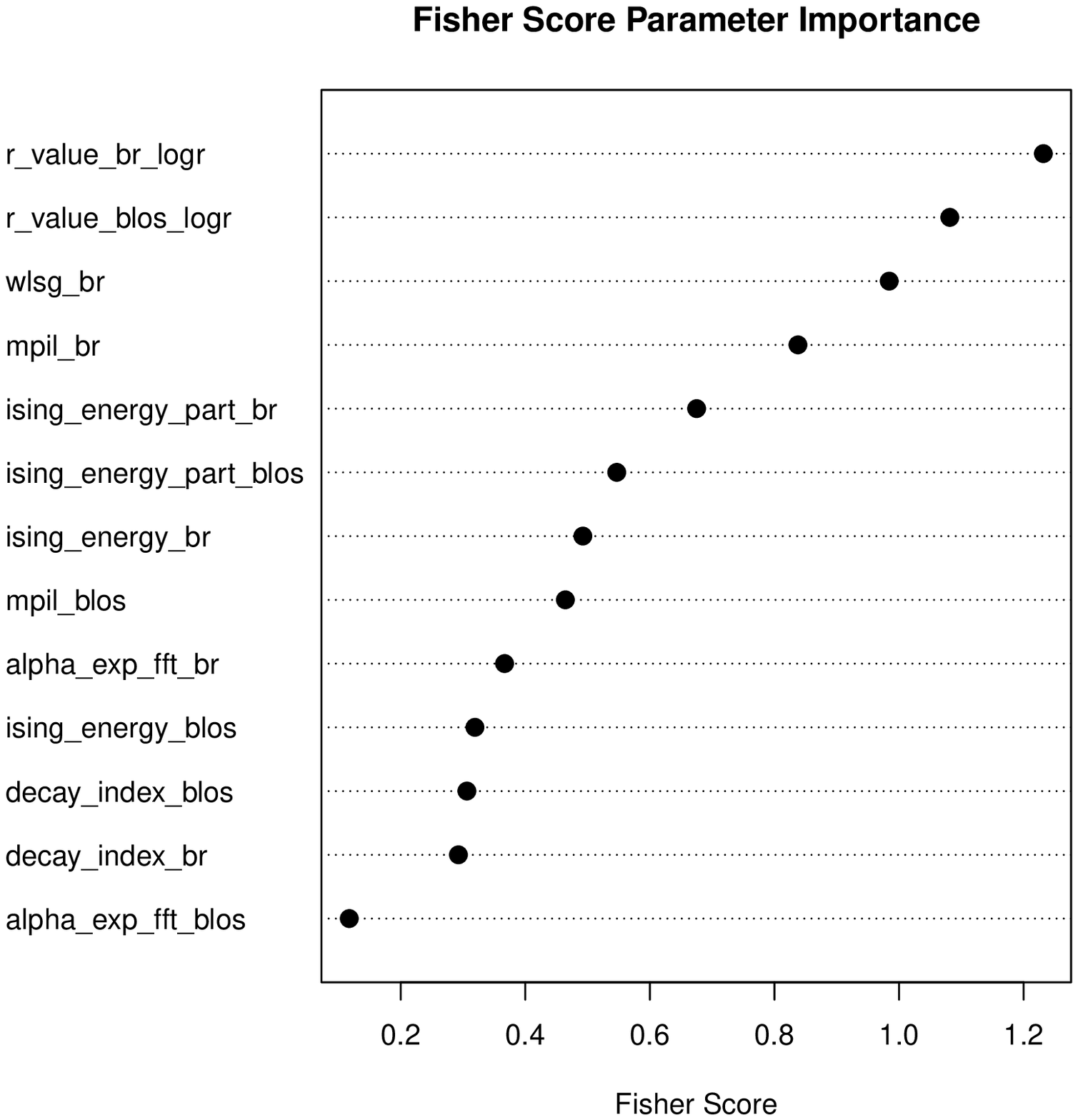}
\label{fig:aaaa}}
\subfloat[Subfigure 2 list of figures text][Gini importance, $>$M1 flares]{
\includegraphics[width=0.475\textwidth]{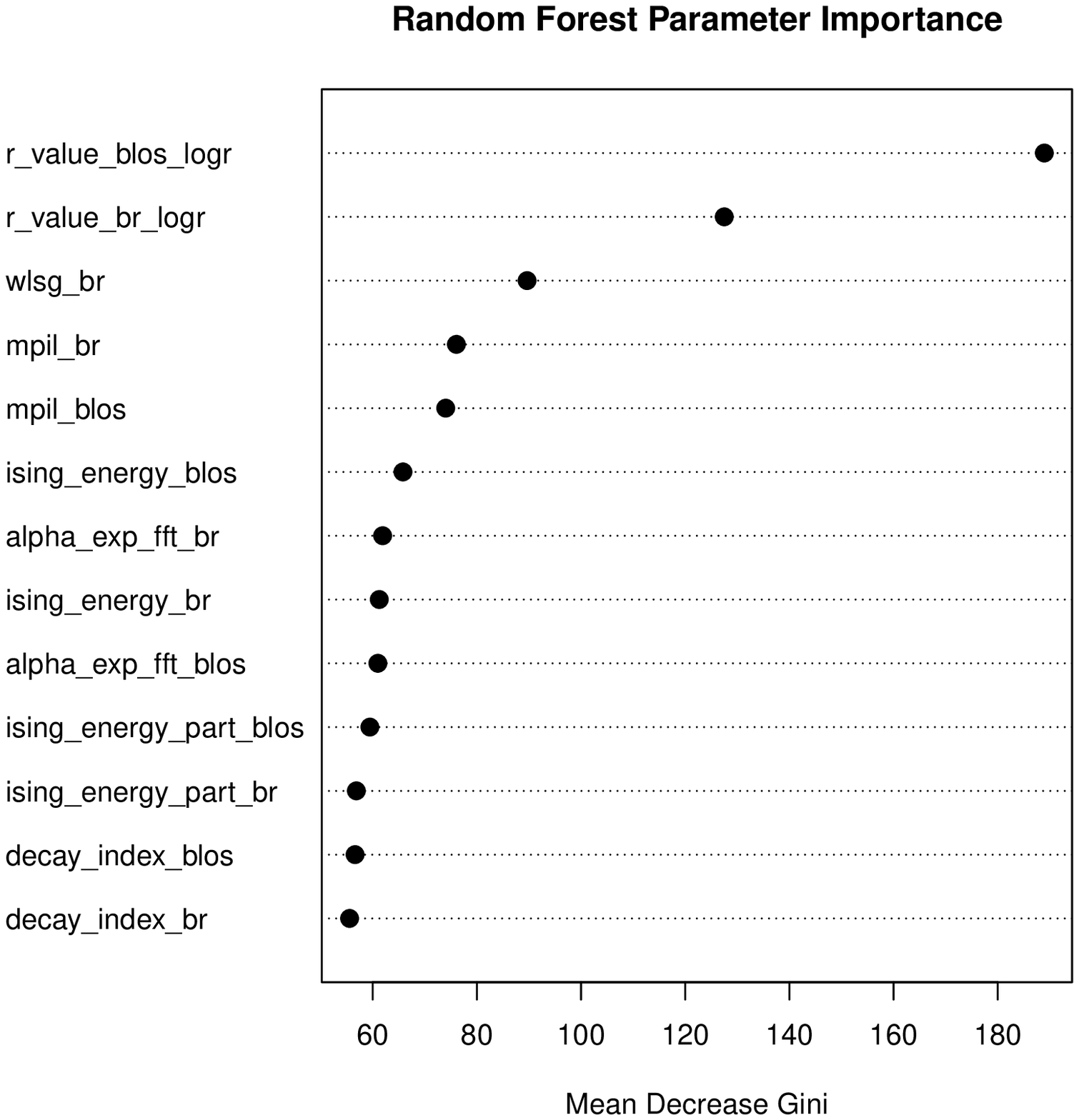}
\label{fig:bbbb}}
\qquad
\subfloat[Subfigure 3 list of figures text][Correlation, $>$M1 flares]{
\includegraphics[width=0.875\textwidth]{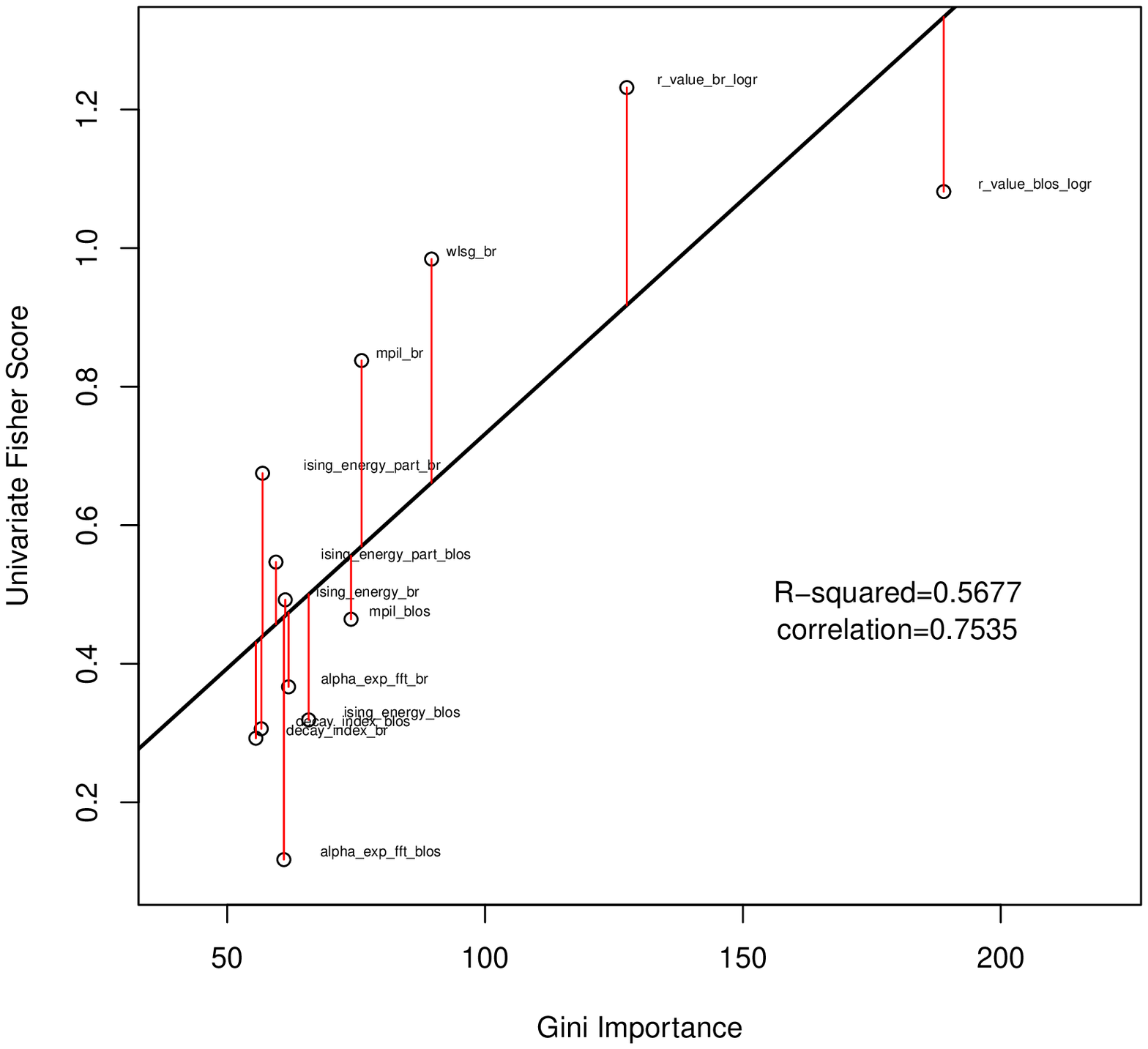}
\label{fig:cccc}}
\caption{Importance of several predictors while predicting $>$M1 flares.}
\label{fig:A2}

\end{figure}

We computed the Fisher score \citep{Bobra:15,Chang:08,Chen:06} and the Gini importance \citep{Breiman:01} for every predictor in the case of $>$M1 and $>$C1 flares. The obtained values for the importance of several predictors are presented in Figures \ref{fig:A1} and  \ref{fig:A2} for $>$C1 and $>$M1 flare prediction, respectively. The Fisher score, $F$, is defined for the \textit{j}th predictor as,

\begin{equation}
F(j) = \frac{ (\bar{x}^{(+)}_{j} - \bar{x}_j)^2 + (\bar{x}^{(-)}_j - \bar{x}_j)^2 }{ \frac{1}{n^+ - 1} \sum\limits_{k=1}^{n^+} ( x^{(+)}_{k,j} - \bar{x}^{(+)}_j)^2   +   \frac{1}{n^- - 1} \sum\limits_{k=1}^{n^-} ( x^{(-)}_{k,j} - \bar{x}^{(-)}_j)^2 } \,.
\label{eq.A1}
\end{equation}

In Equation \ref{eq.A1}, $\bar{x}_j$, $\bar{x}^{(+)}_{j}$ and $\bar{x}^{(-)}_{j}$ are the mean values for the \textit{j}th predictor over the entire sample, the positive class and the negative class, respectively. Furthermore, $n^+$ ($n^-$) are the number of positive (negative) class observations. Also, $x^{(+)}_{k,j}$ $(x^{(-)}_{k,j})$ are the values for the \textit{k}th observation of the \textit{j}th predictor belonging in the positive (negative) class. The higher the value of $F(j)$ the more important the \textit{j}th predictor.

The Gini importance is returned with the randomForest function of the randomForest package in R. The higher the Gini importance of the \textit{j}-th predictor the more important this predictor is. 

We note that the correlation between the two quantities ({\it e.g.} Fisher score and Gini importance) is $r = 0.7441 $ for $>$C1 sflares and $r = 0.7535 $ for $>$M1 flares, respectively. So, the two methods qualitatively agree on describing which predictors are the most important regarding flare prediction, in both classes of flare prediction. Also, by looking at Figure \ref{fig:A1} we see that for $>$C1 flares, the top three ranked predictors for both Fisher Score and Gini importance are: the two versions of Schrijver's $R$ and ${\rm WL}_{\rm SG}$. Regarding the $>$M1 flares, from Figure \ref{fig:A2} the top four ranked predictors for either Fisher score or Gini importance are: the two versions of Schrijver's $R$, ${\rm WL}_{\rm SG}$ and ${\rm TLMPIL}_{\rm Br}$. In Appendix A the terminology for every predictor is explained in Table \ref{tab:AbbrevPred}.


\begin{table}[ht]
\caption{Abbreviations for predictors used in main text (Symbol1) and in Figures \ref{fig:A1} \textbf{and} \ref{fig:A2} (Symbol2).}
\label{tab:AbbrevPred}
\begin{tabular}{lll}
\hline
\multicolumn{3}{c}{Abbreviations for Predictors}\\
  \hline
 Symbol1 & Symbol2 & Description \\
 ${\rm log}{\rm R}$ & r\_value\_logr & Schrijver's $R$ value \\ 
 {\rm FSPI} & alpha\_exp\_fft & Fourier spectral power index \\ 
 {\rm TLMPIL} & mpil & Magnetic polarity inversion line \\ 
 {\rm DI} & decay\_index & Decay index \\ 
 ${{\rm WL}_{\rm SG}}$ & wlsg & Gradient-weighted integral length of the neutral line \\ 
 {\rm IsinEn1} & ising\_energy & Ising Energy original \\ 
 {\rm IsinEn2} & ising\_energy\_part & Ising Energy partitioned \\ 
 \hline
\end{tabular}
\end{table}


\clearpage

\section{Prediction Models Resulting From Ranking the Predictors}

\begin{figure}[h]
\centering
\subfloat[Subfigure 1 list of figures text][RF, ranking, $>$C1 flares ]{
\includegraphics[width=0.475\textwidth]{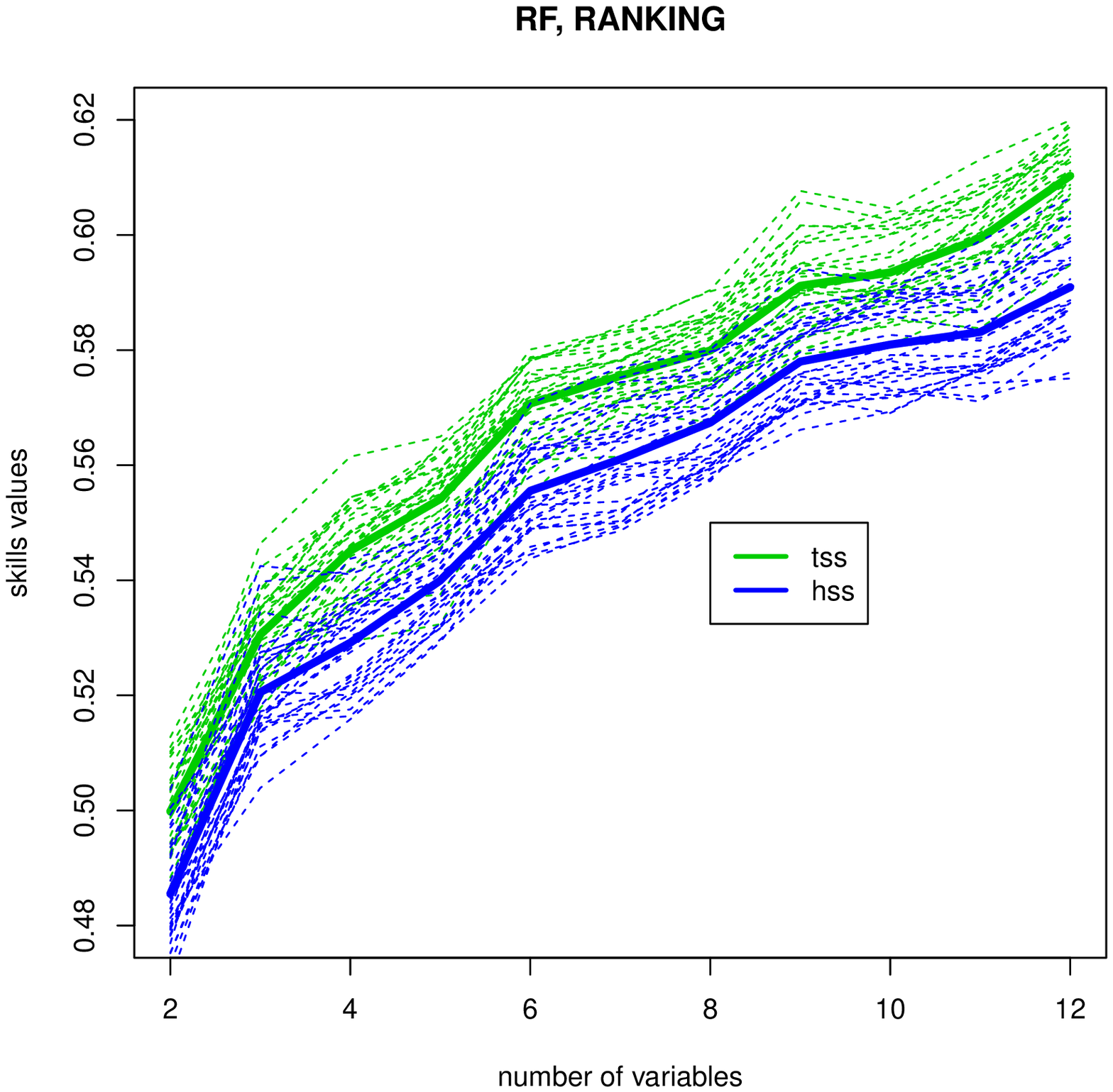}
\label{fig:aaaa}}
\subfloat[Subfigure 2 list of figures text][SVM, ranking, $>$C1 flares]{
\includegraphics[width=0.475\textwidth]{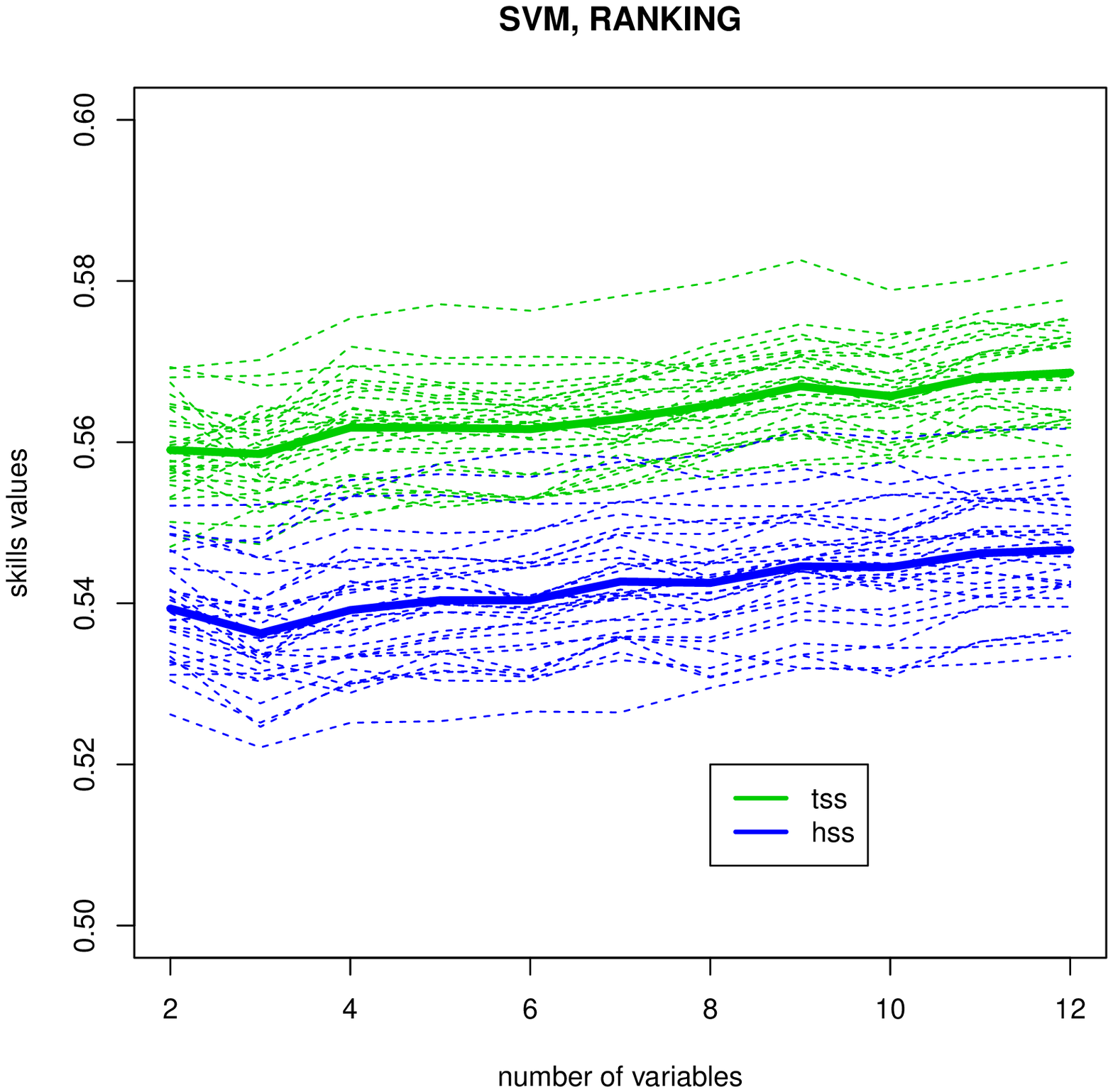}
\label{fig:bbbb}}
\qquad
\subfloat[Subfigure 3 list of figures text][MLP, ranking, $>$C1 flares]{
\includegraphics[width=0.475\textwidth]{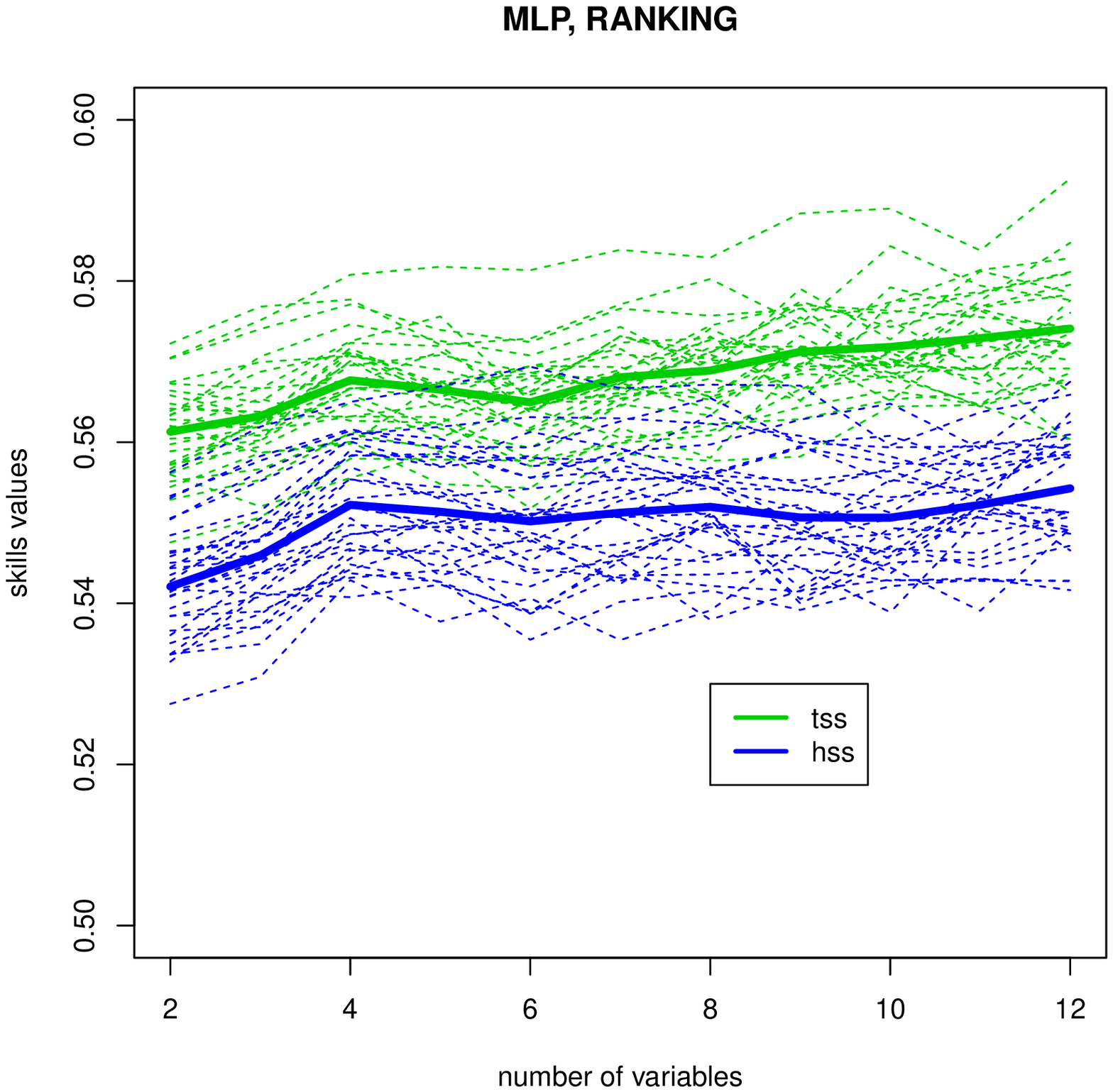}
\label{fig:cccc}}
\subfloat[Subfigure 4 list of figures text][LM, ranking, $>$C1 flares]{
\includegraphics[width=0.475\textwidth]{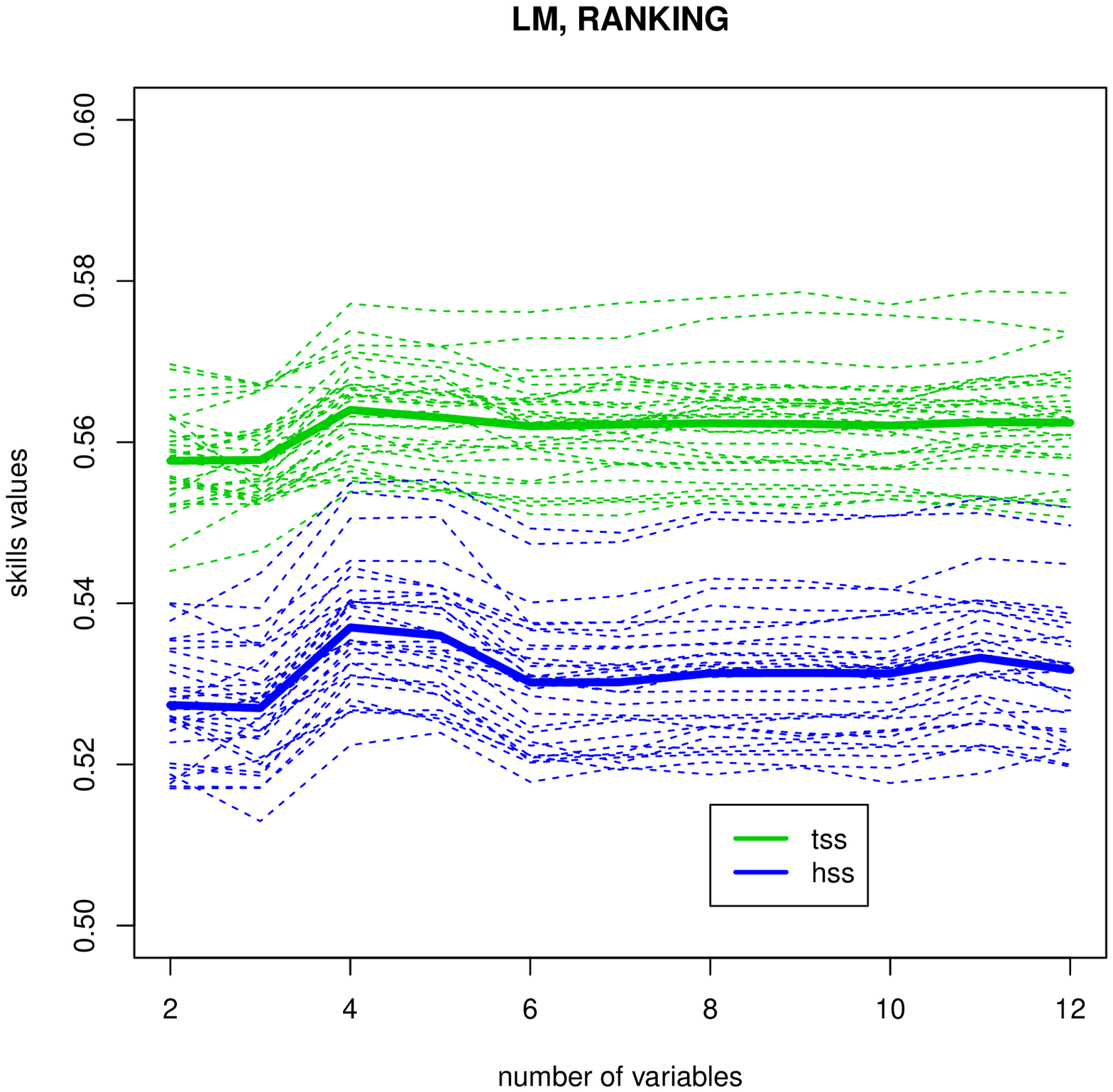}
\label{fig:dddd}}
\qquad
\subfloat[Subfigure 5 list of figures text][PR, ranking, $>$C1 flares]{
\includegraphics[width=0.475\textwidth]{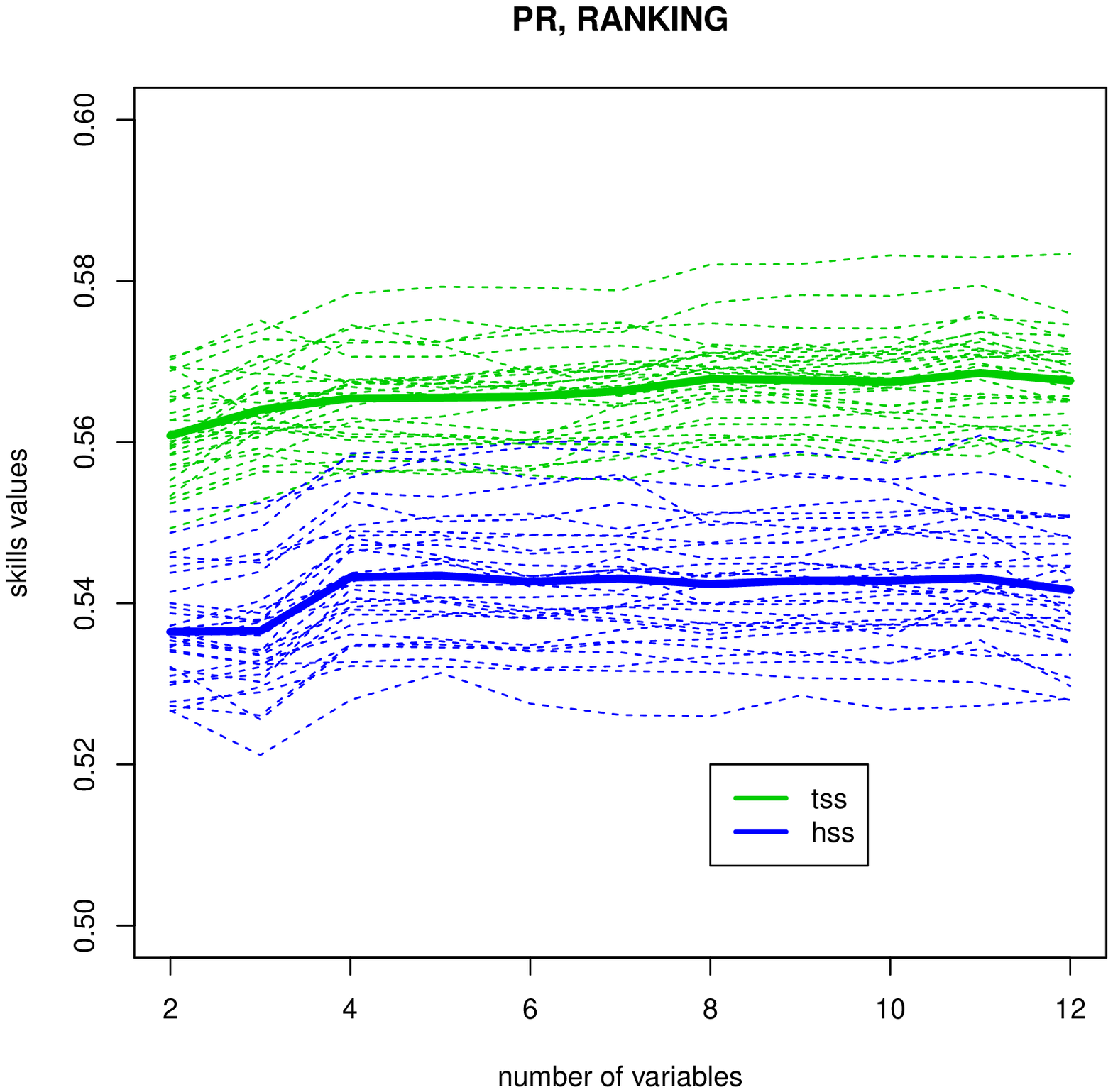}
\label{fig:eeee}}
\subfloat[Subfigure 6 list of figures text][LG, ranking, $>$C1 flares]{
\includegraphics[width=0.475\textwidth]{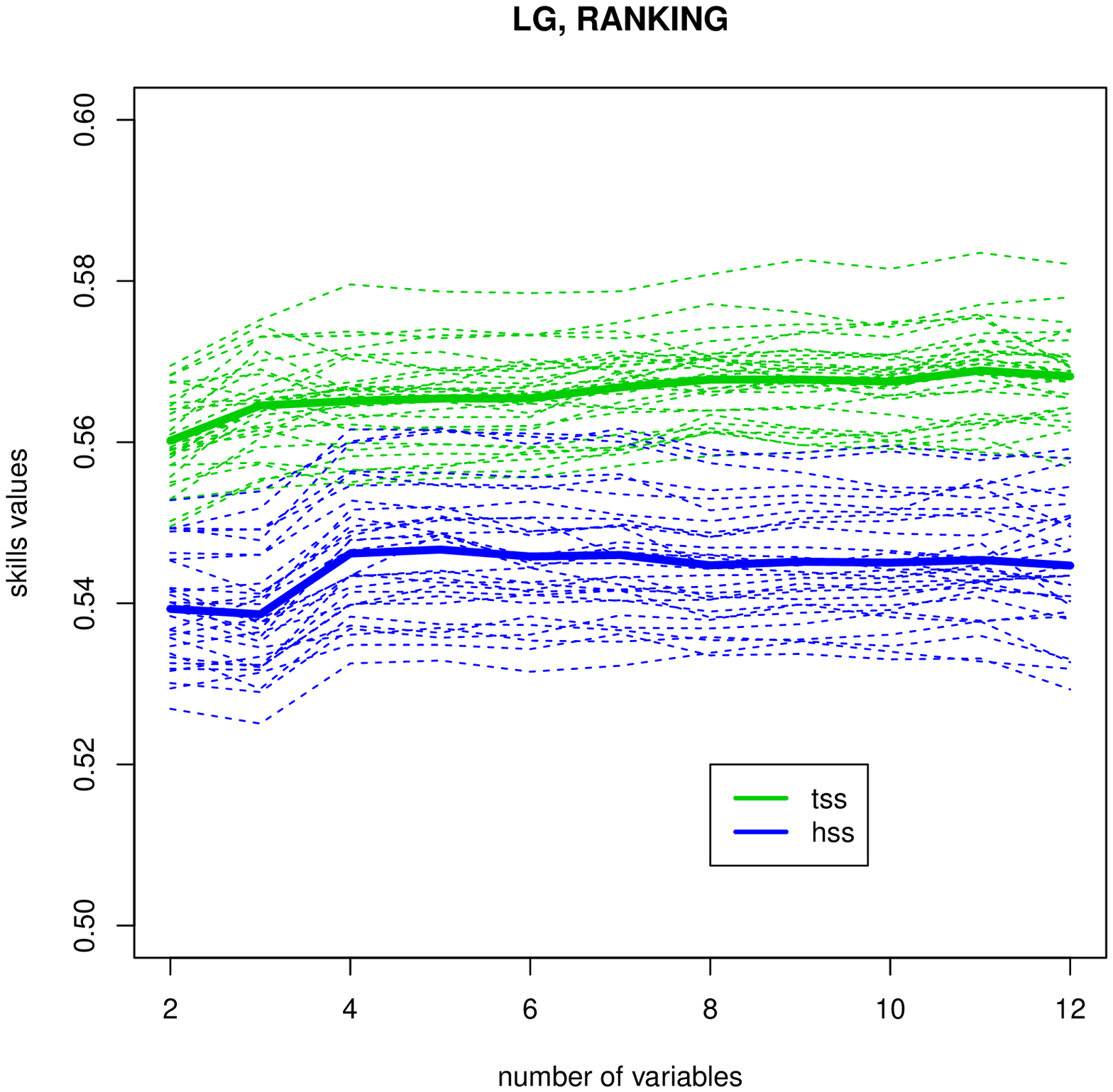}
\label{fig:ffff}}
\caption{Out-of-sample skill scores ({\rm TSS} and {\rm HSS}) for the three ML prediction methods anf the three statistics methods during the ranking procedure for $>$C1 flares. The thick continuous lines depict the averages of the skill scores over 30 randomized runs.}
\label{fig:B1}

\end{figure}

\begin{figure}[h]
\centering
\subfloat[Subfigure 1 list of figures text][RF, ranking, $>$M1 flares]{
\includegraphics[width=0.475\textwidth]{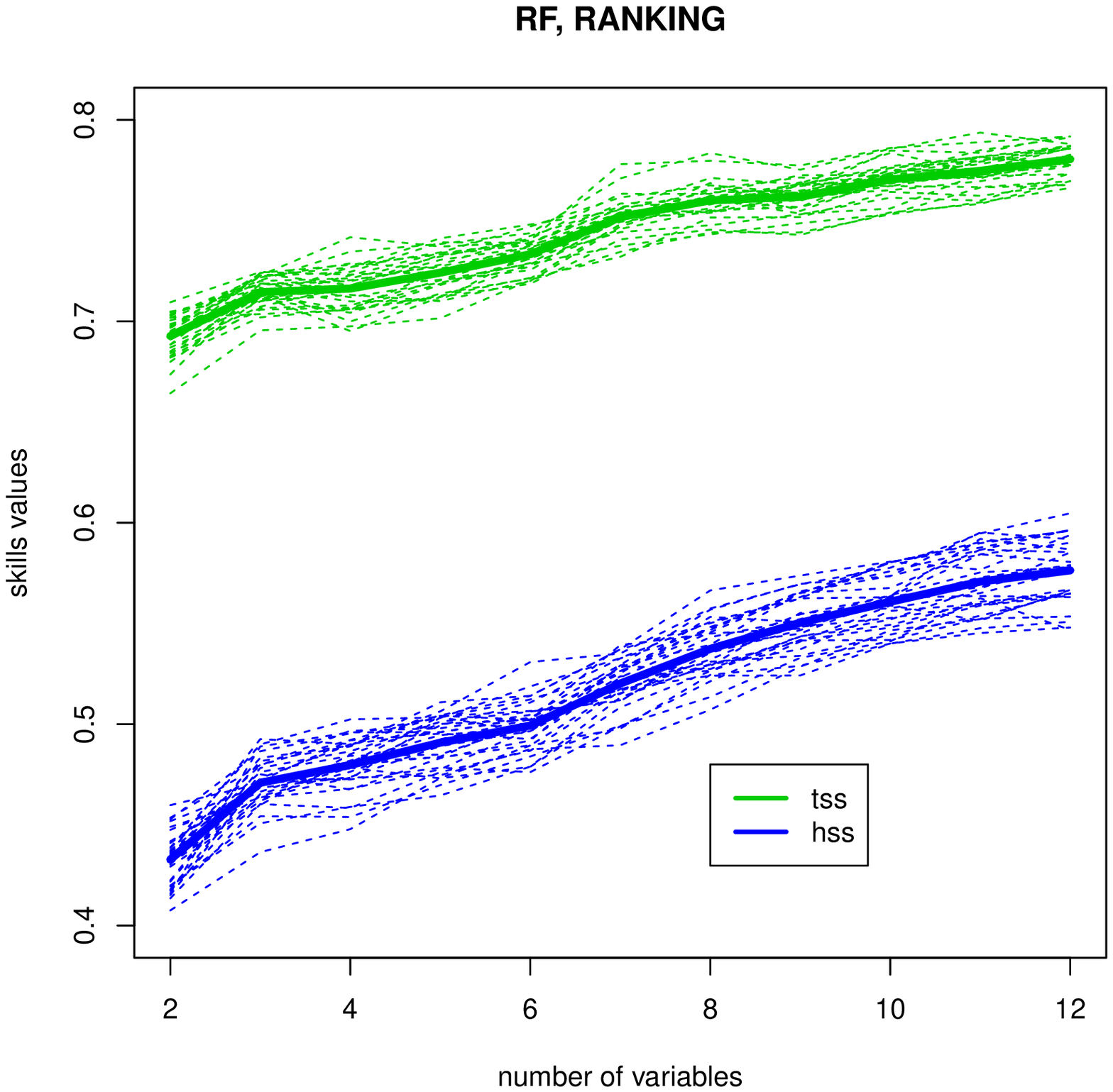}
\label{fig:aaaa}}
\subfloat[Subfigure 2 list of figures text][SVM, ranking, $>$M1 flares]{
\includegraphics[width=0.475\textwidth]{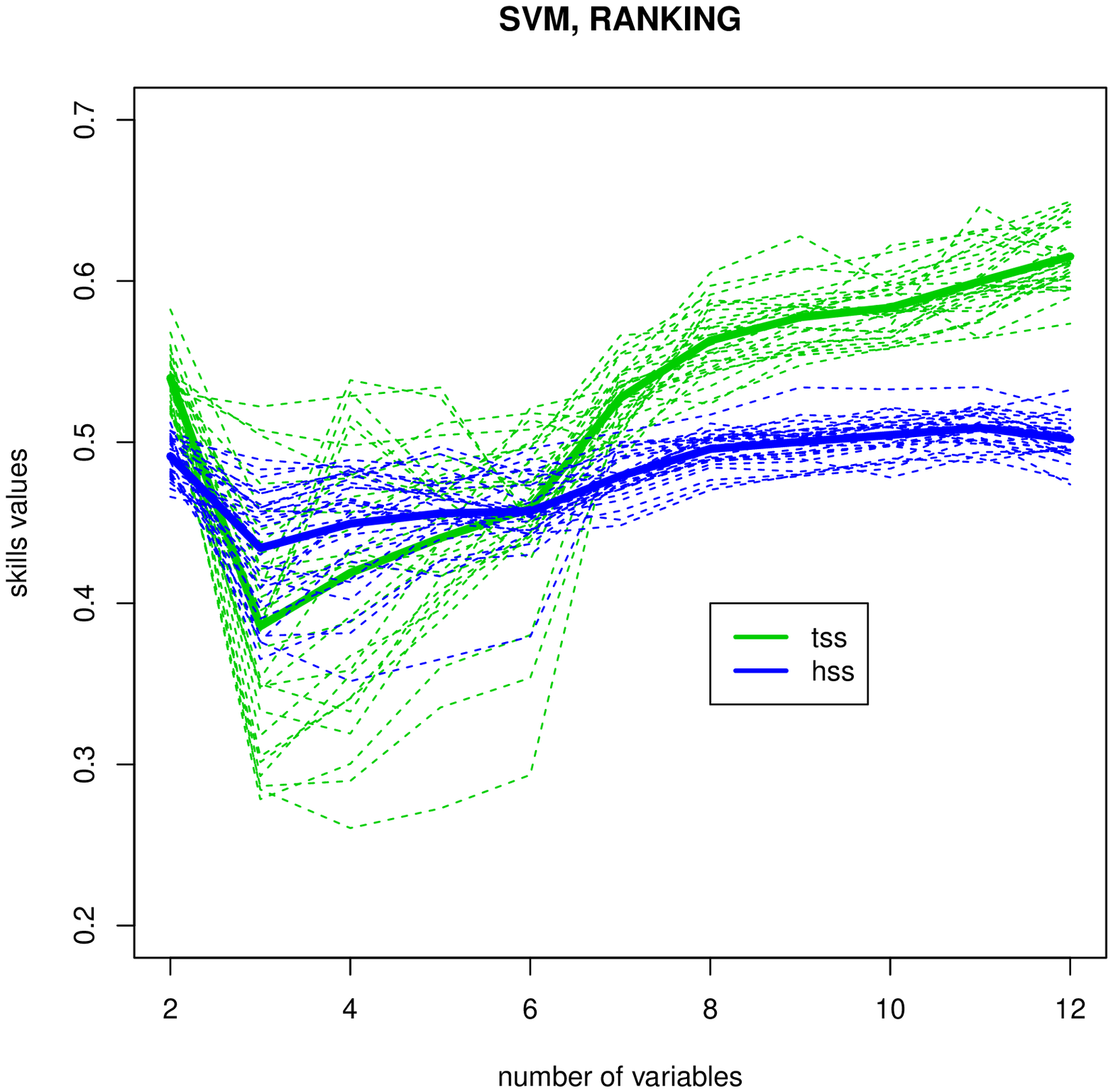}
\label{fig:bbbb}}
\qquad
\subfloat[Subfigure 3 list of figures text][MLP, ranking, $>$M1 flares]{
\includegraphics[width=0.475\textwidth]{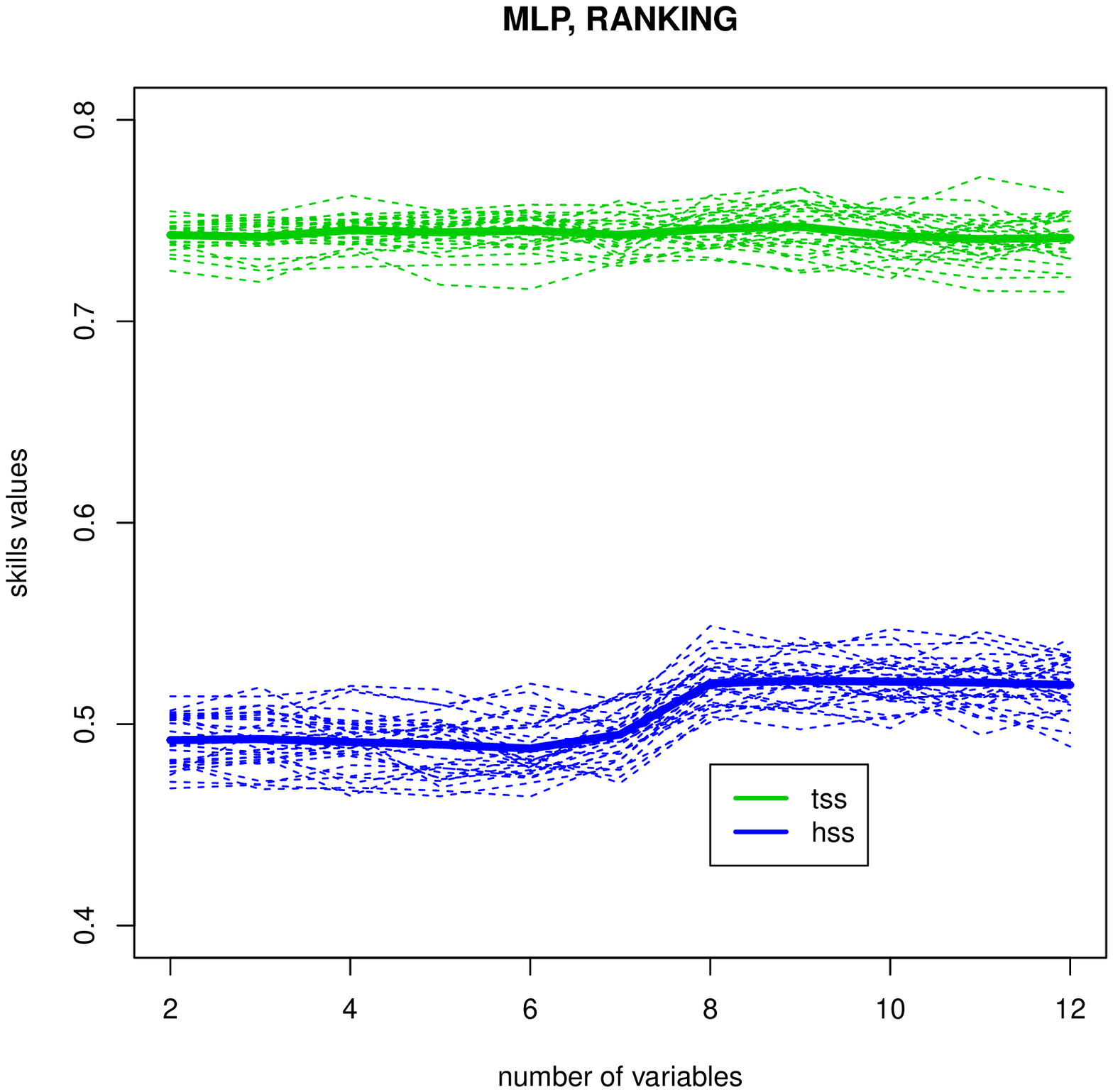}
\label{fig:cccc}}
\subfloat[Subfigure 3 list of figures text][LM, ranking, $>$M1 flares]{
\includegraphics[width=0.475\textwidth]{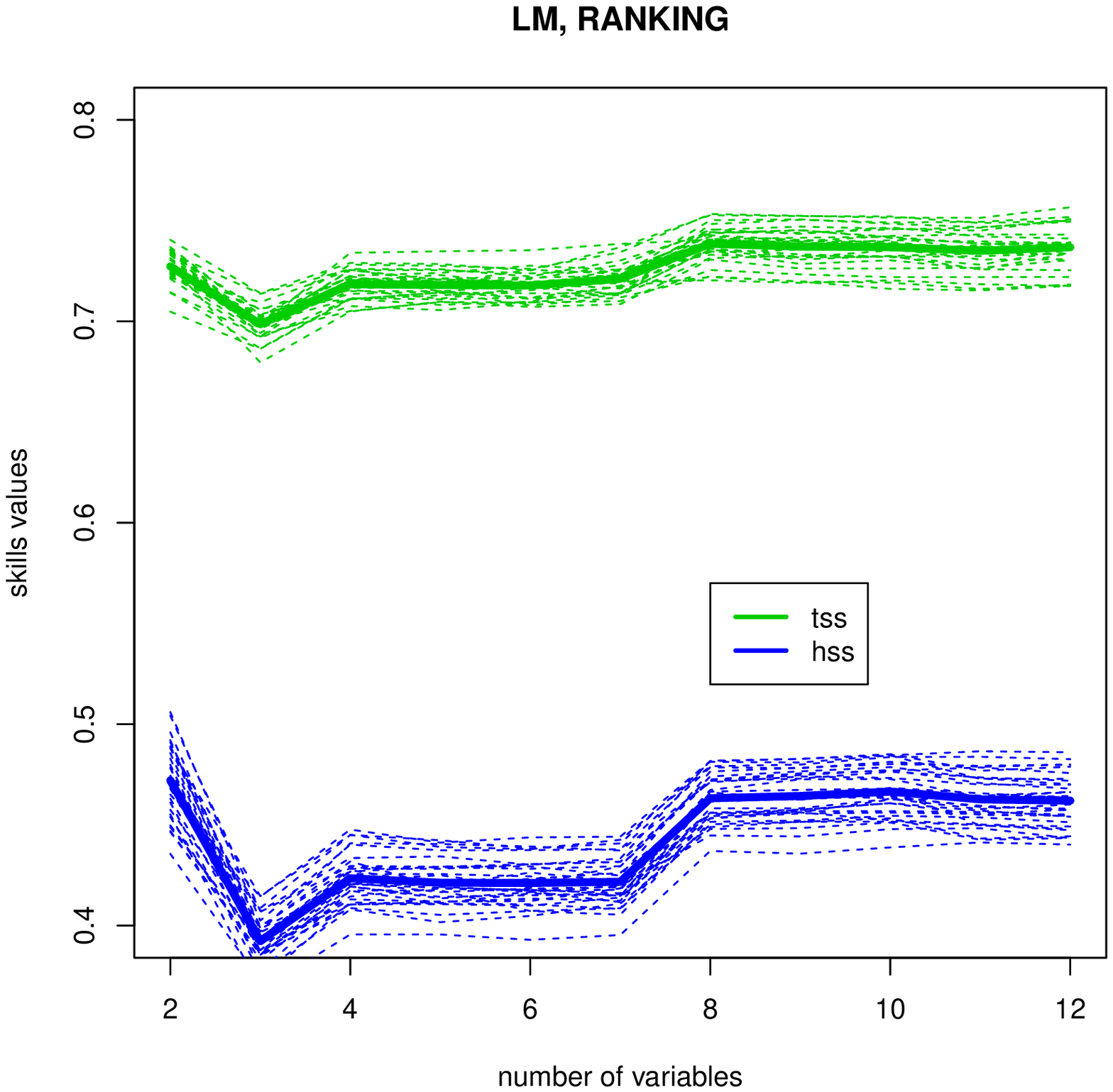}
\label{fig:dddd}}
\qquad
\subfloat[Subfigure 3 list of figures text][PR, ranking, $>$M1 flares]{
\includegraphics[width=0.475\textwidth]{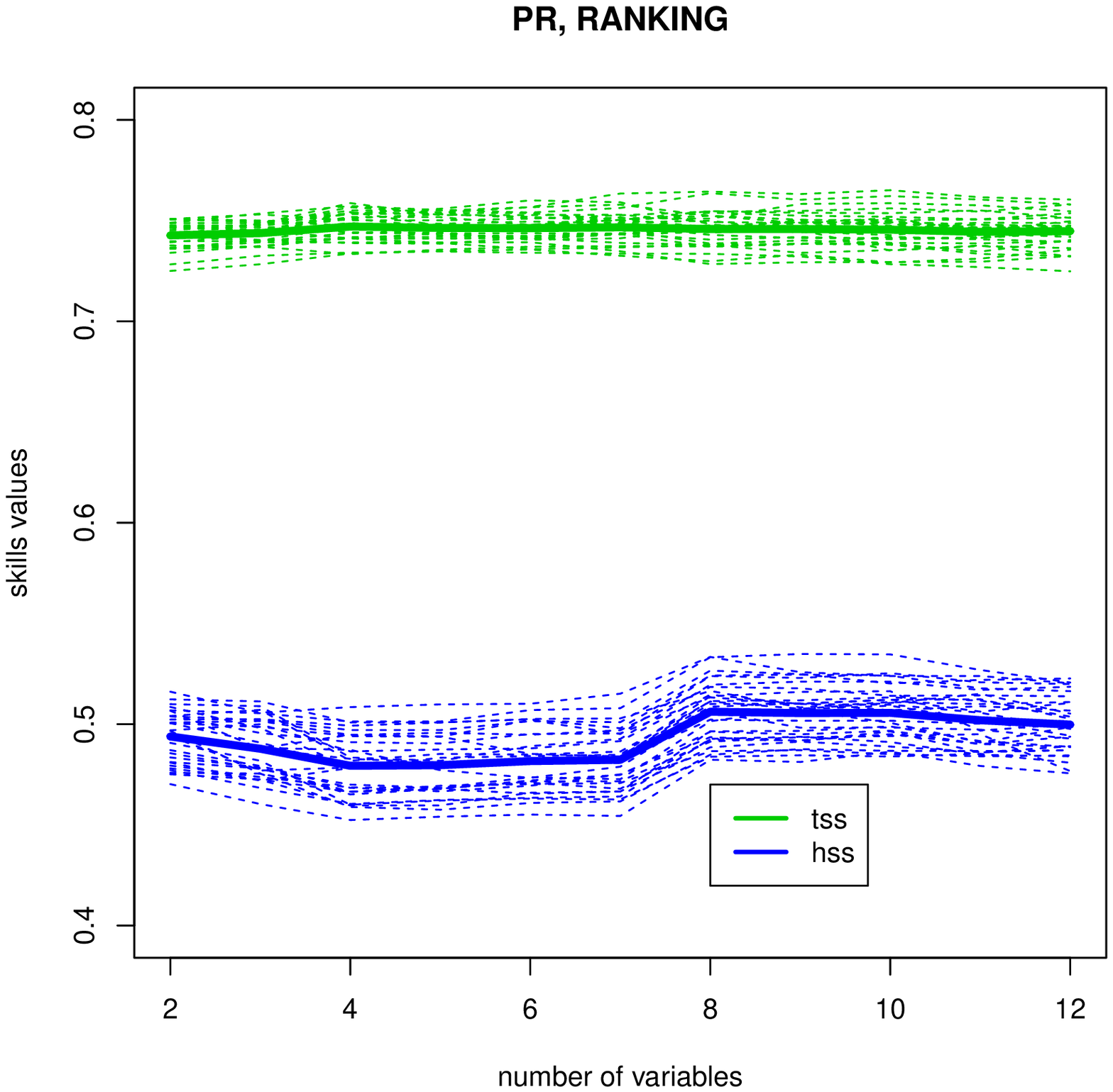}
\label{fig:eeee}}
\subfloat[Subfigure 3 list of figures text][LG, ranking, $>$M1 flares]{
\includegraphics[width=0.475\textwidth]{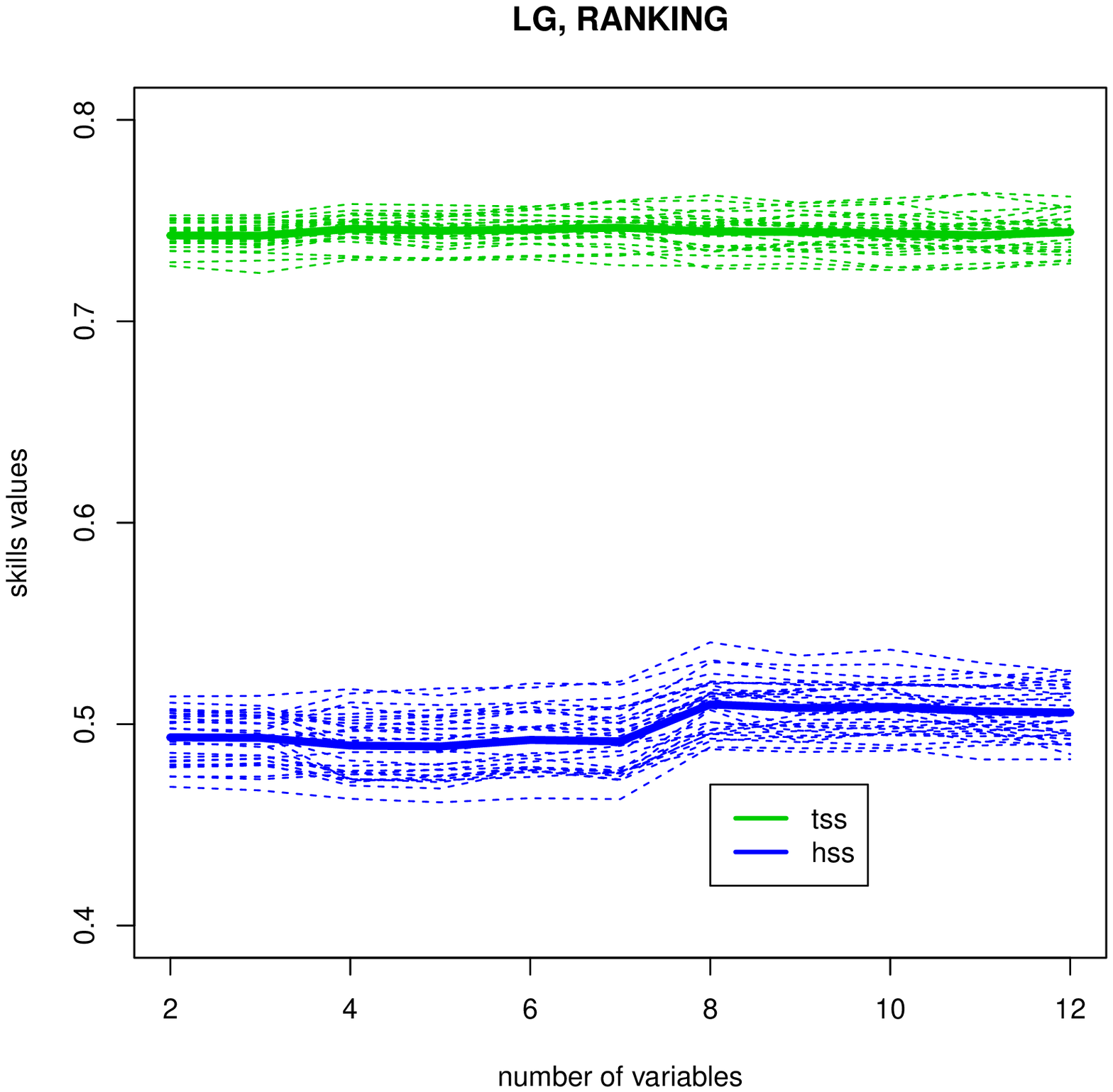}
\label{fig:ffff}}
\caption{Same as in Figure \ref{fig:B1} but for $>$M1 flares.}
\label{fig:B2}

\end{figure}

We employed a backward elimination procedure, eliminating gradually predictors according to their Fisher score rank, starting form the model with all $K=13$ predictors included. In every step, we eliminated the least important predictor from the set of currently included predictors. So, we obtained prediction results for models with 2, 3, \dots, 11, 12 predictors included for the ML methods, RF, SVM and MLP and the conventional statistics methods LM, PR and LG. The results of this iterative procedure for flares $>$C1 and $>$M1 are presented in Figures \ref{fig:B1} and \ref{fig:B2}.

Figure \ref{fig:B1} shows that there is a cut-off for the number of parameters included in the RF equal to the 6 most important ones (according to Fisher score in Equation \ref{eq.A1}) above which the RF is advantageous over the other two ML algorithms. For low-dimensional prediction models ({\it e.g.} below 6 included parameters) there is no special advantage in using RF, and MLP or SVM seem a better choice then. This finding shows that among the highly correlated set of predictors, the MLP and SVM perform well using only a handful of them (below 6), yet the RF continues to improve its performance in higher-dimensional settings, when the prediction model includes all 12 most important predictors. There is interest in investigating the performance of RF when the number of (correlated) predictors would be twice or three times that of the present study (24-36 predictors). Would the upwards trend in Figure \ref{fig:B1}a continue to hold when the number of included parameters increases to 24 or 36? We note that RF is the only ML algorithm in the present study which belongs in the category of ``ensemble'' methods. Moreover, in Figure \ref{fig:B1} the performance of the three conventional statistics methods LM, PR and LG is presented. Clearly, the LM presents the worst forecasting ability and also we notice that in general the other two methods, PR and LG, score similar values for the {\rm TSS} and {\rm HSS}. Also, it is noteworthy that the profiles of PR and LG are pretty flat as a function of the number of included predictors, even flatter than the profiles from SVM and MLP.

Likewise, Figure \ref{fig:B2} shows that for low-dimensional settings RF is worse than MLP. The cut-off seems again to be 6 included parameters. Above this value, the RF provides better out-of-sample {\rm TSS} and {\rm HSS} than MLP. There seems to be a problematic region between 3 and 6 parameters included for the SVM, where adding more parameters to the SVM degrades its performance. Above 6 parameters, the SVM performance again improves. Similarly to the $>$C1 class flares case, we again notice in Figure \ref{fig:B2} rather flat profiles for the {\rm TSS} and {\rm HSS} for the conventional statistics methods, with PR and LG showing better behaviour than LM.

One general conclusion is that for very few predictors $K<6$, all methods work the same, so for parsimony the conventional statistics methods could be preferred. This is also true for very small samples $N<2000$ (results available upon request). On the contrary, when $K > 6$ and $N \geq 10,000$ the ML methods and especially the RF are better.

We note that in Appendix B, the MLP has always 4 hidden nodes and the SVM has $\gamma$ and {\rm cost} parameters analogously to the full $K=13$ SVM model for $>$C1 and $>$M1 flares cases.

\clearpage

\clearpage

\clearpage

\section{Validation Results When Predictions Are Issued Only Once a Day (at Midnight)}

We present here forecasting results in the following scenario:
\begin{enumerate}[i)]
\item
The training is perfomed as in the main scenario.
\item
The testing is performed only for the observations in the testing set of the main scenario which correspond to a time of 00:00 UT. To achieve this, we filter for the observations in the previous testing set with midnightStatus=TRUE.

\end{enumerate}

\noindent This method of training-testing is called the ``hybrid method'' where training is done with a cadence of 3\;h and a forecast window of 24\;h, and testing is done with a cadence of 24\;h and a forecast window of 24\;h. The hybrid method is preferable over doing a training phase with cadence of 24\;h, which would result in under-trained models, due to the limited sample size during training.

Tables \ref{tab:AuxSkillScoresM1.C1} and \ref{tab:AuxSkillScoresCclass.C2} are analogous to Tables \ref{tab:AuxSkillScoresM1} and \ref{tab:AuxSkillScoresCclass} of the main scenario, but for midnight (so once a day) only predictions. For completeness, we remind that Table \ref{tab:AuxSkillScoresM1.C1} is for $>$M1 flare prediction and Table \ref{tab:AuxSkillScoresCclass.C2} is for $>$C1 flare prediction.

By comparing Table \ref{tab:AuxSkillScoresM1.C1} to Table \ref{tab:AuxSkillScoresM1} we see that BS and AUC do not change much on average when we move from the baseline scenario to the midnight prediction scenario. Nevertheless, the associated uncertainty increases in the case of midnight only predictions, since the size of the testing set is smaller (only one, rather than eight, predictions per day). More significant differences are observed for BSS since the associated climatology is also different. Nevertheless, the finding that RF is the best overall method continues to hold.

Similar conclusions can be drawn for the $>$C1 flare prediction case, so through Table \ref{tab:AuxSkillScoresCclass.C2} and Table \ref{tab:AuxSkillScoresCclass}. Here, noticeably, not even the BSS changes significantly, since the underlying climatology seems similar in both cases. This is because, contrary to $>$M1 class flares, with a mean frequency of $\approx$5\%, $>$C1 class flares show a mean frequency of $\approx$25\%.

Finally, Tables \ref{tab:MonteCarloM1.C3} and \ref{tab:MonteCarloC.C4} present the skill scores {\rm ACC}, {\rm TSS} and {\rm HSS} for the midnight prediction scenario analogously to Tables \ref{tab:MonteCarloM1} and \ref{tab:MonteCarloC} for the baseline scenario. For completeness, we notice that Table \ref{tab:MonteCarloM1.C3} is for $>$M1 flare prediction and Table \ref{tab:MonteCarloC.C4} is for $>$C1 flare prediction. We see that on average the issuing of midnight only predictions does not change much the {\rm ACC}, {\rm TSS} and {\rm HSS} with respect to the probability threshold. For example, on $>$M1 flares midnight-only predictions, the RF provides {\rm ACC}=0.93$\pm$0.01, {\rm TSS}=0.73$\pm$0.04 and {\rm HSS}=0.47$\pm$ 0.03 for probability threshold 15\%. Also, for $>$C1 flares midnight-only predictions, the RF yields {\rm ACC}=0.85$\pm$0.01, {\rm TSS}=0.63$\pm$0.02 and {\rm HSS}=0.61$\pm$0.02 for probability threshold 35\%.

\begin{table}[ht]
\caption{Same as Table \ref{tab:AuxSkillScoresM1} but for predictions issued only at midnight.}
\label{tab:AuxSkillScoresM1.C1}
\begin{tabular}{cccccc}
\hline
\multicolumn{6}{c}{BS}\\
 \hline
 MLP & LM & PR & LG & RF & SVM \\
 0.0320 & 0.0328 & 0.0305 & 0.0305 & 0.0262 & 0.0333 \\ 
 (0.0031) & (0.0025) & (0.0025) & (0.0025) & (0.0022) & (0.0032) \\
  \hline
  \multicolumn{6}{c}{AUC}\\
  \hline
 MLP & LM & PR & LG & RF & SVM \\
 0.9342 & 0.9245 & 0.9419 & 0.9412 & 0.9558 & 0.8361 \\ 
 (0.0131) & (0.0111) & (0.0087) & (0.0089) & (0.0081) & (0.0339) \\ 
  \hline
  \multicolumn{6}{c}{BSS}\\
  \hline
 MLP & LM & PR & LG & RF & SVM \\
 0.2311 & 0.2122 & 0.2681 & 0.2686 & 0.3722 & 0.2007 \\ 
 (0.0671) & (0.0316) & (0.0391) & (0.0421) & (0.0397) & (0.0538) \\ 
   \hline
\end{tabular}
\end{table}

\begin{table}[ht]
\caption{Same as Table \ref{tab:AuxSkillScoresCclass}, but for predictions issued only at midnight.}
\label{tab:AuxSkillScoresCclass.C2}
\begin{tabular}{cccccc}
\hline
\multicolumn{6}{c}{BS}\\
  \hline
 MLP & LM & PR & LG & RF & SVM \\
 0.1142 & 0.1273 & 0.1181 & 0.1169 & 0.1023 & 0.1187 \\ 
 (0.0041) & (0.0034) & (0.0037) & (0.0038) & (0.0037) & (0.0041) \\ 
   \hline
   \multicolumn{6}{c}{AUC}\\
   \hline
 MLP & LM & PR & LG & RF & SVM \\
 0.8771 & 0.8673 & 0.8696 & 0.8696 & 0.9004 & 0.8620 \\ 
 (0.0078) & (0.0079) & (0.0079) & (0.0079) & (0.0074) & (0.0086) \\ 
  \hline
  \multicolumn{6}{c}{BSS}\\
  \hline
 MLP & LM & PR & LG & RF & SVM \\
 0.3970 & 0.3281 & 0.3767 & 0.3826 & 0.4597 & 0.3735 \\ 
 (0.0190) & (0.0143) & (0.0163) & (0.0169) & (0.0169) & (0.0172) \\ 
   \hline
\end{tabular}
\end{table}

\begin{ltable}

\caption{Same as Table \ref{tab:MonteCarloM1} but for predictions issued only at midnight.}
\label{tab:MonteCarloM1.C3}
{\tiny
\begin{tabular}{|lr|rrr|rrr|rrr|rrr|rrr|rrr|}
  \hline
 &  &  & MLP &  &  & LM &  &  & PR &  &  & LG &  &  & RF &  &  & SVM &  \\
Par & \% & ACC & TSS & HSS & ACC & TSS & HSS & ACC & TSS & HSS & ACC & TSS & HSS & ACC & TSS & HSS & ACC & TSS & HSS \\
  \hline
$val_{0}$ & 0.00 & 0.18 & 0.14 & 0.01 & 0.33 & 0.29 & 0.04 & 0.00 & 0.00 & 0.00 & 0.00 & 0.00 & 0.00 & 0.50 & 0.47 & 0.07 & 0.00 & 0.00 & 0.00 \\ 
  $val_{5}$ & 0.05 & 0.90 & 0.70 & 0.37 & 0.76 & 0.69 & 0.19 & 0.84 & 0.77 & 0.28 & 0.85 & 0.77 & 0.30 & 0.85 & 0.78 & 0.31 & 0.93 & 0.58 & 0.42 \\ 
  $val_{10}$ & 0.10 & 0.93 & 0.66 & 0.43 & 0.88 & 0.72 & 0.32 & 0.90 & 0.73 & 0.37 & 0.90 & 0.71 & 0.38 & 0.91 & 0.78 & 0.41 & 0.95 & 0.47 & 0.44 \\ 
  $val_{15}$ & 0.15 & 0.94 & 0.62 & 0.46 & 0.92 & 0.59 & 0.38 & 0.93 & 0.65 & 0.42 & 0.93 & 0.64 & 0.43 & 0.93 & 0.73 & 0.47 & 0.96 & 0.42 & 0.45 \\ 
  $val_{20}$ & 0.20 & 0.95 & 0.58 & 0.47 & 0.94 & 0.45 & 0.38 & 0.94 & 0.56 & 0.43 & 0.94 & 0.54 & 0.42 & 0.95 & 0.67 & 0.52 & 0.96 & 0.38 & 0.43 \\ 
  $val_{25}$ & 0.25 & 0.95 & 0.55 & 0.48 & 0.95 & 0.36 & 0.37 & 0.95 & 0.49 & 0.44 & 0.95 & 0.50 & 0.45 & 0.96 & 0.61 & 0.53 & 0.96 & 0.35 & 0.42 \\ 
  $val_{30}$ & 0.30 & 0.95 & 0.52 & 0.48 & 0.95 & 0.30 & 0.36 & 0.96 & 0.43 & 0.44 & 0.96 & 0.45 & 0.45 & 0.96 & 0.54 & 0.53 & 0.96 & 0.33 & 0.40 \\ 
  $val_{35}$ & 0.35 & 0.96 & 0.49 & 0.47 & 0.96 & 0.22 & 0.29 & 0.96 & 0.35 & 0.40 & 0.96 & 0.38 & 0.43 & 0.96 & 0.48 & 0.52 & 0.96 & 0.31 & 0.39 \\ 
  $val_{40}$ & 0.40 & 0.96 & 0.46 & 0.47 & 0.96 & 0.19 & 0.27 & 0.96 & 0.31 & 0.39 & 0.96 & 0.33 & 0.39 & 0.96 & 0.42 & 0.50 & 0.96 & 0.29 & 0.38 \\ 
  $val_{45}$ & 0.45 & 0.96 & 0.43 & 0.46 & 0.96 & 0.16 & 0.23 & 0.96 & 0.26 & 0.35 & 0.96 & 0.30 & 0.39 & 0.97 & 0.38 & 0.48 & 0.96 & 0.28 & 0.36 \\ 
  $val_{50}$ & 0.50 & 0.96 & 0.40 & 0.45 & 0.96 & 0.13 & 0.21 & 0.96 & 0.21 & 0.31 & 0.96 & 0.26 & 0.35 & 0.97 & 0.34 & 0.45 & 0.96 & 0.26 & 0.35 \\ 
  $val_{55}$ & 0.55 & 0.96 & 0.37 & 0.43 & 0.96 & 0.09 & 0.16 & 0.96 & 0.18 & 0.27 & 0.96 & 0.21 & 0.31 & 0.97 & 0.29 & 0.41 & 0.96 & 0.25 & 0.34 \\ 
  $val_{60}$ & 0.60 & 0.96 & 0.34 & 0.41 & 0.96 & 0.07 & 0.11 & 0.96 & 0.14 & 0.23 & 0.96 & 0.17 & 0.26 & 0.96 & 0.25 & 0.37 & 0.96 & 0.23 & 0.33 \\ 
  $val_{65}$ & 0.65 & 0.96 & 0.32 & 0.40 & 0.96 & 0.06 & 0.10 & 0.96 & 0.11 & 0.18 & 0.96 & 0.14 & 0.22 & 0.96 & 0.21 & 0.32 & 0.96 & 0.22 & 0.31 \\ 
  $val_{70}$ & 0.70 & 0.96 & 0.29 & 0.38 & 0.96 & 0.05 & 0.10 & 0.96 & 0.09 & 0.15 & 0.96 & 0.11 & 0.18 & 0.96 & 0.16 & 0.26 & 0.96 & 0.20 & 0.30 \\ 
  $val_{75}$ & 0.75 & 0.96 & 0.26 & 0.36 & 0.94 & 0.04 & 0.08 & 0.96 & 0.07 & 0.13 & 0.96 & 0.09 & 0.15 & 0.96 & 0.12 & 0.20 & 0.96 & 0.19 & 0.28 \\ 
  $val_{80}$ & 0.80 & 0.96 & 0.23 & 0.33 & 0.87 & 0.02 & 0.04 & 0.96 & 0.05 & 0.09 & 0.96 & 0.07 & 0.12 & 0.96 & 0.09 & 0.15 & 0.96 & 0.17 & 0.26 \\ 
  $val_{85}$ & 0.85 & 0.96 & 0.20 & 0.30 & 0.66 & 0.01 & 0.02 & 0.95 & 0.04 & 0.06 & 0.95 & 0.04 & 0.06 & 0.95 & 0.07 & 0.12 & 0.96 & 0.15 & 0.23 \\ 
  $val_{90}$ & 0.90 & 0.96 & 0.17 & 0.25 & 0.50 & 0.01 & 0.01 & 0.91 & 0.02 & 0.03 & 0.91 & 0.02 & 0.03 & 0.93 & 0.04 & 0.07 & 0.96 & 0.12 & 0.20 \\ 
  $val_{95}$ & 0.95 & 0.96 & 0.12 & 0.19 & 0.48 & 0.01 & 0.01 & 0.54 & 0.00 & 0.00 & 0.65 & 0.00 & 0.00 & 0.53 & 0.01 & 0.01 & 0.96 & 0.08 & 0.14 \\ 
  $val_{100}$ & 1.00 & 0.00 & 0.00 & 0.00 & 0.45 & 0.01 & 0.01 & 0.00 & 0.00 & 0.00 & 0.00 & 0.00 & 0.00 & 0.00 & 0.00 & 0.00 & 0.00 & 0.00 & 0.00 \\ 
   \hline
\end{tabular}
}

\end{ltable}


\begin{ltable}

\caption{Same as Table \ref{tab:MonteCarloC}, but for predictions issued only at midnight.}
\label{tab:MonteCarloC.C4}
{\tiny
\begin{tabular}{|lr|rrr|rrr|rrr|rrr|rrr|rrr|}
  \hline
 &  &  & MLP &  &  & LM &  &  & PR &  &  & LG &  &  & RF &  &  & SVM &  \\
Par & \% & ACC & TSS & HSS & ACC & TSS & HSS & ACC & TSS & HSS & ACC & TSS & HSS & ACC & TSS & HSS & ACC & TSS & HSS \\
\hline
$val_{0}$ & 0.00 & 0.00 & 0.00 & 0.00 & 0.40 & 0.18 & 0.10 & 0.00 & 0.00 & 0.00 & 0.00 & 0.00 & 0.00 & 0.27 & 0.02 & 0.01 & 0.00 & 0.00 & 0.00 \\ 
  $val_{5}$ & 0.05 & 0.53 & 0.35 & 0.22 & 0.44 & 0.23 & 0.14 & 0.48 & 0.29 & 0.17 & 0.50 & 0.31 & 0.19 & 0.52 & 0.35 & 0.22 & 0.29 & 0.05 & 0.02 \\ 
  $val_{10}$ & 0.10 & 0.67 & 0.51 & 0.37 & 0.51 & 0.32 & 0.20 & 0.59 & 0.41 & 0.28 & 0.61 & 0.44 & 0.30 & 0.65 & 0.50 & 0.35 & 0.47 & 0.26 & 0.16 \\ 
  $val_{15}$ & 0.15 & 0.73 & 0.56 & 0.44 & 0.58 & 0.40 & 0.26 & 0.66 & 0.49 & 0.35 & 0.68 & 0.51 & 0.38 & 0.72 & 0.57 & 0.43 & 0.75 & 0.57 & 0.47 \\ 
  $val_{20}$ & 0.20 & 0.77 & 0.59 & 0.49 & 0.65 & 0.49 & 0.35 & 0.73 & 0.55 & 0.43 & 0.74 & 0.56 & 0.45 & 0.77 & 0.61 & 0.50 & 0.80 & 0.59 & 0.53 \\ 
  $val_{25}$ & 0.25 & 0.80 & 0.59 & 0.52 & 0.73 & 0.55 & 0.44 & 0.78 & 0.57 & 0.49 & 0.78 & 0.58 & 0.50 & 0.81 & 0.63 & 0.55 & 0.82 & 0.57 & 0.55 \\ 
  $val_{30}$ & 0.30 & 0.81 & 0.58 & 0.54 & 0.79 & 0.57 & 0.50 & 0.80 & 0.57 & 0.53 & 0.81 & 0.57 & 0.53 & 0.83 & 0.64 & 0.59 & 0.83 & 0.55 & 0.55 \\ 
  $val_{35}$ & 0.35 & 0.83 & 0.57 & 0.56 & 0.82 & 0.56 & 0.53 & 0.82 & 0.56 & 0.54 & 0.82 & 0.56 & 0.54 & 0.85 & 0.63 & 0.61 & 0.84 & 0.54 & 0.56 \\ 
  $val_{40}$ & 0.40 & 0.83 & 0.55 & 0.56 & 0.83 & 0.52 & 0.54 & 0.84 & 0.54 & 0.55 & 0.84 & 0.54 & 0.56 & 0.86 & 0.60 & 0.62 & 0.84 & 0.51 & 0.55 \\ 
  $val_{45}$ & 0.45 & 0.84 & 0.53 & 0.55 & 0.84 & 0.48 & 0.53 & 0.84 & 0.51 & 0.55 & 0.84 & 0.52 & 0.56 & 0.86 & 0.58 & 0.61 & 0.84 & 0.49 & 0.54 \\ 
  $val_{50}$ & 0.50 & 0.84 & 0.51 & 0.55 & 0.83 & 0.42 & 0.48 & 0.84 & 0.48 & 0.54 & 0.84 & 0.50 & 0.55 & 0.86 & 0.55 & 0.60 & 0.84 & 0.47 & 0.53 \\ 
  $val_{55}$ & 0.55 & 0.84 & 0.48 & 0.54 & 0.82 & 0.33 & 0.41 & 0.84 & 0.44 & 0.51 & 0.84 & 0.46 & 0.52 & 0.86 & 0.51 & 0.58 & 0.84 & 0.45 & 0.52 \\ 
  $val_{60}$ & 0.60 & 0.84 & 0.45 & 0.51 & 0.81 & 0.28 & 0.35 & 0.83 & 0.38 & 0.45 & 0.83 & 0.41 & 0.48 & 0.85 & 0.47 & 0.55 & 0.84 & 0.42 & 0.49 \\ 
  $val_{65}$ & 0.65 & 0.83 & 0.42 & 0.49 & 0.79 & 0.20 & 0.27 & 0.82 & 0.33 & 0.41 & 0.82 & 0.36 & 0.43 & 0.85 & 0.43 & 0.51 & 0.83 & 0.37 & 0.45 \\ 
  $val_{70}$ & 0.70 & 0.83 & 0.37 & 0.45 & 0.78 & 0.15 & 0.21 & 0.81 & 0.28 & 0.35 & 0.81 & 0.31 & 0.39 & 0.84 & 0.38 & 0.47 & 0.82 & 0.34 & 0.42 \\ 
  $val_{75}$ & 0.75 & 0.82 & 0.32 & 0.40 & 0.77 & 0.12 & 0.17 & 0.80 & 0.22 & 0.29 & 0.80 & 0.26 & 0.33 & 0.82 & 0.33 & 0.41 & 0.82 & 0.32 & 0.39 \\ 
  $val_{80}$ & 0.80 & 0.81 & 0.27 & 0.35 & 0.77 & 0.10 & 0.14 & 0.78 & 0.17 & 0.23 & 0.79 & 0.20 & 0.26 & 0.81 & 0.27 & 0.35 & 0.81 & 0.28 & 0.36 \\ 
  $val_{85}$ & 0.85 & 0.80 & 0.22 & 0.29 & 0.77 & 0.09 & 0.13 & 0.78 & 0.14 & 0.19 & 0.78 & 0.15 & 0.21 & 0.80 & 0.21 & 0.28 & 0.80 & 0.23 & 0.30 \\ 
  $val_{90}$ & 0.90 & 0.78 & 0.14 & 0.20 & 0.76 & 0.07 & 0.10 & 0.77 & 0.11 & 0.15 & 0.77 & 0.11 & 0.16 & 0.78 & 0.14 & 0.20 & 0.79 & 0.17 & 0.24 \\ 
  $val_{95}$ & 0.95 & 0.73 & 0.06 & 0.08 & 0.76 & 0.06 & 0.08 & 0.76 & 0.07 & 0.10 & 0.76 & 0.07 & 0.10 & 0.76 & 0.07 & 0.10 & 0.77 & 0.12 & 0.16 \\ 
  $val_{100}$ & 1.00 & 0.00 & 0.00 & 0.00 & 0.76 & 0.05 & 0.07 & 0.00 & 0.00 & 0.00 & 0.00 & 0.00 & 0.00 & 0.00 & 0.00 & 0.00 & 0.00 & 0.00 & 0.00 \\ 
   \hline
\end{tabular}
}

\end{ltable}

\section{Concluding Remarks on ML versus Statistical Methods for Flare Forecasting}

In order to assess the overall forecasting ability of ML {\it vs.} Statistical approaches in our dataset and problem definition, we employ the weighted-sum (WS) multicriteria ranking approach \citep{Greco:16}, using a composite index (CI) defined in Equation \ref{eq.D1}:

\begin{equation}
               {\rm CI}=\frac{1}{3}({\rm \frac{ACC - ACC_{min}}{ACC_{max} - ACC_{min}}}+{\rm \frac{TSS - TSS_{min}}{TSS_{max} - TSS_{min}}}+{\rm \frac{HSS - HSS_{min}}{HSS_{max} - HSS_{min}}}) \,.
\label{eq.D1}
\end{equation}

The {\rm CI} value is computed for $6 \times 21 = 126$ probabilistic classifiers using the set of methods \{MLP, LM, PR, LG, RF, SVM\} and a probability threshold grid of 5\%. Then, the 126 probabilistic classifiers are ranked in non-increasing values of the CI index. Notice the normalization which is done for {\rm ACC}, {\rm TSS} and {\rm HSS}, so that each metric over the set of alternatives takes values in the range $[0,1]$. The normalization is useful because the range of values for {\rm ACC} is different from the range of values for {\rm TSS} and {\rm HSS}. Also, notice that ${\rm {ACC}_{min}}$ is the minimum of {\rm ACC} over all 126 alternative models. Likewise, ${\rm {ACC}_{max}}$ is the maximum {\rm ACC} obtained over all 126 alternative models. Similar facts hold for ${\rm {TSS}_{min}}$, ${\rm {TSS}_{max}}$, ${\rm {HSS}_{min}}$ and ${\rm {HSS}_{max}}$. Analytically, Table \ref{tab:D1.CMXranking} presents the results of the multicriteria ranking approach for all methods used using various probability thresholds, especially for the $>$C1 flare forecasting case. Table \ref{tab:D2.MXranking} conveys a similar ranking of all methods developed in this paper, but for the $>$M1 flare prediction.

\begin{figure}[h]
\centering
\subfloat[Subfigure 1 list of figures text][ALL methods, $>$C1 flares]{
\includegraphics[width=0.475\textwidth]{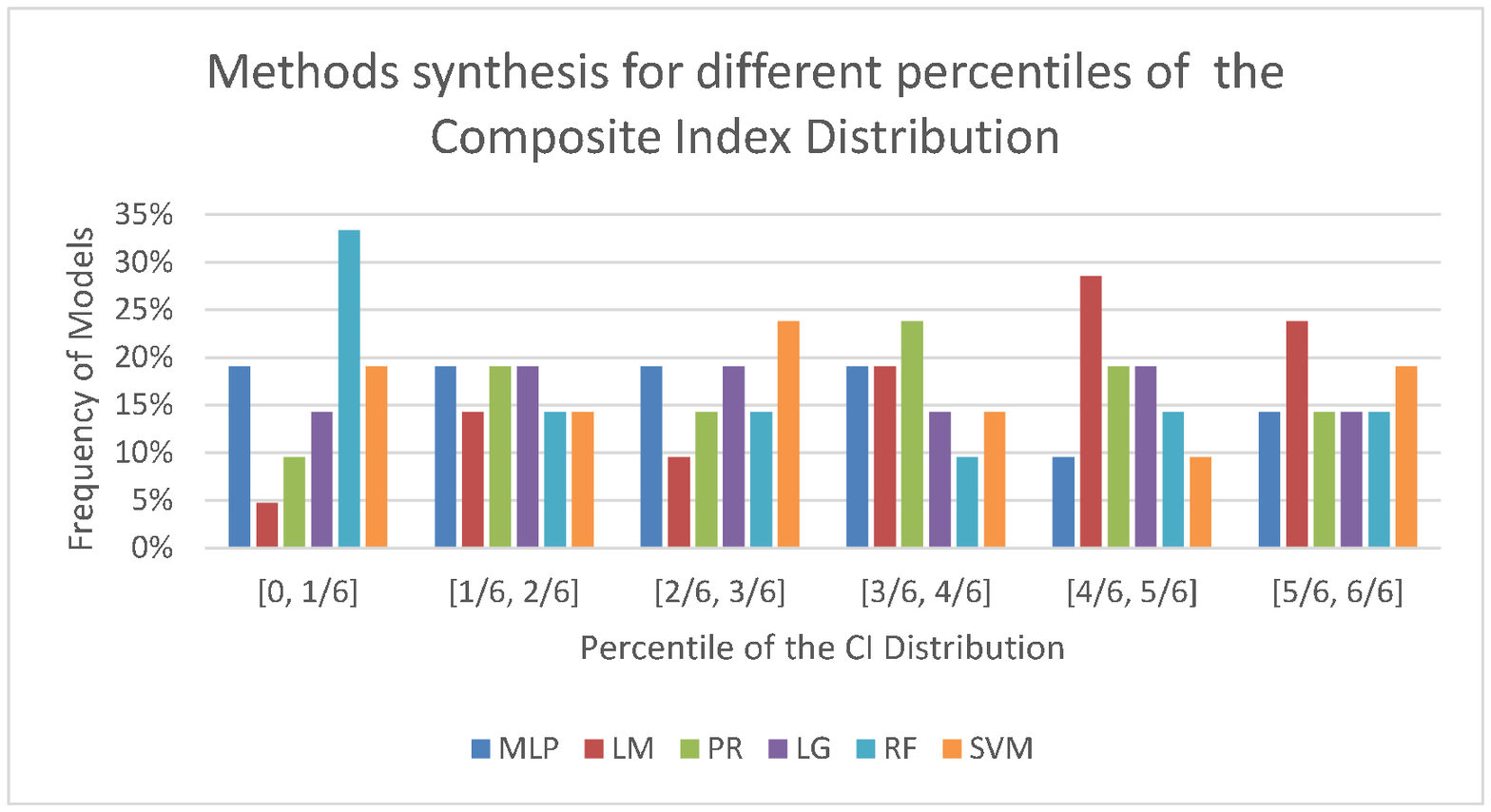}
\label{fig:aaaa}}
\subfloat[Subfigure 2 list of figures text][ML {\it vs.} statistical, $>$C1 flares]{
\includegraphics[width=0.475\textwidth]{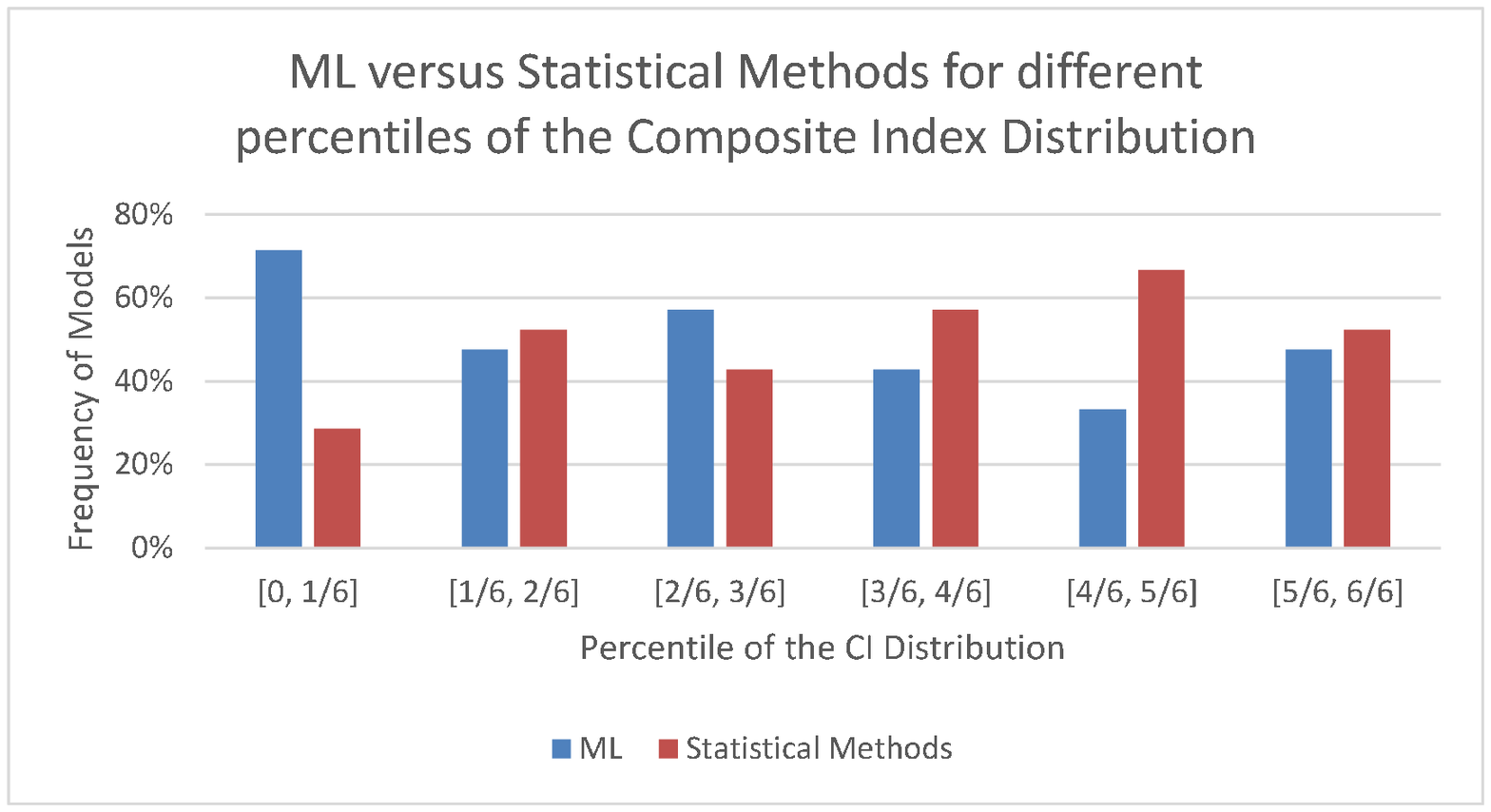}
\label{fig:bbbb}}
\qquad
\subfloat[Subfigure 3 list of figures text][ALL methods, $>$M1 flares]{
\includegraphics[width=0.475\textwidth]{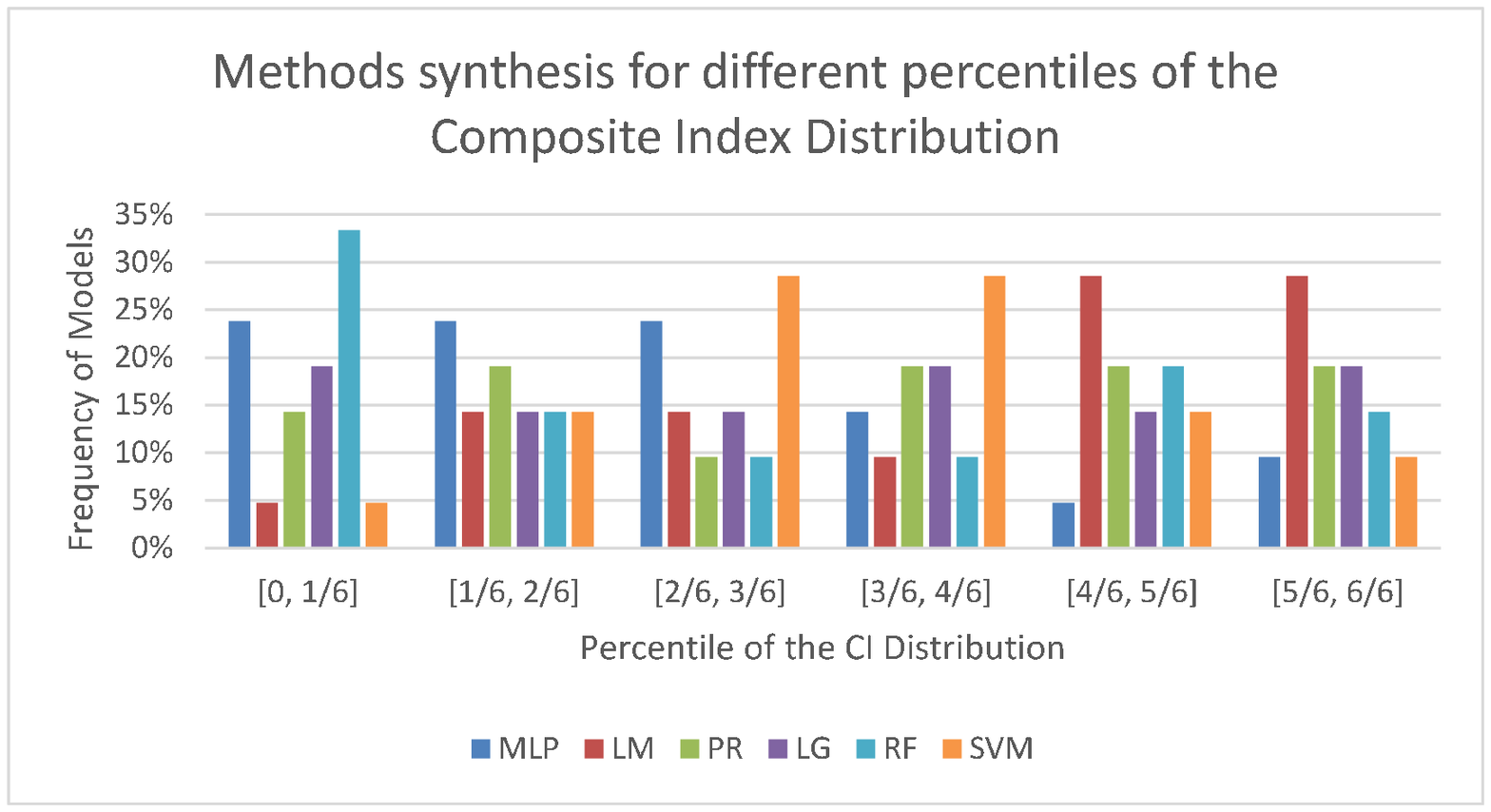}
\label{fig:cccc}}
\subfloat[Subfigure 3 list of figures text][ML {\it vs.} statistical, $>$M1 flares]{
\includegraphics[width=0.475\textwidth]{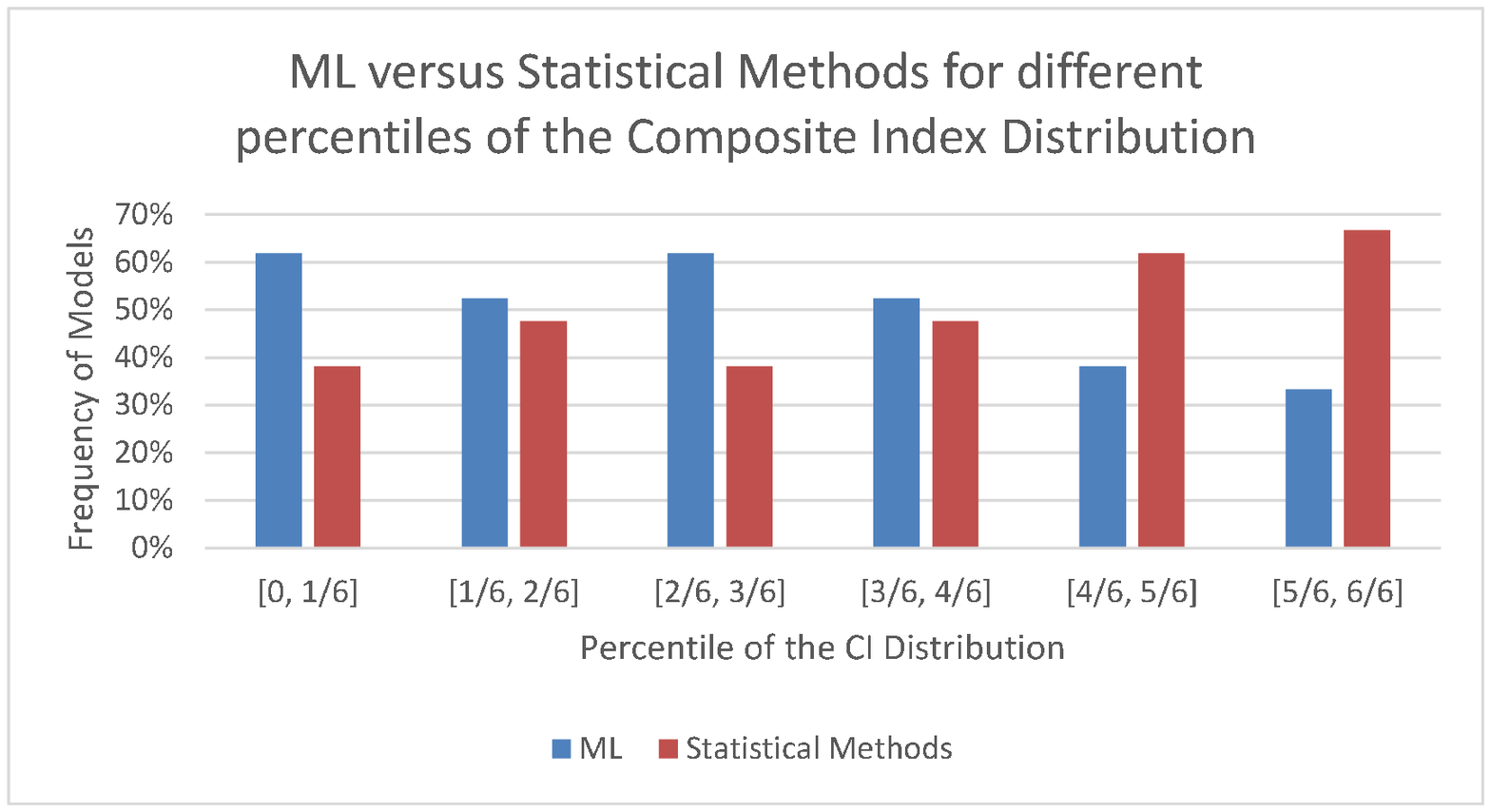}
\label{fig:dddd}}
\caption{Descriptive statistics on the frequency with which every forecasting method for any probability threshold presents itself to the top 100$\tau$ \% percentile of the CI distribution. Panels (a) and (c) describe frequencies for all methods and panels (b) and (d) group the results by category of methods ({\it e.g.} {\rm ML} {\it vs.} statistical methods). For example, for $>$C1 flares in panel (a), notice that the top 16.6\% methods are dominated by RF with a frequency of 7/21=33\%. Likewise, for $>$M1 flares in panel (c), notice that the top 16.6\% methods are again dominated by RF with a frequency of 7/21=33\%.}
\label{fig:D1}

\end{figure}

Figure \ref{fig:D1} summarizes the results shown in Tables \ref{tab:D1.CMXranking} and \ref{tab:D2.MXranking}, so that the differences between {\rm ML} and Statistical methods are highlighted ({\it e.g.} see Figure \ref{fig:D1}b and \ref{fig:D1}d). Similarly, conclusions for the merit of all methods developed in this paper can be drawn in Figures \ref{fig:D1}a and \ref{fig:D1}c. The top $100\tau$ percentile methods are the ones ranked in the corresponding positions of Tables \ref{tab:D1.CMXranking} and \ref{tab:D2.MXranking}. For example, the top 16.6\%(1/6) methods are the ones ranked in positions 1-21. For small values of $\tau$ one gets the best methods designated as the top $100\tau$\% methods. From Figure \ref{fig:D1} we see that both for $>$C1 and $>$M1 flares, the RF has the greatest frequency in the top 16.6\% percentile of methods, with a frequency of 33.3\%. This means, that in Tables \ref{tab:D1.CMXranking} and \ref{tab:D2.MXranking}, in positions 1-21, the RF method appears 7 times, in each Table. Also, in Figure \ref{fig:D1}b we see that for $>$C1 flares the top 16.6\% methods are of type {\rm ML} with a frequency 71\% ({\it versus} 29\% for Statistical Methods). Similarly, in Figure \ref{fig:D1}d {\rm ML} dominates in the top 16.6\% methods with a frequency 62\% ({\it versus} 38\% for Statistical Methods).

\begin{table}[htbp]
  \centering
  \caption{Ranking of all models with ML and statistical methods with the multicriteria WS method with respect to the three criteria: {\rm ACC}, {\rm TSS} and {\rm HSS} and using a weight vector equal to $w=[1/3, 1/3, 1/3]$ for $>$C1 flare forecasting. The methods are ranked in decreasing order of {\rm CI}, for varying levels of probability thresholds used, using a grid of 5\% for the probability thresholds. We notice what the six best positions of the ranking are covered by the RF method for various probability thresholds. }
    \begin{tabular}{rlrrlrrlr}
    \hline
    \multicolumn{1}{l}{rank} & model & \multicolumn{1}{l}{{\rm CI}} & \multicolumn{1}{l}{rank} & model & \multicolumn{1}{l}{{\rm CI}} & \multicolumn{1}{l}{rank} & model & \multicolumn{1}{l}{{\rm CI}} \\
    \hline
    \multicolumn{9}{c}{PANEL A  $>$C1 flares } \\
    \hline
    1     & RF-val35 & 0.983 & 22    & PR-val30 & 0.919 & 43    & MLP-val60 & 0.856 \\
    2     & RF-val40 & 0.983 & 23    & LG-val45 & 0.914 & 44    & SVM-val55 & 0.849 \\
    3     & RF-val45 & 0.971 & 24    & LM-val40 & 0.907 & 45    & SVM-val15 & 0.843 \\
    4     & RF-val30 & 0.970 & 25    & MLP-val50 & 0.907 & 46    & LG-val20 & 0.842 \\
    5     & RF-val50 & 0.953 & 26    & PR-val45 & 0.907 & 47    & PR-val55 & 0.840 \\
    6     & RF-val25 & 0.938 & 27    & MLP-val25 & 0.906 & 48    & LM-val25 & 0.835 \\
    7     & LG-val35 & 0.935 & 28    & SVM-val40 & 0.906 & 49    & PR-val20 & 0.828 \\
    8     & MLP-val35 & 0.934 & 29    & LM-val30 & 0.898 & 50    & SVM-val60 & 0.824 \\
    9     & PR-val35 & 0.933 & 30    & RF-val60 & 0.898 & 51    & MLP-val65 & 0.821 \\
    10    & MLP-val40 & 0.932 & 31    & LG-val25 & 0.892 & 52    & MLP-val15 & 0.821 \\
    11    & SVM-val25 & 0.932 & 32    & LG-val50 & 0.889 & 53    & LG-val60 & 0.818 \\
    12    & LG-val40 & 0.930 & 33    & SVM-val45 & 0.887 & 54    & RF-val15 & 0.810 \\
    13    & RF-val55 & 0.929 & 34    & PR-val25 & 0.886 & 55    & RF-val70 & 0.809 \\
    14    & SVM-val30 & 0.928 & 35    & RF-val20 & 0.886 & 56    & LM-val50 & 0.803 \\
    15    & MLP-val30 & 0.925 & 36    & MLP-val55 & 0.885 & 57    & SVM-val65 & 0.789 \\
    16    & PR-val40 & 0.925 & 37    & PR-val50 & 0.877 & 58    & PR-val60 & 0.787 \\
    17    & LM-val35 & 0.923 & 38    & MLP-val20 & 0.875 & 59    & MLP-val70 & 0.778 \\
    18    & MLP-val45 & 0.923 & 39    & LM-val45 & 0.869 & 60    & LG-val65 & 0.768 \\
    19    & LG-val30 & 0.921 & 40    & SVM-val50 & 0.867 & 61    & SVM-val70 & 0.751 \\
    20    & SVM-val35 & 0.920 & 41    & LG-val55 & 0.861 & 62    & LG-val15 & 0.748 \\
    21    & SVM-val20 & 0.920 & 42    & RF-val65 & 0.859 & 63    & RF-val75 & 0.747 \\
    \hline
    \multicolumn{9}{c}{PANEL B $>$C1 flares} \\
    \hline
    64    & PR-val65 & 0.738 & 85    & LG-val80 & 0.593 & 106   & PR-val95 & 0.420 \\
    65    & LM-val55 & 0.735 & 86    & LM-val15 & 0.592 & 107   & LG-val95 & 0.417 \\
    66    & MLP-val75 & 0.727 & 87    & RF-val85 & 0.587 & 108   & RF-val95 & 0.410 \\
    67    & PR-val15 & 0.724 & 88    & PR-val80 & 0.566 & 109   & LM-val90 & 0.405 \\
    68    & MLP-val10 & 0.720 & 89    & SVM-val90 & 0.552 & 110   & MLP-val95 & 0.400 \\
    69    & LM-val20 & 0.718 & 90    & LM-val70 & 0.537 & 111   & LM-val95 & 0.387 \\
    70    & LG-val70 & 0.715 & 91    & LG-val85 & 0.533 & 112   & LM-val5 & 0.372 \\
    71    & SVM-val75 & 0.709 & 92    & PR-val85 & 0.515 & 113   & LM-val100 & 0.370 \\
    72    & RF-val10 & 0.699 & 93    & MLP-val90 & 0.510 & 114   & SVM-val10 & 0.363 \\
    73    & PR-val70 & 0.682 & 94    & RF-val5 & 0.506 & 115   & LM-val0 & 0.291 \\
    74    & RF-val80 & 0.671 & 95    & MLP-val5 & 0.501 & 116   & SVM-val5 & 0.142 \\
    75    & SVM-val80 & 0.667 & 96    & RF-val90 & 0.500 & 117   & RF-val0 & 0.131 \\
    76    & MLP-val80 & 0.666 & 97    & LM-val75 & 0.494 & 118   & MLP-val0 & 0.000 \\
    77    & LM-val60 & 0.664 & 98    & LM-val10 & 0.488 & 119   & MLP-val100 & 0.000 \\
    78    & LG-val75 & 0.657 & 99    & LG-val90 & 0.484 & 120   & PR-val0 & 0.000 \\
    79    & PR-val75 & 0.622 & 100   & SVM-val95 & 0.484 & 121   & PR-val100 & 0.000 \\
    80    & LG-val10 & 0.621 & 101   & PR-val90 & 0.476 & 122   & LG-val0 & 0.000 \\
    81    & SVM-val85 & 0.613 & 102   & LM-val80 & 0.461 & 123   & LG-val100 & 0.000 \\
    82    & MLP-val85 & 0.596 & 103   & LG-val5 & 0.441 & 124   & RF-val100 & 0.000 \\
    83    & PR-val10 & 0.594 & 104   & LM-val85 & 0.431 & 125   & SVM-val0 & 0.000 \\
    84    & LM-val65 & 0.593 & 105   & PR-val5 & 0.427 & 126   & SVM-val100 & 0.000 \\
    \hline
    \end{tabular}%
  \label{tab:D1.CMXranking}%
\end{table}%

\begin{table}[htbp]
  \centering
  \caption{Same as Table \ref{tab:D1.CMXranking} but for $>$M1 flare forecasting.}
    \begin{tabular}{rlrrlrrlr}
    \hline
    \multicolumn{1}{l}{rank} & model & \multicolumn{1}{l}{{\rm CI}} & \multicolumn{1}{l}{rank} & model & \multicolumn{1}{l}{{\rm CI}} & \multicolumn{1}{l}{rank} & model & \multicolumn{1}{l}{{\rm CI}} \\
    \hline
    \multicolumn{9}{c}{PANEL A  $>$M1 flares } \\
    \hline
    1     & RF-val20 & 0.938 & 22    & PR-val25 & 0.839 & 43    & SVM-val25 & 0.774 \\
    2     & RF-val25 & 0.932 & 23    & MLP-val5 & 0.836 & 44    & MLP-val55 & 0.773 \\
    3     & RF-val15 & 0.927 & 24    & MLP-val35 & 0.836 & 45    & LG-val40 & 0.770 \\
    4     & RF-val30 & 0.912 & 25    & SVM-val10 & 0.834 & 46    & SVM-val30 & 0.756 \\
    5     & RF-val10 & 0.890 & 26    & LG-val30 & 0.827 & 47    & MLP-val60 & 0.754 \\
    6     & RF-val35 & 0.883 & 27    & LM-val20 & 0.823 & 48    & LM-val30 & 0.749 \\
    7     & MLP-val15 & 0.870 & 28    & LM-val10 & 0.822 & 49    & PR-val40 & 0.749 \\
    8     & MLP-val20 & 0.867 & 29    & MLP-val40 & 0.820 & 50    & LG-val45 & 0.743 \\
    9     & MLP-val10 & 0.866 & 30    & PR-val30 & 0.817 & 51    & SVM-val35 & 0.741 \\
    10    & MLP-val25 & 0.859 & 31    & SVM-val15 & 0.816 & 52    & RF-val55 & 0.736 \\
    11    & PR-val15 & 0.856 & 32    & RF-val45 & 0.815 & 53    & MLP-val65 & 0.735 \\
    12    & PR-val20 & 0.856 & 33    & RF-val5 & 0.809 & 54    & SVM-val40 & 0.725 \\
    13    & LG-val15 & 0.854 & 34    & MLP-val45 & 0.805 & 55    & PR-val45 & 0.713 \\
    14    & LG-val20 & 0.851 & 35    & LG-val35 & 0.799 & 56    & MLP-val70 & 0.713 \\
    15    & RF-val40 & 0.851 & 36    & LG-val5 & 0.796 & 57    & LG-val50 & 0.709 \\
    16    & MLP-val30 & 0.849 & 37    & SVM-val20 & 0.792 & 58    & SVM-val45 & 0.709 \\
    17    & LM-val15 & 0.849 & 38    & MLP-val50 & 0.790 & 59    & LM-val5 & 0.699 \\
    18    & SVM-val5 & 0.848 & 39    & PR-val5 & 0.786 & 60    & LM-val35 & 0.694 \\
    19    & PR-val10 & 0.845 & 40    & LM-val25 & 0.786 & 61    & RF-val60 & 0.693 \\
    20    & LG-val10 & 0.843 & 41    & PR-val35 & 0.782 & 62    & SVM-val50 & 0.692 \\
    21    & LG-val25 & 0.840 & 42    & RF-val50 & 0.778 & 63    & MLP-val75 & 0.689 \\
    \hline
    \multicolumn{9}{c}{PANEL B  $>$M1 flares } \\
    \hline
    64    & PR-val50 & 0.679 & 85    & LM-val50 & 0.571 & 106   & LM-val80 & 0.424 \\
    65    & SVM-val55 & 0.679 & 86    & SVM-val85 & 0.550 & 107   & LG-val90 & 0.413 \\
    66    & LG-val55 & 0.676 & 87    & PR-val70 & 0.546 & 108   & PR-val90 & 0.410 \\
    67    & MLP-val80 & 0.663 & 88    & LM-val55 & 0.546 & 109   & LM-val85 & 0.404 \\
    68    & SVM-val60 & 0.662 & 89    & RF-val75 & 0.543 & 110   & RF-val90 & 0.392 \\
    69    & PR-val55 & 0.648 & 90    & LG-val75 & 0.539 & 111   & LM-val90 & 0.381 \\
    70    & LM-val40 & 0.646 & 91    & MLP-val95 & 0.539 & 112   & PR-val95 & 0.378 \\
    71    & RF-val65 & 0.645 & 92    & LM-val60 & 0.521 & 113   & LG-val95 & 0.372 \\
    72    & SVM-val65 & 0.644 & 93    & PR-val75 & 0.516 & 114   & LM-val95 & 0.364 \\
    73    & LG-val60 & 0.644 & 94    & SVM-val90 & 0.512 & 115   & LM-val100 & 0.352 \\
    74    & MLP-val85 & 0.633 & 95    & LG-val80 & 0.506 & 116   & RF-val95 & 0.334 \\
    75    & SVM-val70 & 0.625 & 96    & RF-val80 & 0.492 & 117   & LM-val0 & 0.251 \\
    76    & PR-val60 & 0.612 & 97    & LM-val65 & 0.489 & 118   & MLP-val0 & 0.108 \\
    77    & LG-val65 & 0.607 & 98    & PR-val80 & 0.483 & 119   & MLP-val100 & 0.000 \\
    78    & SVM-val75 & 0.604 & 99    & LG-val85 & 0.466 & 120   & PR-val0 & 0.000 \\
    79    & LM-val45 & 0.603 & 100   & LM-val70 & 0.461 & 121   & PR-val100 & 0.000 \\
    80    & MLP-val90 & 0.594 & 101   & SVM-val95 & 0.450 & 122   & LG-val0 & 0.000 \\
    81    & RF-val70 & 0.593 & 102   & PR-val85 & 0.444 & 123   & LG-val100 & 0.000 \\
    82    & SVM-val80 & 0.579 & 103   & RF-val85 & 0.442 & 124   & RF-val100 & 0.000 \\
    83    & PR-val65 & 0.576 & 104   & LM-val75 & 0.441 & 125   & SVM-val0 & 0.000 \\
    84    & LG-val70 & 0.572 & 105   & RF-val0 & 0.433 & 126   & SVM-val100 & 0.000 \\
    \hline
    \end{tabular}%
  \label{tab:D2.MXranking}%
\end{table}%

\clearpage


\bibliographystyle{spr-mp-sola}
\bibliography{Bib_FCAST1}  

\IfFileExists{\jobname.bbl}{} {\typeout{}
\typeout{****************************************************}
\typeout{****************************************************}
\typeout{** Please run "bibtex \jobname" to obtain} \typeout{**
the bibliography and then re-run LaTeX} \typeout{** twice to fix
the references !}
\typeout{****************************************************}
\typeout{****************************************************}
\typeout{}}

\end{article} 

\end{document}